\documentclass[prc,aps,amsmath,amssymb,superscriptaddress,twocolumn,showpacs,floatfix,a4paper]{revtex4-2}

\usepackage{graphicx,colordvi}
\usepackage{dcolumn}
\usepackage{bm}
\usepackage{threeparttable}
\usepackage{xspace}
\usepackage{gensymb}
\usepackage{cases}
\usepackage{textcomp}
\usepackage{tabularx,booktabs}
\usepackage{epstopdf}
\usepackage{xcolor}
\usepackage{physics} 
\usepackage{mathtools, amssymb, amsthm, amsmath}

\usepackage{appendix}
\usepackage{csquotes}

\usepackage{xcolor}
\usepackage{tcolorbox}

\usepackage[dvipsnames]{xcolor}
\usepackage{textcase}

\begin{document}

\title{Microscopic description of the fission process including intrinsic excitations \\
Part I: $^{240}$Pu adiabatic and asymmetric fission path within the Schr\"odinger Collective Intrinsic Model}

\author{P. Carpentier}
\affiliation{%
 CEA, DAM, DIF, F-91297 Arpajon cedex, France
}%
\affiliation{%
  Universit\'e Paris-Saclay, CEA, LMCE, 91680, Bruyères-le-Châtel, France
}%

\author{N. Pillet}%
\affiliation{%
 CEA, DAM, DIF, F-91297 Arpajon cedex, France
}%
\affiliation{%
 Universit\'e Paris-Saclay, CEA, LMCE, 91680, Bruyères-le-Châtel, France
}%

\author{R.N. Bernard}%
\affiliation{%
CEA, DES, IRESNE, DER, SPRC, LEPh, 13115 Saint-Paul-lès-Durance, France
}%

\author{L.M. Robledo}%
\affiliation{%
Center for Computational Simulation, Universidad Polit\'ecnica de 
Madrid, Campus Montegancedo, 28660 Boadilla del Monte, Madrid, Spain
}%
\affiliation{Departamento  de F\'{\i}sica Te\'orica and CIAFF, 
Universidad Aut\'onoma de Madrid, 28049-Madrid, Spain}%

\author{D. Lacroix}
\affiliation{Universit\'e Paris-Saclay, CNRS/IN2P3, IJCLab, 91405 Orsay, France}

\author{N. Dubray}%
\affiliation{%
 CEA, DAM, DIF, F-91297 Arpajon cedex, France
}%
\affiliation{%
  Universit\'e Paris-Saclay, CEA, LMCE, 91680, Bruyères-le-Châtel, France
}%

\author{D. Regnier}%
\affiliation{%
 CEA, DAM, DIF, F-91297 Arpajon cedex, France
}%
\affiliation{%
 Universit\'e Paris-Saclay, CEA, LMCE, 91680, Bruyères-le-Châtel, France
}%

\author{W. Younes}%
\affiliation{%
Nuclear Science Division, Lawrence Berkeley National Laboratory, Berkeley, California 94720, USA
}%

\date{\today}

\begin{abstract}
This article is the first in a trilogy \cite{trilogy1,trilogy2,trilogy3} aimed at presenting the first practical implementation of the Schrödinger Collective-Intrinsic Model (SCIM) applied to nuclear fission. Within the SCIM framework, the many-body wave function explicitly couples collective motion to intrinsic excitations, necessitating sets of Hartree-Fock-Bogoliubov (HFB) configurations that remain continuous and regular across a broad deformation range, from the ground state to scission and beyond.

This paper focuses on constructing adiabatic HFB paths suitable for subsequent SCIM dynamical calculations. Standard constrained adiabatic paths often suffer from discontinuities and irregularities, which prevent the direct application of the formalism.
To address these challenges, we implement two recently proposed overlap-based protocols, the \enquote{Link} and \enquote{Drop} methods, and combine them into a new numerical procedure. This procedure generates a one-dimensional adiabatic path for the asymmetric fission of $^{240}$Pu. Additionally, we introduce a new collective coordinate, noted $c_\#$, derived from the sequence of overlaps between neighboring HFB states. This coordinate provides a more accurate measure of the distance along the path than the quadrupole moment alone. A comparison with the exact Gaussian Overlap Approximation confirms that the resulting adiabatic kernels exhibit properties consistent with the assumptions of the SCIM formalism.

The regularized path is then analyzed in the scission region. We identify characteristic structures in the proton and neutron chemical potentials, a pronounced neutron enrichment of the neck at scission, and fragment particle-number distributions displaying a strong odd-even staggering in the proton sector. Finally, using a microscopic fragment-separation procedure formulated in the canonical basis, we extract static scission properties including fragment deformation energies and both Coulomb and nuclear contributions to the fragment interaction energy. These results establish the adiabatic foundations required for future SCIM calculations with intrinsic excitations and provide a microscopic characterization of the scission region in $^{240}$Pu.
\end{abstract}

\maketitle

\section{Introduction}\label{sec1}

Nuclear fission is a highly collective process in which large-amplitude deformations of the nuclear mean field drive a nucleus toward scission, typically resulting in two fragments and, more rarely, three. Given the extensive shape rearrangements involved across a wide range of deformations, this process inherently requires theoretical frameworks capable of describing collective motion beyond the small-amplitude approximation.

Many microscopic approaches currently used to model fission are based on self-consistent mean-field theories \cite{REVIEW,Shalal}. These methods explicitly account for interacting nucleons while reducing the complexity of the full many-body quantum problem to a tractable set of effective degrees of freedom, a necessity for heavy nuclei such as actinides. Within this framework, the nuclear interaction is described using phenomenological effective forces, adjusted to reproduce selected nuclear properties. The most widely employed families of interactions include Skyrme interactions \cite{Sky1,Sky2,Sky3,Sky4}, Gogny interactions \cite{D1S1,D1S2,D1S3,TDGCM3,D2NatChap,DGGeoffrey}, and covariant energy density functionals \cite{ring2007,agbemava2014,yang2020}.

Among microscopic descriptions of fission, two major theoretical approaches can be distinguished. First, time-dependent mean-field methods, such as Time-Dependent Hartree-Fock (TDHF) \cite{SiSi,Seki} and its extensions that include pairing correlations \cite{SCAMPS,RBUL}, provide a fully dynamical description of nuclear evolution. While these approaches naturally incorporate one-body dissipation mechanisms during fission, they remain limited in their ability to describe the distribution of possible fragmentations. In particular, they struggle to satisfactorily account for fragment mass and charge yields or total kinetic energy distributions.
Second, approaches based on the Generator Coordinate Method (GCM), originally introduced by Hill and Wheeler \cite{hillwheeler1} and later extended by Griffin 
and Wheeler \cite{griffinwheeler1}, were specifically designed to describe collective motion through a superposition of constrained mean-field states characterized 
by collective coordinates. The time-dependent extension of this formalism, known as the Time-Dependent Generator Coordinate Method (TDGCM), has emerged as one of 
the most successful microscopic approaches for predicting fission fragment yields \cite{TDGCM3,bergerTDGCM,DynaGou,DynaRe,Rever}. In this framework, collective 
dynamics are governed by the evolution of collective amplitudes associated with deformed configurations, enabling a quantum description of the propagation toward 
scission. Despite these successes, the majority of TDGCM applications still rely on the adiabatic approximation: only the collective evolution of the lowest-energy 
configurations is explicitly considered, while intrinsic excitations are neglected. However, several experimental observables, such as structures in fragment 
kinetic-energy distributions and the reduction of odd-even staggering with increasing excitation energy, suggest that intrinsic excitations and pair breaking 
already play a role during the descent from saddle to scission. Furthermore, without an explicit treatment of fragment excitation energy, standard TDGCM 
approaches cannot provide fully consistent predictions for observables such as prompt neutron multiplicities.

These limitations have motivated ongoing efforts to bridge the gap between collective and intrinsic descriptions. Within TDHF-based approaches, recent developments have improved the treatment of collective phenomena \cite{LastVret,TaniMu,Bli}. For TDGCM-based theories, several extensions have been proposed to include dissipation through couplings with intrinsic excitations, including the work of Dietrich et al. \cite{dietrich2010microscopic}, Bernard et al. \cite{TBernard,Bernardetal}, and, more recently, Bertsh et al. \cite{Hagi} and Zhao  \textit{et al.} \cite{OldVret}. Although these formalisms establish a theoretical foundation for coupling collective and intrinsic degrees of freedom, no complete numerical implementation has yet been achieved for at least the first two.

The primary objective of this work is to provide the first application of the Schrödinger Collective-Intrinsic Model (SCIM), introduced in Ref. \cite{Bernardetal}. This framework extends TDGCM by explicitly incorporating intrinsic excitations into the many-body wave function and by introducing couplings between collective and single-particle degrees of freedom, couplings expected to play a major role during the descent from saddle to scission. This implementation has required overcoming several theoretical and numerical challenges, many of which were not anticipated in the original formal developments. In particular, special attention was devoted to ensuring the continuity and regularity of both adiabatic and excited HFB configurations along the fission path. Additionally, the selection and organization of excited states proved particularly delicate due to discontinuities arising in constrained calculations \cite{DRdisco}.

This article is the first of a trilogy dedicated to the implementation of the SCIM approach \cite{trilogy1,trilogy2,trilogy3}. It is devoted to the construction of regular adiabatic paths suitable for subsequent dynamical calculations. Section \ref{sec2} briefly reviews the SCIM formalism, along with the Hartree-Fock-Bogoliubov (HFB) ingredients required for its formulation using a two-center harmonic oscillator (HO) representation. 
Section \ref{sec3} presents the construction of discontinuity-free adiabatic paths using the newly developed \enquote{Link} and \enquote{Drop} methods \cite{carpentier2024}, as well as the definition of a new collective variable. A comparison with the exact Gaussian Overlap Approximation is also discussed. Finally, in section \ref{sectrue4}, we analyze the one-dimensional (1D) asymmetric fission path of $^{240}$Pu from the ground state to scission and beyond, including some properties of the fragments at scission. Section \ref{sec4} provides conclusions and perspectives.

\section{The Schr\"odinger Collective Intrinsic Model}\label{sec2}

\subsection{Reminder of the SCIM formalism}

We briefly review the key ideas underlying the SCIM equations to clarify its assumptions and establish the notation. Building on the pioneering works of \cite{kerman1,didong1,muther1,holzwarth1,tajima1}, the SCIM many-body wave function is introduced as a generalization of the GCM ansatz, explicitly incorporating intrinsic excitations:
\begin{eqnarray}\label{eq:SCIM1}
\ket{\Psi_{  SCIM}} = \int dq^n f_0(q) \ket{\Phi_0(q)} +  \quad \quad \quad \quad \nonumber \\ \sum_{i=1}^{N}
 \int dq^n f_i(q) \ket{\Phi_i(q)}
\end{eqnarray}
where the mean-field wave functions $\ket{\Phi_0(q)}$ and $\ket{\Phi_i(q)}$ represent the adiabatic and intrinsic excited HFB states, respectively. 

As in the GCM formalism, we assume the continuity hypothesis with respect to the collective coordinates $q_i$ holds. We apply the variational principle to the total energy $\mathcal{E}_{  SCIM}$, defined as:
\begin{eqnarray}\label{eq:SCIM2}
\mathcal{E}_{  SCIM} = \frac{\bra{\Psi_{  SCIM}} \hat{H} \ket{\Psi_{  SCIM}}}{\bra{\Psi_{  SCIM}}\ket{\Psi_{  SCIM}}}
\end{eqnarray}
where $\hat{H} = \hat T + \hat V$ is the Hamiltonian that describes the fissioning system. It contains both a kinetic and a nuclear interaction terms.
In the present study, this last term is taken as the phenomenological two-body Gogny interaction. 

The variational principle leads to:
\begin{eqnarray}\label{eq:SCIM3}
\delta \left[ \sum_{j=0}^{N}\sum_{i=0}^{N} \int dq^n \int dq'^n f_j^*(q) \times \quad \quad \quad \quad \quad \quad \quad \quad \right. \\ \left. \bra{\Phi_{j}(q)}|\hat H-\mathcal{E}_{  SCIM}\ket{\Phi_{i}(q')} f_i(q') \right] = 0. \nonumber 
\end{eqnarray}
In order to derive the collective-intrinsic Hamiltonian, Eq.\eqref{eq:SCIM3} is first rewritten using the center-of-mass and relative collective coordinates $\bar q$ and $s$, whose expressions are respectively:
\begin{eqnarray}\label{eq:SCIM4}
\bar q  = \frac{q+q'}{2}, \qquad s = \frac{q-q'}{2}.
\end{eqnarray}

Here, the relative coordinate $s$ differs by a factor of two from the convention used in Ref. \cite{bernard1}, a choice that simplifies the subsequent derivations. This yields:
\begin{eqnarray}\label{eq:SCIM5}
\displaystyle \delta \big(\frac{\displaystyle 1}{\displaystyle 2^n} \sum_j\sum_i \int d \bar{q}^{n}  \int ds^{n}  f_j^*(\bar q - s) \times \quad \quad \quad  \\\bra{\Phi_{j}(\bar q - s)} \hat{H} - \mathcal{E}_{  SCIM} \ket{\Phi_{i}(\bar q + s)}f_i(\bar q + s) \big) = 0. \nonumber
\end{eqnarray}
A Taylor expansion of the functions $f_i(\bar q+s)$ around $\bar q$ gives:
\begin{equation}\label{eq:SCIM6}
\begin{array}{lcl}
\displaystyle f_i(\bar q + s) &=&  \displaystyle \sum_{\alpha_1 = 0}^{+\infty}...\sum_{\alpha_n = 0}^{+\infty} \frac{\displaystyle s_1^{\alpha_1}...s_n^{\alpha_n}}{\displaystyle \alpha_1!...\alpha_n!} ~ \frac{\displaystyle \partial^{\alpha_1 +...+\alpha_n}}{\displaystyle \partial q_1^{\alpha_1} ... \partial q_n^{\alpha_n}} ~ f_i(\bar q) \\ 
&=&  \displaystyle e^{\displaystyle s \frac{\partial}{\partial q}} f_i(\bar q).
\end{array}
\end{equation}
The quantities $\displaystyle \big(s \frac{\partial}{\partial q}\big)^k = \big(s_1 \frac{\partial}{\partial q_1} + ... + s_n \frac{\partial}{\partial q_n}\big)^k$ appearing in Eq.\eqref{eq:SCIM6} are expanded using the multinomial theorem:
\begin{eqnarray}\label{eq:SCIM7}
\big(s \frac{\partial}{\partial q}\big)^k = \sum_{\alpha_1 + ... + \alpha_n = k} \begin{pmatrix}
k \\ \alpha_1,...,\alpha_n \end{pmatrix} s_1^{\alpha_1}\frac{\partial^{\alpha_1}}{\partial q_1^{\alpha_1}}...\frac{\partial^{\alpha_n}}{\partial q_n^{\alpha_n}}.
\end{eqnarray}
The integrals in Eq. \eqref{eq:SCIM5} can then be expressed as:
\begin{eqnarray}\label{eq:SCIM8}
\displaystyle \int d\bar{q}^n  \int ds^n  f_j^*(\bar q - s) \times \quad \quad \quad \quad \notag \\ \bra{\Phi_{j}(\bar q - s)}\hat{H} - \mathcal{E}_{SCIM} \ket{\Phi_{i}(\bar q + s)}f_i(\bar q + s) \notag \\ \displaystyle = \int d\bar{q}^n  \int ds^n \sum_{k = 0}^{+\infty}\frac{(-1)^k}{k!}
 \sum_{\alpha_1 + ... + \alpha_n = k}  \begin{pmatrix}
k \\ \displaystyle \alpha_1,...,\alpha_n                                                   \end{pmatrix}  \\ \displaystyle \nonumber s_1^{\alpha_1}\frac{\displaystyle \partial^{\alpha_1}}{\displaystyle \partial q_1^{\alpha_1}}...\frac{\displaystyle \partial^{\alpha_n}}{\displaystyle \partial q_n^{\alpha_n}}[f_j^*(\bar q)] \times \quad \quad \quad \quad\\ \bra{\Phi_{j}(\bar q - s)} \hat H - \mathcal{E}_{  SCIM} \ket{\Phi_{i}(\bar q + s)}e^{\displaystyle s \frac{\partial}{\partial q}}f_i(\bar q).\nonumber 
\end{eqnarray}
After repeated integrations by parts, assuming that all relevant quantities and their derivatives vanish at infinity, we finally obtains:
\begin{eqnarray}\label{eq:SCIM9}
\delta\big(\frac{1}{2^n}\sum_j \sum_i \int ds^n \int d\bar q^n f_j^*(\bar q)e^{\displaystyle s \frac{\partial}{\partial q}} \times \quad \quad \quad \\ \bra{\Phi_{j}(\bar q - s)} \hat{H} -\mathcal{E}_{SCIM} \ket{\Phi_{i}(\bar q + s)}e^{\displaystyle s \frac{\partial}{\partial q}}f_i(\bar q)\big) = 0. \nonumber 
\end{eqnarray}
Then, applying variations with respect to $f_j^*(\bar q)$ yield:
\begin{eqnarray}\label{eq:SCIM10}
\displaystyle \sum_i \int ds^n e^{\displaystyle s \frac{\partial}{\partial q}}\bra{\Phi_{j}(\bar q - s)}\hat H - \mathcal{E}_{SCIM} \ket{\Phi_{i}(\bar q + s)} \times \nonumber \\ e^{\displaystyle s \frac{\partial}{\partial q}}f_i(\bar q) = 0. \quad \quad \quad \quad \quad \quad \quad 
\end{eqnarray}
for all $\bar q$ and $j$. For compactness purposes, we introduce the SCIM Hamiltonian and norm kernels:
\begin{eqnarray}\label{eq:SCIM11}
\begin{cases}
\displaystyle  \mathcal{H}_{ji}(\bar q,s) = \bra{\Phi_{j}(\bar q - s)}\hat H \ket{\Phi_{i}(\bar q + s)}\\
\displaystyle  \mathcal{N}_{ji}(\bar q,s) = \bra{\Phi_{j}(\bar q - s)}\ket{\Phi_{i}(\bar q + s)}
\end{cases}
\end{eqnarray}
Their explicit calculation are discussed in Appendices \ref{overlap} and \ref{hamiltonkernel}, in the case of a two-center HO oscillator basis. Then, Eq. \eqref{eq:SCIM10} becomes:
\begin{eqnarray}\label{eq:SCIM12}
\displaystyle \sum_i \int ds^n e^{\displaystyle s \frac{\partial}{\partial q}}
 \left( \mathcal{H}_{ji}(\bar q,s) \qquad \qquad \qquad \qquad \right.\\ \left. - \mathcal{E}_{SCIM} \mathcal{N}_{ji}(\bar q,s) \right)
e^{\displaystyle s \frac{\partial}{\partial q}}f_i(\bar q) = 0.  \nonumber  
\end{eqnarray}
The next step involves evaluating the action of the exponential operators $e^{s \frac{\partial}{\partial q}}$ on these kernels using the Symmetric Ordered Products of Operators (SOPO) technique \cite{rsbook}, as detailed in Appendix \ref{appendixa}. After straightforward manipulations, we obtain:
\begin{eqnarray}\label{eq:SCIM13}
\begin{cases}
\displaystyle e^{\displaystyle s \frac{\partial}{\partial q}} \mathcal{H}_{ji}(\bar q,s) e^{\displaystyle s \frac{\partial}{\partial q}} = \sum_{k=0}^{+\infty}\frac{1}{k!}\left[\mathcal{H}_{ji}(\bar q,s)\left(s\frac{\partial}{\partial q}\right)\right]^{(k)} \\ 
\displaystyle e^{\displaystyle s \frac{\partial}{\partial q}} \mathcal{N}_{ji}(\bar q,s) e^{\displaystyle s \frac{\partial}{\partial q}} = \sum_{k=0}^{+\infty}\frac{1}{k!}\left[\mathcal{N}_{ji}(\bar q,s)\left(s\frac{\partial}{\partial q}\right)\right]^{(k)}.
\end{cases}
\end{eqnarray}
Since the dependence on the relative coordinate $s$ is fully contained in the SOPO expansion, integration over $s$ can be carried out term by term. This leads to the definition of the special Hamiltonian and norm kernels:
\begin{eqnarray}\label{eq:SCIM14}
\begin{cases}
\displaystyle \mathcal{\bar H}_{ji}(\bar q) = \int ds \sum_{k=0}^{+\infty}\frac{1}{k!}\left[\mathcal{H}_{ji}(\bar q,s)\left(s\frac{\partial}{\partial q}\right)\right]^{(k)} \\
\displaystyle \mathcal{\bar N}_{ji}(\bar q) = \int ds \sum_{k=0}^{+\infty}\frac{1}{k!}\left[\mathcal{N}_{ji}(\bar q,s)\left(s\frac{\partial}{\partial q}\right)\right]^{(k)}.\end{cases}
\end{eqnarray}
We deduce for Eq. \eqref{eq:SCIM12} the following expression:
\begin{eqnarray}\label{eq:SCIM15}
\sum_i \mathcal{\bar H}_{ji}(\bar q) f_i(\bar q)  = \mathcal{E}_{SCIM} \sum_i \mathcal{\bar N}_{ji}(\bar q)f_i(\bar q).
\end{eqnarray}
for all $\bar{q}$ and $j$. In a compact notation, it reduces to:
\begin{eqnarray}\label{eq:SCIM16}
\mathcal{\bar H}f = E \mathcal{\bar N} f.
\end{eqnarray}
As in the GCM framework, we assume that the inverse square root of the norm kernel $\mathcal{\bar N}$ can be constructed. The inversion procedure used in the present work is detailed in Appendix \ref{appendixb} for the case of one collective degree of freedom, introducing a truncation at second order in SOPOs. Its numerical implementation will be discussed in the third article of the trilogy \cite{trilogy3}, section II.A.

Introducing the transformed collective amplitudes $g$
\begin{eqnarray}\label{eq:SCIM19}
g = \mathcal{\bar N}^{1/2}f,
\end{eqnarray}
and defining the collective-intrinsic Hamiltonian $\mathcal{H}_{SCIM}$
\begin{eqnarray}\label{eq:SCIM18}
\mathcal{H}_{SCIM} = \mathcal{\bar N}^{-1/2}~\mathcal{\bar H}~\mathcal{\bar N}^{-1/2},
\end{eqnarray}
we show that Eq.\eqref{eq:SCIM16} reduces to a standard eigenvalue problem:
\begin{eqnarray}\label{eq:SCIM17}
\mathcal{H}_{SCIM} ~g = \mathcal{E}_{SCIM} ~g.
\end{eqnarray}
In the SCIM approach, we assume a truncation at second order in SOPOs, such that the Hamiltonian $\mathcal{H}_{SCIM}$ takes the form:
\begin{eqnarray}\label{hscim}
\mathcal{H}_{SCIM} \left(\bar q\right) = V_{SCIM}(\bar q) + \left[D_{SCIM}(\bar q)\frac{\partial}{\partial q}\right]^{(1)}  \\ +
\left[B_{SCIM}(\bar q)\frac{\partial}{\partial q}\right]^{(2)}. \nonumber 
\end{eqnarray}
The zeroth-order term $V_{SCIM}$ and second-order term $B_{SCIM}$ are analogous to the potential and inertia terms obtained in TDGCM within the Gaussian Overlap Approximation (GOA). The first-order term $D_{SCIM}$ is specific to the SCIM. It is referred to as the dissipation tensor and accounts for dissipative dynamical correlations arising from the coupling to intrinsic excitations. The explicit expressions for these dynamical ingredients are provided in Appendix \ref{appendixc} for the case of a single collective degree of freedom.\\

Extending the static SCIM approach to its dynamical formulation \cite{younes1} is straightforward. Specifically, we assume that the time dependence of the many-body wave function is introduced solely through the weights $f$, such that:
\begin{eqnarray}\label{eq:TDGCM1}
\ket{\Psi_{  SCIM}(t)} = \int dq^n f_0(q,t) \ket{\Phi_0(q)} +  \quad \quad \quad \quad \nonumber \\ \sum_{i=1}^{N}
 \int dq^n f_i(q,t) \ket{\Phi_i(q)},
\end{eqnarray}
where the weights $f_i(q,t)$ are determined by applying a variational principle to the integral action $\mathcal{A}(t,t')$ between two times $t$ and $t'$, built from the the SCIM wave function $\ket{\Psi_{  SCIM}(t)}$:
\begin{eqnarray}\label{eq:TDGCM2}
\mathcal{A} (t,t') = \int_{t}^{t'} \frac{\bra{\Psi_{  SCIM}(t'')}}{\bra{\Psi_{  SCIM}(t'')}\ket{\Psi_{  SCIM}(t'')}^{1/2}}   \notag \\ \left( \hat{H} - i \hbar \frac{\partial}{\partial t''}\right) 
\frac{\ket{\Psi_{  SCIM}(t'')}}{\bra{\Psi_{  SCIM}(t'')}\ket{\Psi_{  SCIM}(t'')}^{1/2}}.
\end{eqnarray}
Proceeding as in the static case and assuming that $\ket{\Psi_{\mathrm{SCIM}}(t'')}$ is normalized, we finally obtain a local Schrödinger-like equation, which can be written in the following compact form:
\begin{eqnarray}\label{dynamicsscim}
\mathcal{H}_{SCIM}(\bar q) g(\bar q, t) = i \hbar\frac{\partial}{\partial t}g(\bar q,t),
\end{eqnarray}
where the time-dependent quantities $g(\bar q, t)$ are defined by:
\begin{eqnarray}\label{eq:SCIM19_bis}
\begin{pmatrix}
    g_0(\bar q,t) \\ \vdots \\ g_N(\bar q,t) 
\end{pmatrix} = \mathcal{\bar N}^{1/2}\begin{pmatrix}
    f_0(\bar q,t) \\ \vdots \\ f_N(\bar q,t) 
\end{pmatrix}.
\end{eqnarray}
Solving Eq. \eqref{dynamicsscim} not only provides access to fission lifetimes and fragment yields corrected for excitation effects, but also allows the evaluation of observables directly related to the energy balance at scission. In particular, the explicit inclusion of intrinsic excitations paves the way for a more consistent evaluation of total kinetic energies and prompt neutron multiplicities.

\subsection{HFB approach with a 2-center representation for fission studies}

The HFB method \cite{Bogo1,Bogo2,Bogo3,Bogo4,RaS} is a fundamental component of the SCIM framework, as it provides the static reference states used in the definition of the wave function \eqref{eq:TDGCM1}. Its principle involves describing the nucleus using a quasiparticle (QP) product wave function $\ket{\Phi_{HFB}}$ that naturally incorporates pairing correlations as:
\begin{eqnarray}\label{ctwo_0}
 \ket{\Phi_{HFB}} = \mathcal{N}\prod_i \xi_i \ket{0}
\end{eqnarray}
with
\begin{eqnarray}\label{ctwo_0b}
\forall i, \; \xi_i \ket{\Phi_{HFB}} = 0.
\end{eqnarray}
Here, $\ket{0}$ denotes the particle vacuum, while the set $\{\xi_i\}$ represents the QP annihilation operators. The factor $\mathcal{N}$ is the normalization constant of the HFB wave function. Eq. (\ref{ctwo_0b}) explicitly shows that $\ket{\Phi_{HFB}}$ is the vacuum associated with the QP annihilation operators. 

Using an orthonormal basis associated with the particle creation and annihilation operators ${c_k^+}$ and ${c_k}$, respectively, the QP operators are defined through the Bogoliubov transformation:
\begin{eqnarray}\label{ctwo_6}
\xi_i = \sum_{k} U^*_{ki} c_k + V^*_{ki} c_k^+ 
\end{eqnarray}
such that
\begin{eqnarray}\label{ctwo_6b}
\begin{pmatrix} \xi \\ \xi^+ \end{pmatrix} =  \begin{pmatrix} U^+ & V^+ \\ V^T & U^T \end{pmatrix} \begin{pmatrix} c \\ c^+ \end{pmatrix}.
\end{eqnarray}
The transformation defined in Eq. \eqref{ctwo_6b} is the most general linear mixing of particle creation and annihilation operators. The matrix constructed from $U$ and $V$ is known as the Bogoliubov transformation matrix $B$:
\begin{eqnarray}\label{bogtrans}
B =  \begin{pmatrix} U^+ & V^+ \\ V^T & U^T \end{pmatrix}.
\end{eqnarray}
For this transformation to be canonical, i.e. invertible and preserving fermionic anticommutation relations, the matrix $B$ must be unitary. This imposes the following relations between the matrices $U$ and $V$:
\begin{eqnarray}\label{HFB_conditions}
\begin{cases}
 U^+U + V^+V = I \\
 UU^+ + V^*V^T = I
\end{cases} \qquad
\begin{cases}
 V^TU + U^TV = 0\\
 UV^+ + V^*U^T = 0.
\end{cases}
\end{eqnarray}
In Eq. (\ref{ctwo_0}), the unknowns are the matrix elements of $U$ and $V$, submitted to the constraints of Eq. (\ref{HFB_conditions}). These matrices are determined variationally by minimizing the energy functional:
\begin{eqnarray}
\mathcal{E} = \frac{\bra{\Phi_{HFB}}\hat H \ket{\Phi_{HFB}}}{\bra{\Phi_{HFB}}\ket{\Phi_{HFB}}}.
\end{eqnarray}
In general, the HFB method is linked to a symmetry breaking context. Thus, the Hamiltonian $\hat H$ is replaced by an Hamiltonian
$\hat H_c$ containing contraints to obtain an HFB solution with the good particle numbers on average and associated with specific values of some multipole moments.
\begin{eqnarray}\label{ctwo_56}
 \hat H_c = \hat H + \mu_N \hat{N}+ \mu_Z \hat{Z}+ \hat Q_\alpha \sum_\alpha \lambda_\alpha \hat Q_\alpha, 
\end{eqnarray}
Where the quantities $\mu_i$ are the chemical potentials related to the neutron and proton number, and $\lambda_i$ are the Lagrange multipliers for the desired constrained collective coordinates including the center of mass $Q_{10}$.

\subsubsection{Symmetries}

Within mean-field based approaches, the quality of the description typically improves when more symmetries are broken at the mean-field level and subsequently restored. However, in practice, symmetry breaking and restoration come with a significant numerical cost. Thus, the choice of symmetries must strike a balance between physical accuracy and computational feasibility.

In fission studies, pairing correlations play a central role. At the mean-field level, they are most conveniently incorporated using the HFB formalism. As directly follows from Eq.~(\ref{ctwo_6}), the HFB wave function does not have a well-defined particle number and thus intrinsically breaks particle-number symmetry. Although proton and neutron numbers are conserved on average through specific constraints, their associated fluctuations may remain significant.

Additionally, fission predominantly occurs along a symmetry axis, except possibly near the first barrier. This motivates the conservation of axial symmetry in the present work. The HFB states are thus chosen as eigenstates of the angular momentum projection operator $\hat J_z$, associated with the quantum number $\Omega$.
Time-reversal symmetry $\hat T$ is also imposed throughout the calculations. In addition to significantly reducing the numerical complexity, this assumption naturally restricts the present framework to even-even nuclei, whose ground-state wave functions exhibit time-reversal invariance.
Finally, experimental observations indicate that fission can proceed through either symmetric or asymmetric fragmentations. To account for both possibilities, parity symmetry is allowed to be broken.

Consistent with these symmetry assumptions, QP operators with a given $\Omega$ are constructed as linear combinations of creation and annihilation operators carrying the same $\Omega$. Furthermore, the QP operators ${\xi_i}$ are paired by the time-reversal operator $\hat T$, such that the HFB wave function takes the form:
\begin{eqnarray}
\ket{\Phi_{HFB}} = \mathcal{N}\prod_i \xi_i \bar \xi_i \ket{0}, \qquad \bar \xi_i = \hat T \xi_i \hat T^{-1}.
\end{eqnarray}
The quantities $U$ and $V$ being real matrices, the QP operators can be written as:
\begin{eqnarray}\begin{cases}
\displaystyle \xi^{\Omega}_i = \sum_{k \in \Omega} U_{ki} c_k + V_{\bar ki} \bar c_k^{+}
\\ \displaystyle \bar \xi^{\Omega}_i = \sum_{\bar k \in \Omega} U_{\bar k \bar i} \bar c_k + V_{k \bar i} c_k^{+}.
\end{cases}
\end{eqnarray}
Applying explicitly the time-reversal operator leads to the conditions:
\begin{eqnarray}
\begin{cases} U_{ki} = U_{\bar k \bar i} \\ - V_{\bar k i} = V_{k \bar i} \end{cases}\qquad \forall k, i.
\end{eqnarray}
Therefore, the full HFB transformation
\begin{eqnarray}
\begin{pmatrix} \xi \\ \bar \xi \\ \xi^+ \\ \bar \xi^+ \end{pmatrix} =  \begin{pmatrix} U^T & 0 & 0 & V^T \\ 0 & U^T & - V^T & 0 \\ 0 & V^T & U^T & 0 \\ -V^T & 0 & 0 & U^T   \end{pmatrix} \begin{pmatrix} c \\ \bar c \\ c^+ \\ \bar c^+ \end{pmatrix}
\end{eqnarray}
reduces to:
\begin{eqnarray}
\begin{pmatrix} \xi \\ \bar \xi^+ \end{pmatrix} =  \begin{pmatrix} U^T & V^T \\ - V^T & U \end{pmatrix} \begin{pmatrix} c \\  \bar c^+ \end{pmatrix}.
\end{eqnarray}
The matrices $U$ and $V$ also exhibit an $\Omega$-block structure:
\begin{eqnarray}
U = \begin{pmatrix}U^{\Omega_0} & & 0 \\
 & \ddots & \\ 0 & & U^{\Omega_n} \end{pmatrix} \qquad V = \begin{pmatrix}V^{\Omega_0} & & 0 \\
 & \ddots & \\ 0 & & V^{\Omega_n} \end{pmatrix}.
\end{eqnarray}
Since the proton-neutron pairing is neglected in the present study, creation and annihilation operators carrying different isospin are not mixed. The HFB wave function therefore factorizes into independent neutron and proton parts:
\begin{eqnarray}\label{ctwo_7}
 \ket{\Phi_{HFB}} = \ket{\Phi_{HFB}^{\tau_n}} \otimes \ket{\Phi_{HFB}^{\tau_p}}.
\end{eqnarray}
The corresponding reduced Bogoliubov transformation reads:
\begin{eqnarray}\label{ctwo_8}
\begin{pmatrix}\xi^\tau \\ \bar \xi^{\tau+} \end{pmatrix} = \begin{pmatrix}U^{\tau T} & V^{\tau T} \\ -V^{\tau T} & U^{\tau T} \end{pmatrix} \begin{pmatrix}c^\tau \\ \bar c^{\tau+} \end{pmatrix}.
\end{eqnarray}
Finally, each block matrix $U^{\tau \Omega}$ and $V^{\tau \Omega}$ possesses a specific internal structure given by the Bloch--Messiah theorem \cite{rsbook}:
\begin{eqnarray}\label{blochmessiah}
U^{\tau \Omega} = \bar D^{\tau \Omega} \bar u^{\tau \Omega} \bar C^{\tau \Omega}, \qquad V^{\tau \Omega} = \bar D^{\tau \Omega}\bar v^{\tau \Omega}\bar C^{\tau \Omega}.
\end{eqnarray}
where $\bar D^{\tau \Omega}$ and $\bar C^{\tau \Omega}$ are orthogonal matrices, while $\bar u^{\tau \Omega}$ and $\bar v^{\tau \Omega}$ are diagonal matrices:
\begin{eqnarray}
\bar u^{\tau \Omega} = \begin{pmatrix}
             u_1 & 0 & \ldots & 0\\
             0 & \ddots & 0 & \vdots \\
             \vdots & 0 & \ddots & 0 \\
             0 & \ldots & 0 & u_n \\
             \end{pmatrix} \qquad \text{with} \; 1\geq u_i \geq 0,
\end{eqnarray}
and
\begin{eqnarray}
\bar v^{\tau \Omega} = \begin{pmatrix}
             v_1 & 0 & \ldots & 0\\
             0 & \ddots & 0 & \vdots \\
             \vdots & 0 & \ddots & 0 \\
             0 & \ldots & 0 & v_n \\
             \end{pmatrix} \qquad \text{with} \; 1\geq v_i \geq -1.
\end{eqnarray}

\subsubsection{Single-particle bases and 2-center representation for fission studies}\label{twocenter}

As introduced above, the HFB QPs are linear combinations of creation and annihilation operators associated with an orthonormal single-particle basis. Common choices include spherical, axial, or triaxial HO bases \cite{TBerger}. Such bases are convenient for computing interaction matrix elements and the associated fields, particularly when using the Gogny interaction.
In principle, all orthonormal bases are equivalent. In practice, however, numerical calculations require truncated bases adapted to the system under study. In fission studies, the highly deformed shapes of heavy nuclei demand many states in a truncated axial HO basis, which significantly slows down computations. To mitigate this
effect, the 2-center HO representation was developed by J. F. Berger \cite{TBerger} and is employed in the new HFB3 solver used for our study \cite{dubray,dubraycode}.

Before introducing the 2-center representation, we recall the standard axial HO basis. Its spatial wave functions factorize into perpendicular and longitudinal parts:
\begin{eqnarray}
\psi_{(m,n_\perp,n_z)}(\vec{r},b_r,b_z) = \phi_{(m,n_\perp)}(\vec{r}_\perp,b_r)\varphi_{n_z}(z,b_z),
\end{eqnarray}
where $\phi_{(m,n_\perp)}$ is a cylindrical HO function labeled by the projection $m$ of the angular momentum and the perpendicular principal quantum number $n_\perp$. The function $\varphi_{n_z}$ is a one-dimensional HO function characterized by the quantum $n_z$ representing the quanta number along $z$. Each function depends on an oscillator length, $b_r$ and $b_z$ respectively, defining a specific orthonormal basis. These functions explicitly read:
\begin{eqnarray}
\varphi_{n_z}\left(z,b_z\right) = \frac{1}{\sqrt{2^{n_z}b_z n_z!\sqrt{\pi}}}e^{-\frac{1}{2}(\frac{z}{b_z})^2}H_{n_z} \left(\frac{z}{b_z} \right),
\end{eqnarray}
and
\begin{eqnarray}
\phi_{(m,n_\perp)}\left(\vec{r}_\perp,b_r \right) = \frac{1}{b_r\sqrt{\pi}}\sqrt{\frac{n_\perp !}{(n_\perp + |m|)!}}e^{\displaystyle im\phi}\left(\frac{r_\perp}{b_r}\right)^{\displaystyle |m|} \\ \times L_{n_\perp}^{|m|}\left[(\frac{r_\perp}{b_r})^2 \right]e^{\displaystyle -\frac{1}{2}\left(\frac{r_\perp}{b_r}\right)^{2}}. \quad \quad \quad \nonumber
\end{eqnarray}
Here, $H_{n_z}$ and $L_{n_\perp}^{|m|}$ are Hermite and generalized Laguerre polynomials, respectively. Including spin and isospin degrees of freedom, the full single-particle states $\ket{\Psi_i}$ have for expression:
\begin{eqnarray}\label{ctwo_11}
\ket{\Psi_i} = \ket{\Psi(m_i,n_{\perp_i},n_{z_i})} \otimes \ket{s_i} \otimes \ket{\tau_i}.
\end{eqnarray}

The 2-center HO representation extends a truncated HO basis by translating the longitudinal functions along the $z$-axis by $\pm d$:
\begin{eqnarray}
\displaystyle \psi_{(m,n_\perp,n_z)}\left(\vec{r},b_r,b_z,\pm d\right) = \phi_{(m,n_\perp)}\left(\vec{r}_\perp,b_r\right) \qquad \qquad  \\ \times \varphi_{n_z}\left(z \pm d,b_z\right). \nonumber
\end{eqnarray} 
This set is generally non-orthogonal, hence the \enquote{2-center representation} name rather than \enquote{2-center basis}. 
In the HFB3 solver, it is orthonormalized using the diagonalization of the overlap matrix:
\begin{eqnarray}
O_{ij} = \bra{\psi_i}\ket{\psi_j} = \delta_{s_i s_j} \delta_{\tau_i \tau_j} \int d\vec{r}\, \psi_i^*\left(\vec{r}\right) \psi_j\left(\vec{r}\right).
\end{eqnarray}
Diagonalizing $O$ with an orthogonal matrix $Q$ gives:
\begin{eqnarray}
Q^T O Q = \text{diag}\left(\lambda\right).
\end{eqnarray}
Selecting eigenvalues $\lambda_k > \varepsilon$ allows defining an orthonormal set:
\begin{eqnarray}
\ket{p_k} = \frac{1}{\sqrt{\lambda_k}} \sum_j Q_{jk} \ket{\psi_j} = \sum_j M_{jk} \ket{\psi_j},
\end{eqnarray}
where $M$ is the transition matrix from the 2-center representation to the orthonormal basis. 

A subtlety arises when expressing operators $\hat A$ in the two-center representation. Considering for example one-body operators, their expression in an orthonormal basis is:
\begin{eqnarray}
\hat A = \sum_{kl} \bra{p_k} \hat A \ket{p_l} p_k^+ p_l.
\end{eqnarray}
where $p_k^+$ and $p_k$ are the associated creation and annihilation operators.

Returning to the original 2-center functions, one finds:
\begin{eqnarray}
\hat A = \sum_{ij} \bra{\psi_i} \hat A \ket{\psi_j} \big(\sum_{k} M_{ik} p_k^+ \big) \big(\sum_l M_{jk}
p_l\big).
\end{eqnarray}
Since $M^T$ is not the inverse of $M$: 
\begin{equation}
\sum_k M_{ik} p_k^+ \ne c_i^+.
\end{equation}
However, the matrix $M^{-1}$ satisfies:
\begin{equation}
M^{-1}_{ij} = \sqrt{\lambda_i} Q_{ij}.
\end{equation}
It naturally defines a biorthogonal representation 
with operators $\{\tilde c_i^+,\tilde c_i\}$ and functions $\{\tilde\psi_i\}$ such that
\begin{eqnarray} \bra{\tilde \psi_i}\ket{\psi_j} = \sum_k\sum_l M_{ik} M^{-1}_{lj}\bra{p_k} \ket{p_l} = \delta_{ij}. \end{eqnarray}
The subtlety lies in the fact that an operator $\hat A$, when expressed in the 2-center representation, must be written in terms of creation and annihilation operators belonging to the corresponding biorthogonal basis:
\begin{eqnarray}
\hat A = \sum_{ij} \bra{\psi_i} \hat A \ket{\psi_j} \tilde c_i^+ \tilde c_j.
\end{eqnarray}
In the following, all creation and annihilation operators are denoted generically as $\{c_i^+\}$ and $\{c_i\}$ for simplicity, unless distinctions are necessary.

\subsubsection{The Gogny effective interaction}

\noindent In the present work, we employ the D1S Gogny interaction for our applications \cite{D1S1, D1S2, D1S3, TDGCM3}. This interaction is a local effective two-body nucleon-nucleon force whose finite-range central terms enable a fully self-consistent treatment of pairing correlations within the HFB formalism. Its analytical expression is composed of three terms plus the Coulomb contribution:
\begin{equation}
\begin{array}{lcl}
\hat V &=& \displaystyle \sum_{i=1}^2 e^{-(|\vec{r}_1 - \vec{r}_2|)^2/\mu_i^2}(W_i + B_iP_\sigma - H_iP_\tau -M_iP_\sigma P_\tau)
\\  &+&  \displaystyle t_3(1+x_0 P_\sigma)\delta(\vec{r}_1 - \vec{r}_2)[\rho(\frac{\vec{r}_1 + \vec{r}_2}{2})]^\alpha
\\ &+&  \displaystyle iW_{LS}\overleftarrow{\nabla}_{12}\delta(\vec{r}_1 - \vec{r}_2)\wedge \overrightarrow{\nabla}_{12}.(\vec{\sigma}_1 + \vec{\sigma}_2)
\\ &+&  \displaystyle e^2 \frac{\delta_{\tau_p \tau_p'}}{|\vec{r}_1 -  \vec{r}_2|}
\end{array}
\end{equation}
where the operators $P_\sigma$ and $P_\tau$ denote the spin- and isospin-exchange operators, respectively. From first to the last lines, the four contributions correspond to the central, density-dependent, spin-orbit term, and Coulomb terms.

It should be noted that an additional term, $-\vec{P}^2/ 2MA$, is often included in the Hamiltonian to account for one- and two-body center-of-mass corrections. Furthermore, the Coulomb contribution to the energy has been evaluated using the Slater approximation \cite{Slater}, consistently with the original fitting procedure of the D1S Gogny interaction.

\section{Construction of adiabatic paths}\label{sec3}

In TDGCM based approaches, potential energy surfaces (PES) obtained from constrained HFB calculations are a fundamental building block for describing nuclear large-amplitude collective motion. These surfaces enable a microscopic exploration of the deformation landscape associated with the nucleus collective degrees of freedom and serve as the foundation for constructing the collective dynamics.

In this work, the first implementation of the SCIM formalism is deliberately restricted to a one-dimensional description of the collective motion. The dynamics are thus assumed to follow a single dominant adiabatic trajectory through the multidimensional deformation landscape. Consequently, the construction of continuous and regular adiabatic paths becomes a central aspect of the approach.

Here, the simplest procedure considered, denoted as $\mathcal{P}_{20}$, involves performing constrained HFB calculations at successive values of $Q_{20}$. This type of construction is widely used in the literature and typically yields smooth potential energy paths. In particular, retro-propagation algorithms are useful to built automatically adiabatic PES continuous in energy \cite{DRdisco}. However, even a smooth potential energy path in terms of $Q_{20}$ may still exhibit abrupt reorganizations of the underlying HFB wave functions, associated with strong local distortions of the overlap metric. Such effects are particularly problematic within the SCIM framework, where the collective dynamics explicitly depend on the overlap and Hamiltonian kernels, as well as their derivatives with respect to the collective coordinate.

In this section, we discuss first the state continuity and regularity issues. Then, we detail the two recently proposed protocols \enquote{Link} and \enquote{Drop}
\cite{carpentier2024} which allow to cure discontinuities in states and reach 
scission as well as the Coulomb valley. Finally, a procedure $\mathcal{\tilde P}_{20}$ used to generate continuous PES in state is proposed. The nuclei $^{16}$O and $^{240}$Pu serve as testing cases.

\subsection{Continuity and regularity issues}

\begin{figure}
\centering
\includegraphics[width=1.0\linewidth]{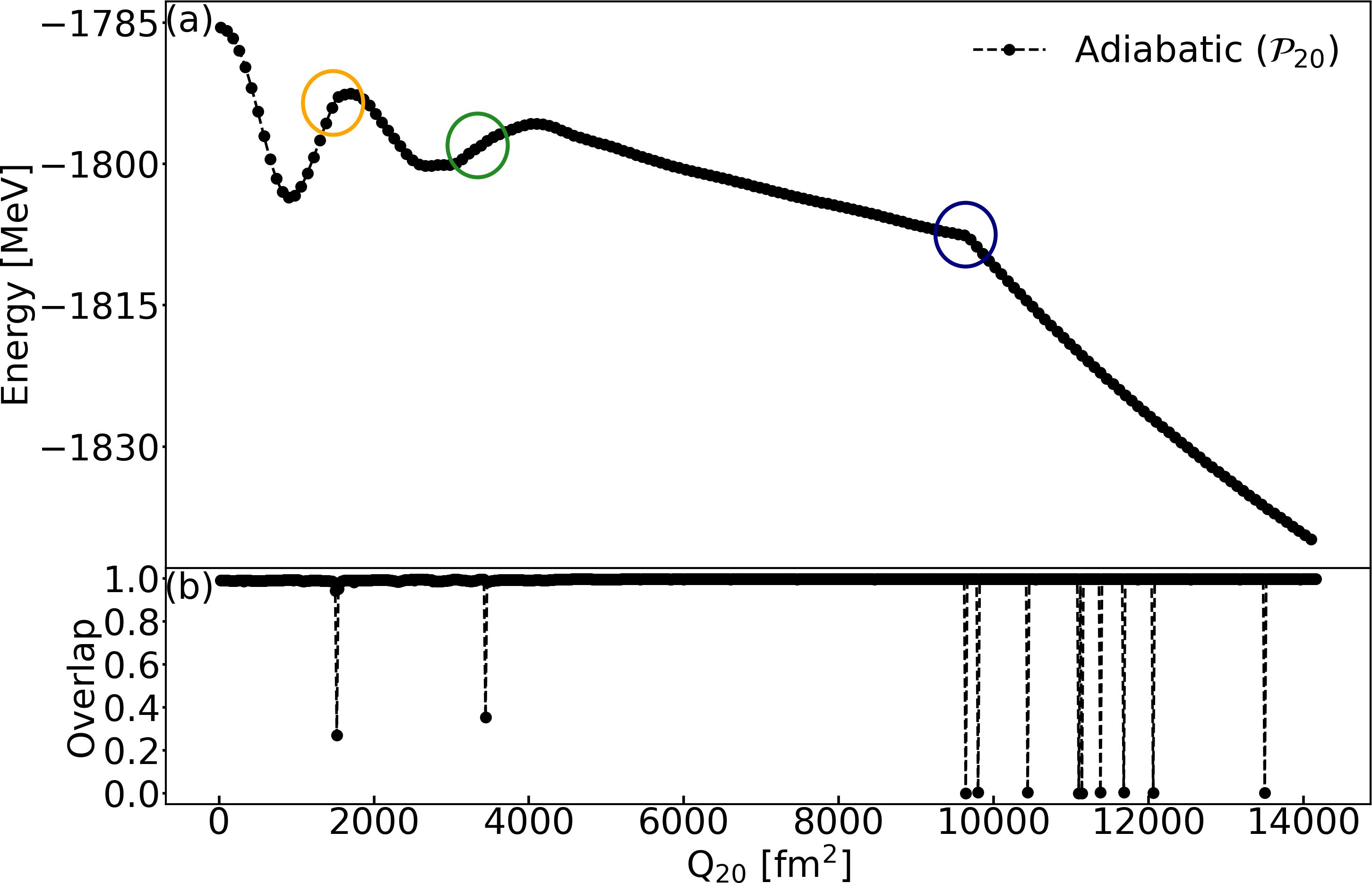}
\caption{Panel (a): Asymmetric path in $^{240}$Pu obtained with the $\mathcal{P}_{20}$ procedure. Colored circles highlight discontinuities appearing in the path. Panel (b): Overlap between neighboring HFB states along the path.}
\label{ctwo_41}
\end{figure}

\begin{figure}
\centering
\includegraphics[width=1.0\linewidth]{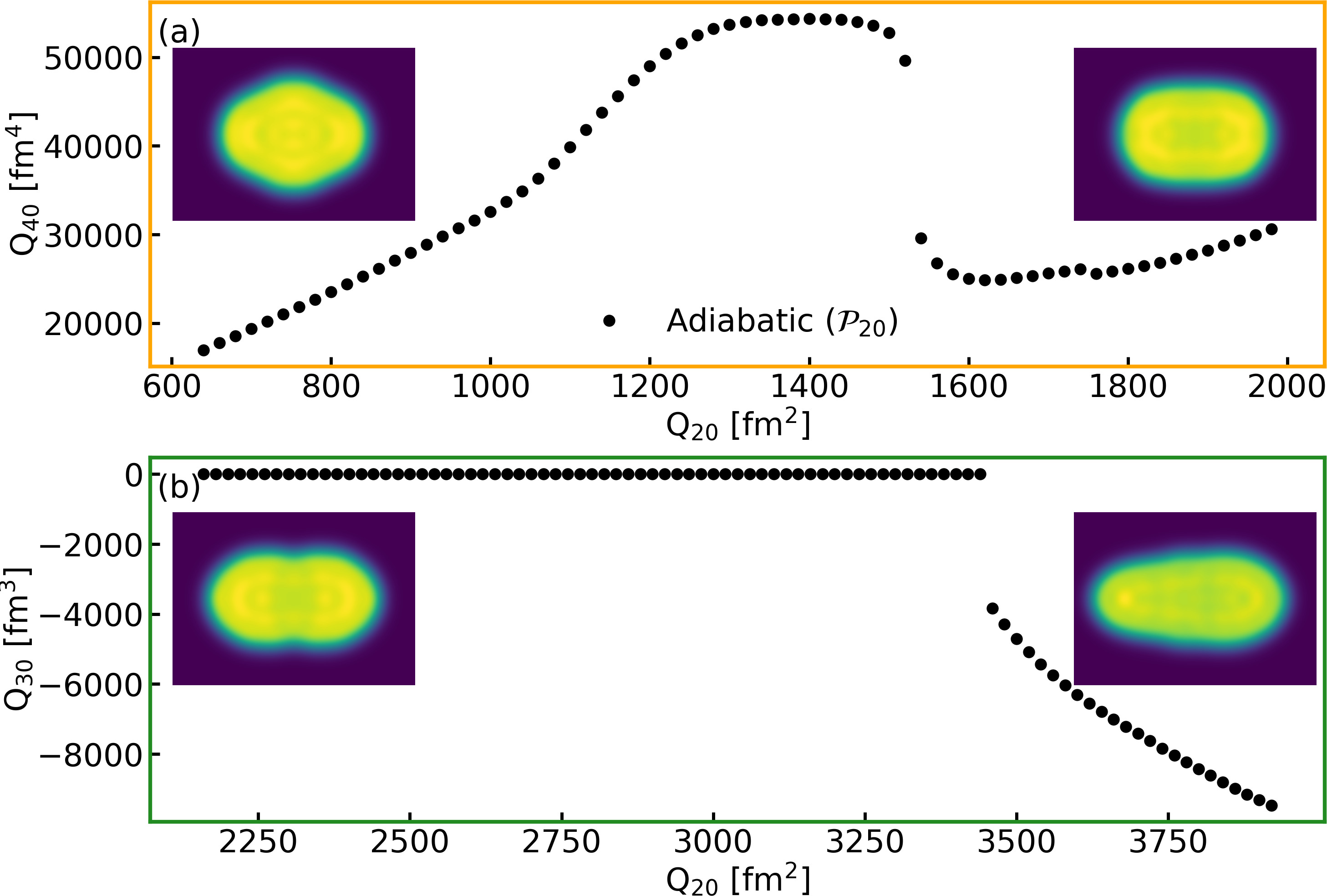}
\caption{Panel (a): Evolution of the hexadecapole moment $Q_{40}$ with respect to the quadrupole deformation $Q_{20}$, in the first barrier area (orange circled discontinuity).
Panel (b): Same as panel (a) but for the octupole moment $Q_{30}$
in the area characterizing the transition to asymmetry (green circled discontinuity). Calculations have been done using the $\mathcal{P}_{20}$ procedure (in black).
Total nucleon densities are shown on both sides of the discontinuities.}
\label{ctwo_43}
\end{figure}

\begin{figure}
\centering
\includegraphics[width=1.0\linewidth]{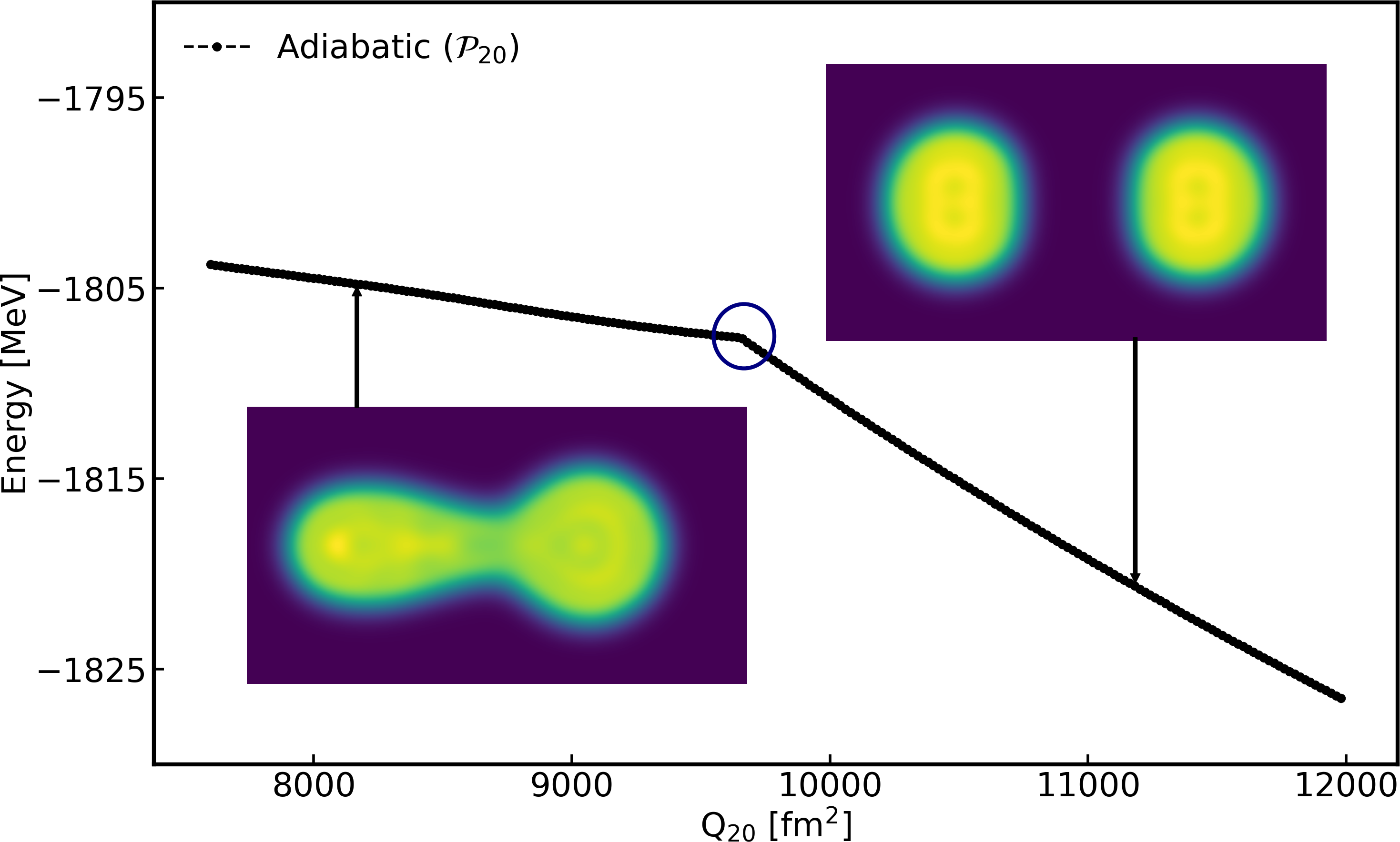}
\caption{Crossing between the fission and fusion valleys obtained with the $\mathcal{P}_{20}$ procedure in $^{240}$Pu. 
Nucleon total densities are displayed on both sides of the discontinuity.}
\label{ctwo_42}
\end{figure}

The first limitation of the $\mathcal{P}_{20}$ procedure relates to the 
continuity of the many-body states along the adiabatic path.
While the energy evolves smoothly as a function of the constrained 
deformation  $Q_{20}$, the underlying HFB states can still exhibit 
abrupt structural rearrangements, inducing discontinuities in the overlap between two neighboring states.
This behavior is illustrated in FIG. \ref{ctwo_41}, panel (a), which shows a continuous in energy
asymmetric fission path (the lowest in energy) of $^{240}$Pu obtained 
using the $\mathcal{P}_{20}$ procedure.
In panel (b), the overlaps between two neighboring HFB states along the path are displayed. A detailed discussion on their evaluation is provided in Appendix \ref{overlap}. We observe values that differ significantly from one.
The origin of these discontinuities stems from the variational nature of the $\mathcal{P}_{20}$ procedure.
While the quadrupole moment $Q_{20}$ is constrained, all other collective degrees of freedom remain unconstrained and are thus free to reorganize in order to further minimize the total energy.
Consequently, the minimization process can drive the system from one local minimum in the energy landscape to another, resulting in abrupt changes in the underlying HFB wave functions along the path.
Such discontinuities typically occur in regions where additional collective degrees of freedom become active, such as in the first barrier (orange circle), during parity breaking (green circle) or near scission (blue circle).
As illustrated in FIG. \ref{ctwo_43}, panel (a), the first discontinuity (the orange circled one) is primarily characterized by an abrupt variation in the hexadecapole moment $Q_{40}$ (black curve), while the second discontinuity (the green circled one) exhibits a rapid change in the octupole moment $Q_{30}$ (black curve).
The total nucleon densities, shown on both sides of these discontinuities, quantify the associated evolution of the fissioning nucleus shape.
A particularly striking example is shown in FIG. \ref{ctwo_42}, where the standard $\mathcal{P}_{20}$ procedure produces a crossing between the fission and fusion valleys (the blue circled discontinuity).
The corresponding local densities reveal abrupt transitions between configurations corresponding to a highly prolate deformed compound nucleus and two already separated oblate fragments, characteristic of Coulomb repulsion in the fusion valley as the nuclei approach each other.
These observations demonstrate that the quadrupole deformation $Q_{20}$ alone is insufficient to globally describe the geometry of the fission path.
In traditional approaches, this limitation is often addressed by introducing additional collective coordinates, such as the neck operator $Q_{\mathrm{neck}}$ proposed in Ref. \cite{QneckWardaFirst}.
However, increasing the number of constrained multipole moments quickly leads to high-dimensional collective spaces and computationally demanding procedures, without necessarily ensuring the continuity required by the SCIM formalism.\\

\begin{figure}
\centering
\includegraphics[width=1.0\linewidth]{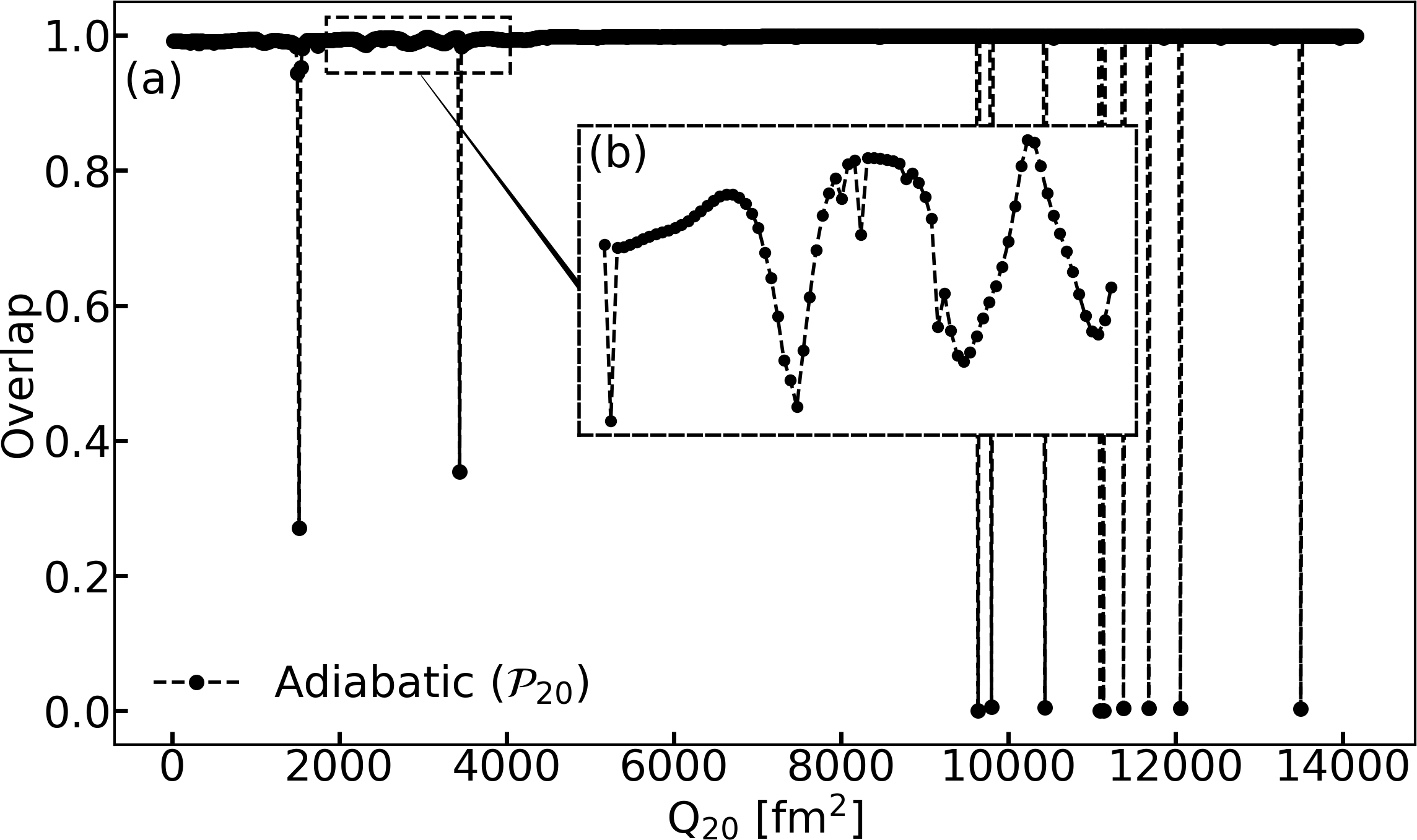}
\caption{Overlap between neighboring HFB states along the asymmetric fission path of $^{240}$Pu, using the $\mathcal{P}_{20}$ procedure.}
\label{ctwo_46}
\end{figure}

Beyond continuity, the SCIM approach also imposes strict regularity requirements on the overlap and Hamiltonian kernels.
Indeed, reducing the Hill-Wheeler equation to an eigenvalue equation involves derivatives of these kernels with respect to the collective coordinate.
As a result, even moderate local irregularities can significantly impact the resulting collective inertia and the associated dynamics (see, e.g., Bonche \textit{et al.} \cite{Bonche1990}).
An important property of the adiabatic kernels is that the Hamiltonian kernel closely follows the overlap kernel, in accordance with the approximate relation:
\begin{equation}\label{ctwo_55_short}
\bra{\Phi(q)} \hat H \ket{\Phi(q')}
\simeq
\bra{\Phi(q)}\ket{\Phi(q')}
E\left(\frac{q+q'}{2}\right).
\end{equation}
The relation \eqref{ctwo_55_short} demonstrates that the regularity of the overlap kernel largely determines the regularity of the entire SCIM formalism (see discussion in section \ref{link}). Thus, the choice of the collective coordinate becomes a critical issue. The limitations of the quadrupole moment $Q_{20}$ as a global collective coordinate, from the perspective of regularity, are illustrated in FIG.~\ref{ctwo_46}.
Although neighboring HFB states are separated by a constant step in $Q_{20}$, the corresponding overlap distances fluctuate significantly along the path.
As a result, kernels expressed in terms of $Q_{20}$ inherit spurious local distortions that arise purely from the parametrization of the path.

The previous discussion demonstrates that constructing adiabatic paths suitable for the SCIM formalism requires more than just a smooth energy landscape.
The path must also remain continuous and sufficiently regular in the space of HFB states.
Rather than indefinitely increasing the number of constrained multipole moments, the strategy adopted in this work involves directly controlling the geometry of the path through overlap constraints.
This idea forms the foundation of the new methods introduced in the following sections, namely the \enquote{Link} and \enquote{Drop} methods \cite{carpentier2024}.
The inclusion of overlap constraints in the HFB formalism is straightforward theoretically.
Projectors are added to the reference states, each associated with new Lagrange multipliers in the constrained Hamiltonian $ \hat H_c$ which becomes:
\begin{eqnarray}\label{ctwo_56_bis}
 \hat H_c = \hat H + \mu_N \hat N + \mu_p \hat P + \sum_\alpha \lambda_\alpha \hat Q_\alpha \\ + \sum_\beta \gamma_\beta \ket{\Phi_\beta}\bra{\Phi_\beta}, \nonumber
\end{eqnarray}
where the set of $\gamma_\beta$ defines the new Lagrange multipliers and $\ket{\Phi_\beta}$ the associated reference states.

Analyzing Eq.(\ref{ctwo_56_bis}), it is obvious that the gradient method is perfectly suited to tackle these new constraints. Indeed, an HFB state can be written as \cite{RaS}:
\begin{eqnarray}
\displaystyle \displaystyle \ket{\Phi(Z)} = \bra{\Phi_0}\ket{\Phi(Z)}\exp\left({\displaystyle \sum_{k<k'}Z_{kk'}\xi_k^+ \xi_{k'}^+} \right)\ket{\Phi_0}.
\end{eqnarray}
Thanks to the Thouless theorem and in the case of time-reversal invariance, the gradient of those constraints simply reads as:
\begin{eqnarray}\label{ctwo_57}
\frac{\partial}{\partial Z_{kk'}} \gamma_\beta \vert \bra{\Phi(Z)}\ket{\Phi_\beta}\vert_{Z=0}^2 \qquad \qquad \qquad \qquad \\ = 2 \sum_\beta \gamma_\beta \bra{\Phi_\beta} \xi_i^+ \bar \xi_j^+ \ket{\Phi_0} \bra{\Phi_\beta} \ket{\Phi_0}. \nonumber 
\end{eqnarray}
The method for evaluating the quantities $\bra{\Phi_\beta} \xi_i^+ \bar \xi_j^+ \ket{\Phi}$ appearing in Eq.~\eqref{ctwo_57} is detailed in the Appendix \ref{overlap2QP}.
The constraints associated with overlaps of HFB wave functions can be treated similarly to other customary constraints.

In the following, we detail the implementation of the Link and Drop methods, whose ultimate goal is to construct a continuous, adiabatic, and asymmetric 1D path suitable for fission applications within the SCIM formalism.
Illustrations will be provided for the rigid nucleus $^{16}$O around its ground state well as well as for the $^{240}$Pu isotope along the full 1D asymmetric path starting from the ground state well up to scission and beyond.

\subsection{Addressing state discontinuities with the Link approach}\label{link}

\begin{figure}
\centering
\includegraphics[width=0.9\linewidth]{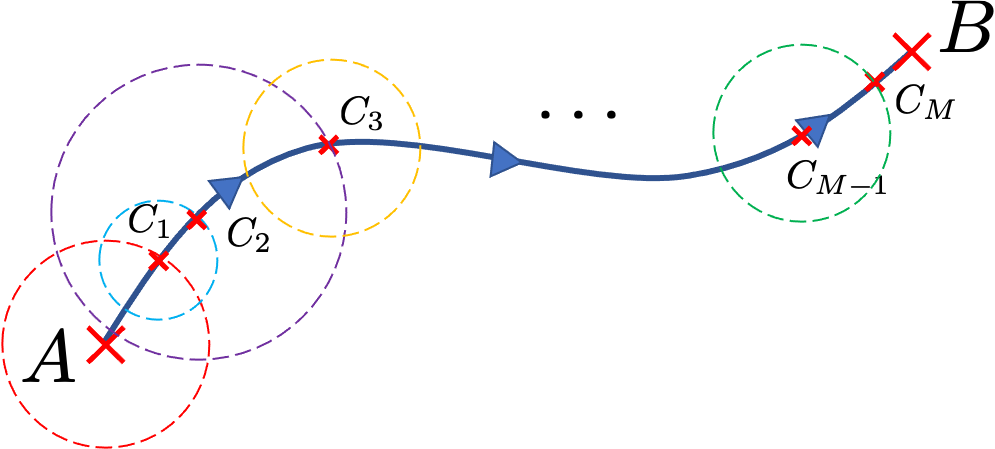}
\caption{Schematic view of the Link method.}
\label{ctwo_58}
\end{figure}

Constructing adiabatic paths suitable for the SCIM formalism requires directly controlling the geometry of the path in the space of HFB states. The first method introduced for this purpose is the \enquote{Link} approach \cite{carpentier2024}.
The principle of this method is to connect two HFB states, $\ket{A}$ and $\ket{B}$, through a sequence of intermediate states whose mutual overlaps, denoted as $x_0$, remain constant up to a given numerical accuracy. Starting from the state $\ket{A}$, each new state is obtained by minimizing the HFB energy under a fixed overlap constraint with the previous state, while simultaneously maximizing the overlap with the target state $\ket{B}$. This procedure is iterated until the final state becomes sufficiently close to $\ket{B}$. A schematic representation of the method is shown in FIG. \ref{ctwo_58}.
Contrary to standard constrained HFB calculations of the $\mathcal{P}_{20}$ procedure, the \enquote{Link} method controls directly the spacing between neighboring many-body states in terms of overlaps rather than in multipole moments. 
The resulting trajectory is therefore constructed according to the intrinsic geometry of the HFB space itself. 

An important question is whether the trajectories reconstructed using the \enquote{Link} procedure remain close to the underlying adiabatic paths produced by the $\mathcal{P}_{20}$ procedure.
To address this, the method was first tested in the $^{16}$O, where no discontinuities are present.

\subsubsection{Validation in the $^{16}$O nucleus}

\begin{figure}
\centering
\includegraphics[width=1.0\linewidth]{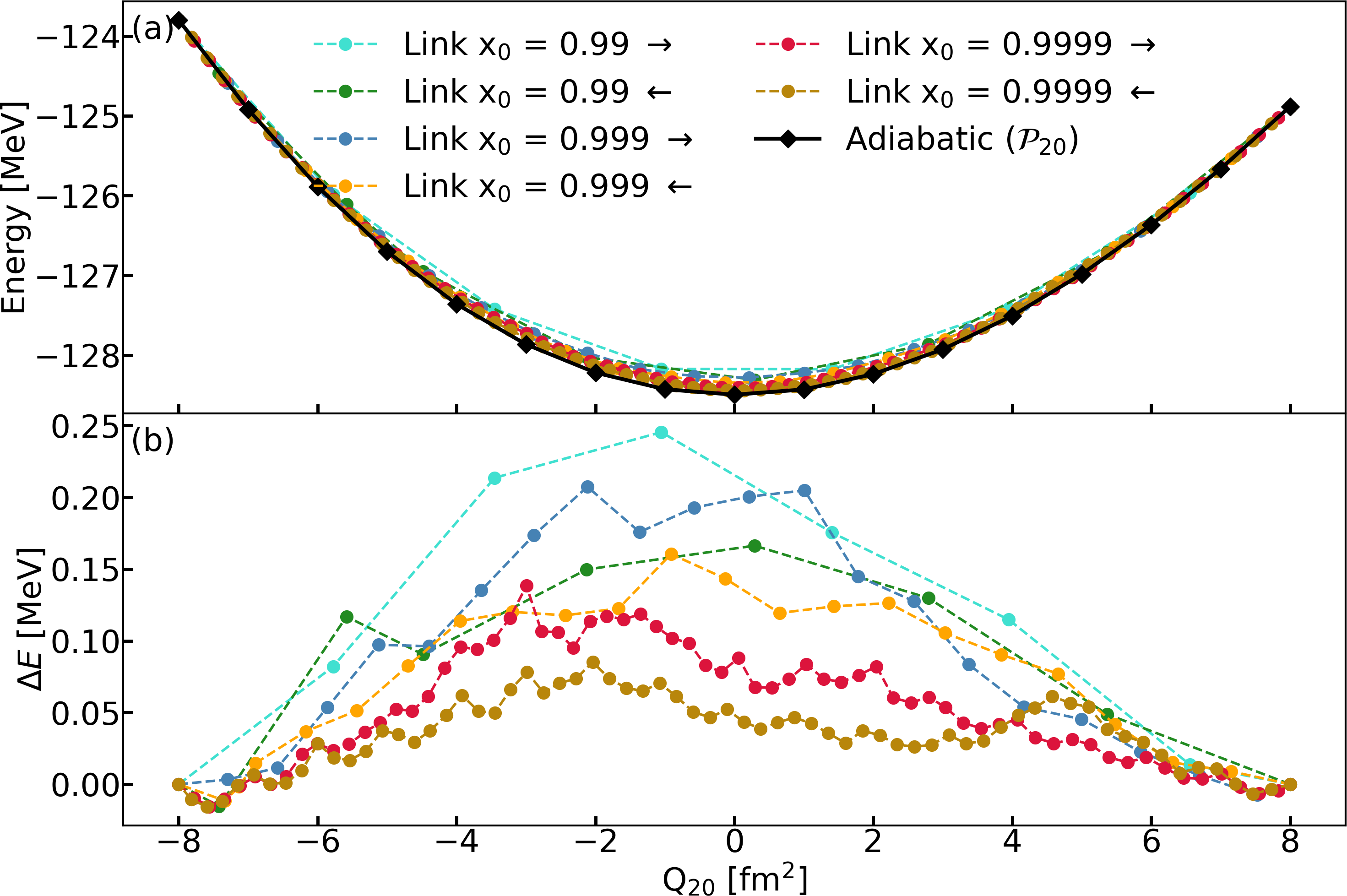}
\caption{Panel (a): PES associated with different paths obtained with the \enquote{Link} method for different values of the parameter $x_0$ and different directions. The adiabatic path produced with the $\mathcal{P}_{20}$ procedure (in black) is also indicated. Panel (b): Energy differences between the different \enquote{Link} PES and the $\mathcal{P}_{20}$ adiabatic PES. Calculations have been done for $^{16}$O.}
\label{ctwo_116}
\end{figure}

The first tests of the \enquote{Link} method were performed in the simple case of the spherical and rigid $^{16}$O nucleus in its ground-state region. 
In this situation, the standard constrained HFB procedure already generates a smooth adiabatic path, making it possible to evaluate directly how closely the \enquote{Link} trajectories reproduce the reference adiabatic evolution.
The corresponding PES obtained for different overlap parameters $x_0$ are shown in FIG.~\ref{ctwo_116}. 
The calculations were performed between two adiabatic configurations characterized by $Q_{20}=-8$ fm$^2$ and $Q_{20}=8$ fm$^2$. 
The notation $\rightarrow$ indicates that the path is generated from oblate deformations toward the prolate ones, while $\leftarrow$ corresponds to the opposite direction.

Several important observations emerge from these calculations.
First, the trajectories generated by the \enquote{Link} method remain very close to the adiabatic path across the entire deformation range.
Second, the agreement systematically improves as the overlap parameter approaches unity.
For $x_0=0.9999$, the maximum energy difference with respect to the adiabatic trajectory is found to be 130 keV in the $\rightarrow$ direction and 75 keV in the $\leftarrow$ one, representing less than 0.1\% of the total binding energy of the nucleus.

\begin{figure}
\centering
\includegraphics[width=1.0\linewidth]{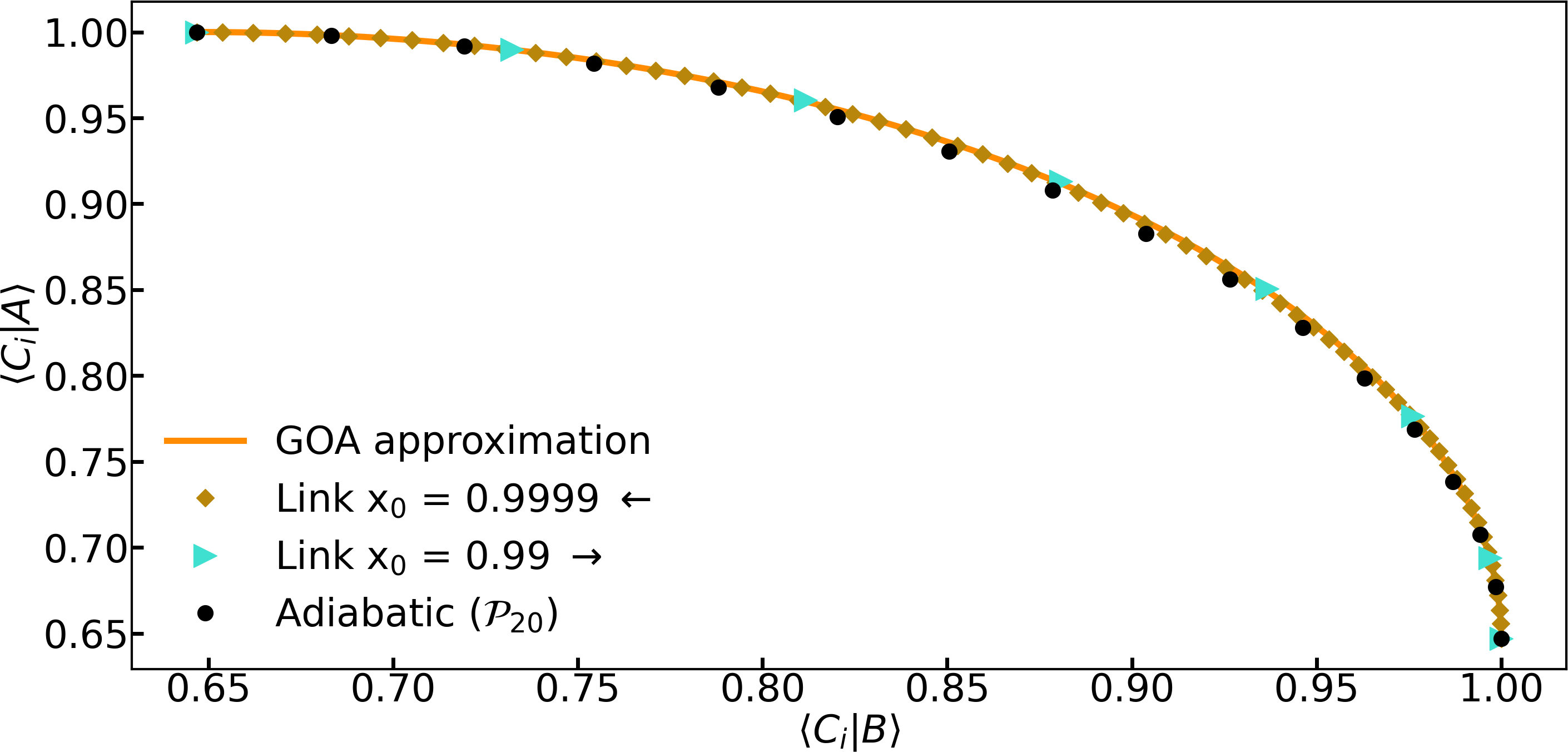}
\caption{Comparison of the distance between the states obtained with the \enquote{Link} method, the adiabatic states (in red) and the GOA approximation in the $^{16}$O.}
\label{ctwo_117}
\end{figure}

Beyond the energy itself, it is also crucial to verify whether the geometry of the reconstructed path remains compatible with the GOA, which plays a central role in collective approaches.
To this end, the states generated by the \enquote{Link} procedure have been plotted according to their overlaps with the initial state $\ket{A}$ and the target state $\ket{B}$.
The corresponding results are displayed in FIG. \ref{ctwo_117}, for two specific values of $x_0$ alongside the predictions of the GOA approximation, given by:
\begin{eqnarray}
   \bra{C_i}\ket{B} = \text{exp}(\ln(\bra{A}\ket{B})(1-\sqrt{\frac{\ln(\bra{C_i}\ket{A})}{\ln(\bra{A}\ket{B})}})^2). 
\end{eqnarray}
The adiabatic state associated with the $\mathcal{P}_{20}$ procedure are also indicated.
One observes that the states generated by the \enquote{Link} method follow very closely the GOA predictions. 
This result indicates that the overlap-based reconstruction preserves the expected local geometry of adiabatic collective motion.
Interestingly, the trajectories obtained with the finest overlap spacings adhere even more closely to the GOA relation than the original adiabatic states themselves.
This behavior suggests that the overlap constraint naturally regularizes the local distribution of many-body states in the HFB space.
These calculations demonstrate that the \enquote{Link} method does not simply interpolate artificially between two configurations. 
Instead, it reconstructs continuous trajectories remaining close both to the adiabatic energy landscape and to the expected overlap geometry of collective motion.

\subsubsection{Crossing discontinuities in $^{240}$Pu}

\begin{figure}
\centering
\includegraphics[width=1.0\linewidth]{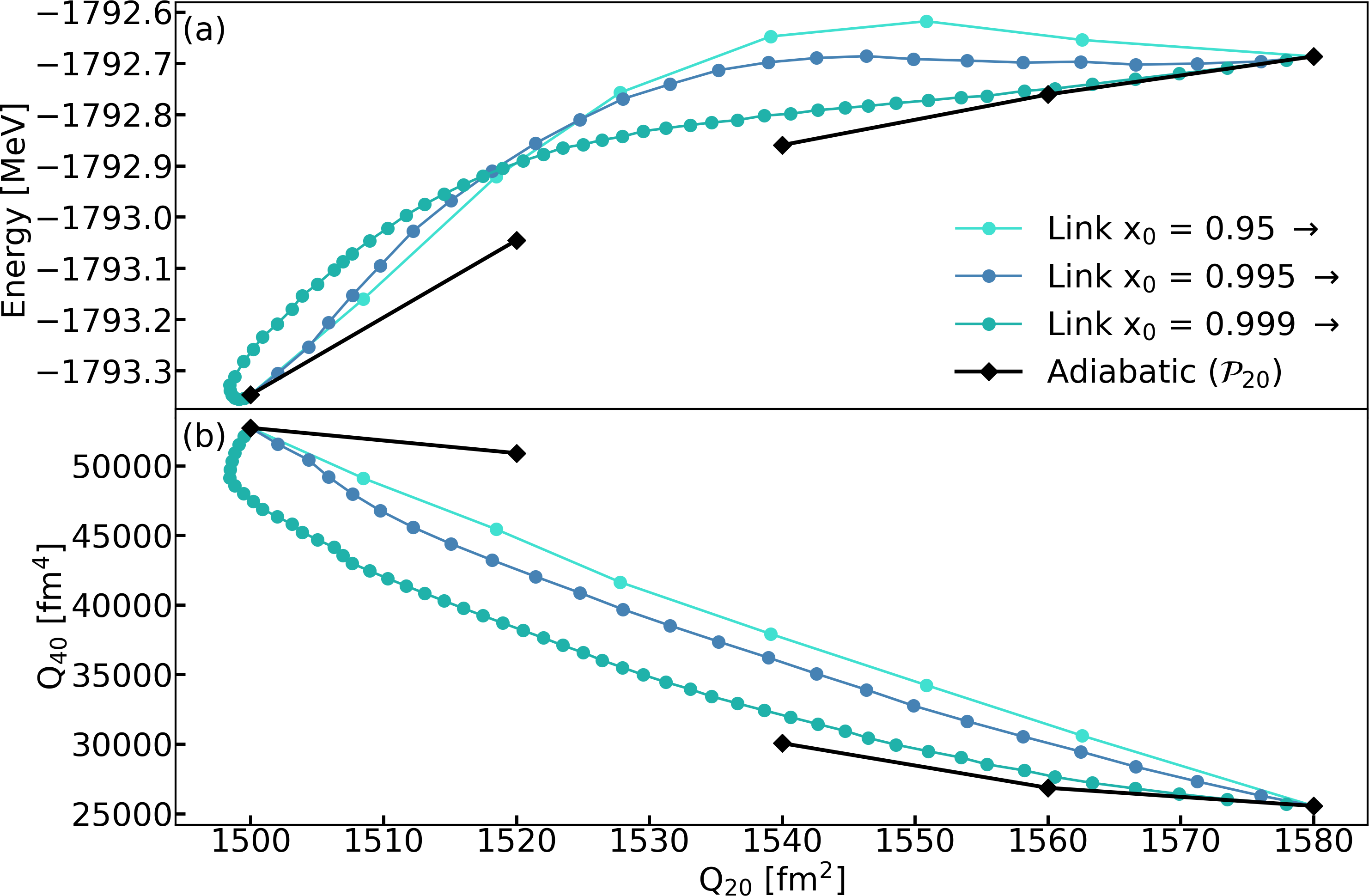}
\caption{Panel (a): PES associated with paths obtained with the \enquote{Link} method using different parameters $x_0$ compared with the adiabatic PES, with respect to the quadrupole deformation $Q_{20}$. Panel (b): Evolution of the hexadecapole deformation $Q_{40}$ with respect to the quadrupole deformation $Q_{20}$. Calculations have been done in the area of the $^{240}$Pu first barrier where the discontinuity in $Q_{40}$ manifests.}
\label{ctwo_118}
\end{figure}

The next step involves applying the \enquote{Link} method to cases where genuine discontinuities appear in standard constrained HFB paths.
A notable example occurs along the first barrier of the asymmetric fission path of $^{240}$Pu, where the standard $\mathcal{P}_{20}$ procedure exhibits a discontinuity associated with the hexadecapole moment $Q_{40}$.
The corresponding calculations are shown in FIG.~\ref{ctwo_118}. 
The \enquote{Link} trajectories obtained for different overlap parameters are compared with the original adiabatic path generated with the standard constrained procedure $\mathcal{P}_{20}$.
One sees that the discontinuity is replaced by a smooth and continuous trajectory connecting the two regions of the PES (panel (a)).
No hidden barrier appears during the reconstruction process and the resulting paths remain close to the original adiabatic landscape $\mathcal{P}_{20}$. 
The evolution of the hexadecapole moment $Q_{40}$ (panel (b)) also becomes continuous, demonstrating that the \enquote{Link} method build continuity simultaneously at the level of both the energy and the intrinsic structure of the HFB states.

\begin{figure}
\centering
\includegraphics[width=1.0\linewidth]{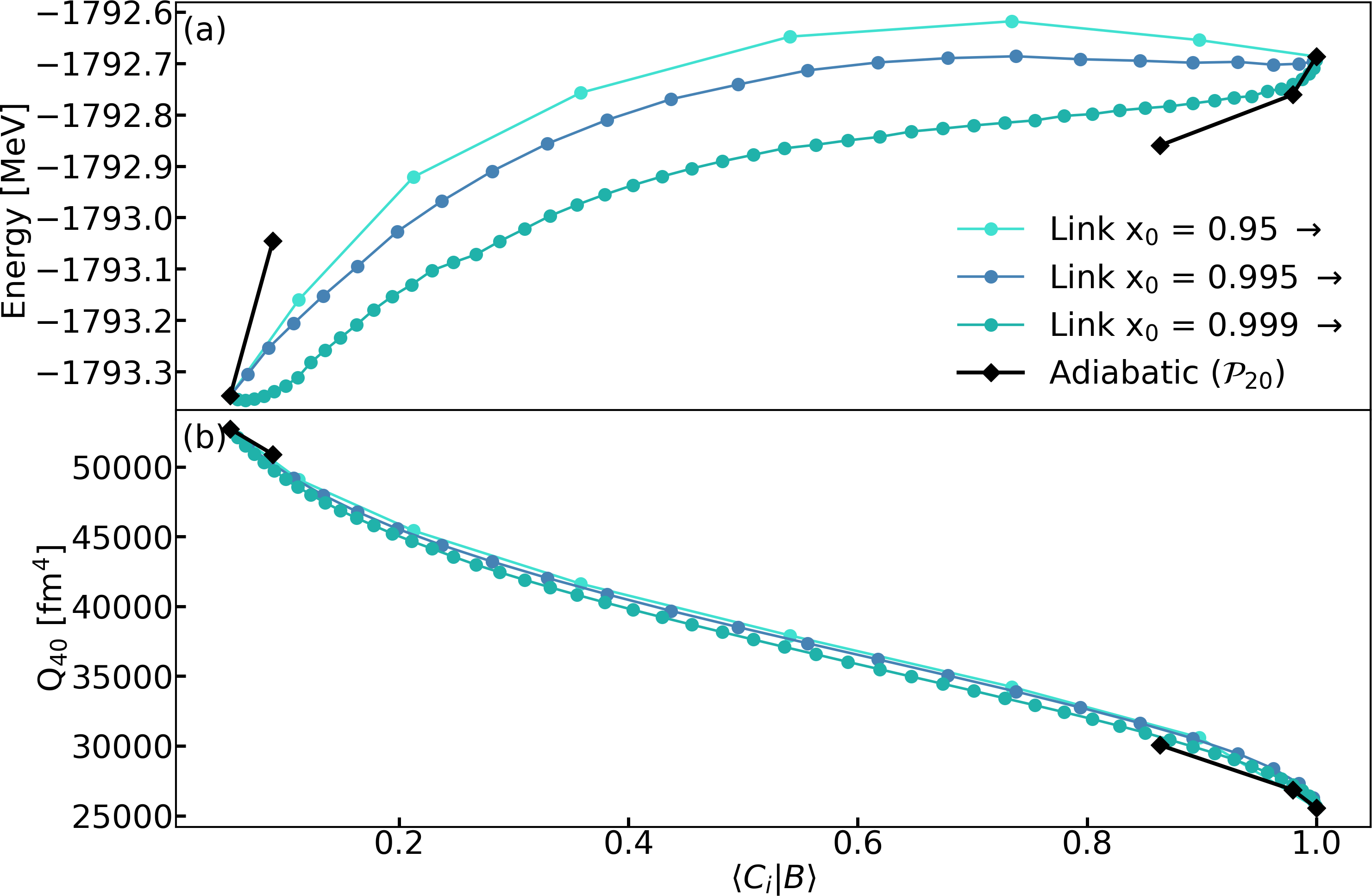}
\caption{Same as FIG. \ref{ctwo_118} but in terms of the overlap with $\ket{B}$ instead of $Q_{20}$.}
\label{ctwo_119}
\end{figure}

At first glance, the trajectories obtained for different overlap parameters appear to differ significantly when plotted as functions of the quadrupole deformation $Q_{20}$.
However, this discrepancy primarily stems from the choice of collective coordinate parametrization, as $Q_{20}$ does not provide a uniform measure of distance in the HFB space.
To illustrate this point, the same trajectories are shown in FIG. \ref{ctwo_119}, now plotted as functions of the overlap with the target state $\ket{B}$.
In this representation, the different trajectories become much more similar and converge toward the same adiabatic behavior as $x_0$ approaches one.
The overlap representation thus provides a far more natural parametrization of the collective motion than the quadrupole deformation itself.

\begin{figure}
\centering \includegraphics[width=1.0\linewidth]{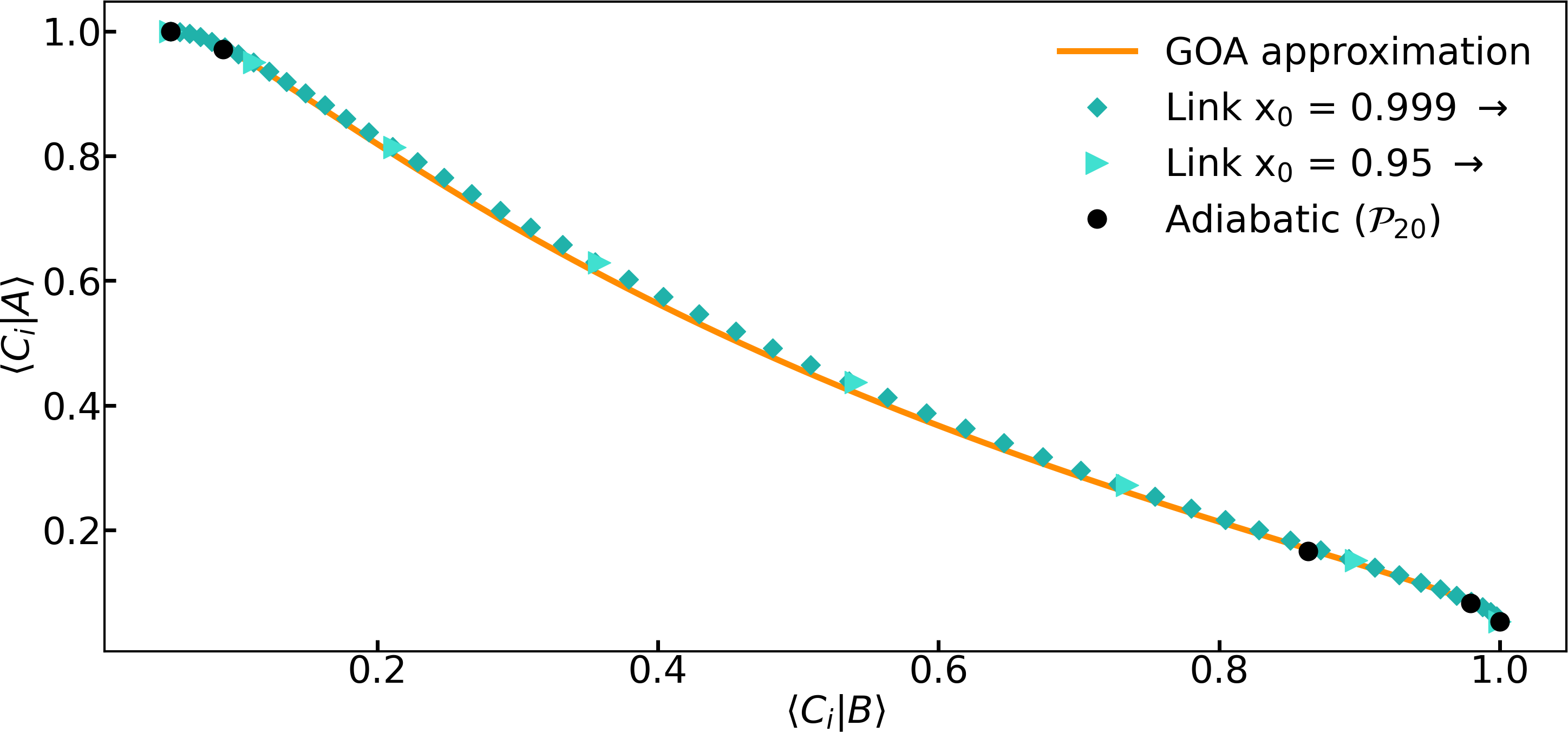} 
\caption{Comparison of the distance between the states obtained with the \enquote{Link} method, the $\mathcal{P}_{20}$ procedure, and the GOA approximation in the $^{240}$Pu.} 
\label{ctwo_120} 
\end{figure}

Furthermore, the overlap geometry of the reconstructed trajectories remains compatible with the GOA approximation, even when crossing the discontinuity.
This property is illustrated in FIG. \ref{ctwo_120}, where the overlap distances associated with the reconstructed paths are compared with the GOA predictions.
The good agreement observed supports the interpretation that the \enquote{Link} method reconstructs physically meaningful adiabatic trajectories rather than artificial interpolations between disconnected configurations.

\subsection{Going through scission with the Drop method}\label{drop}

\begin{figure} 
\centering 
\includegraphics[width=0.3\linewidth]{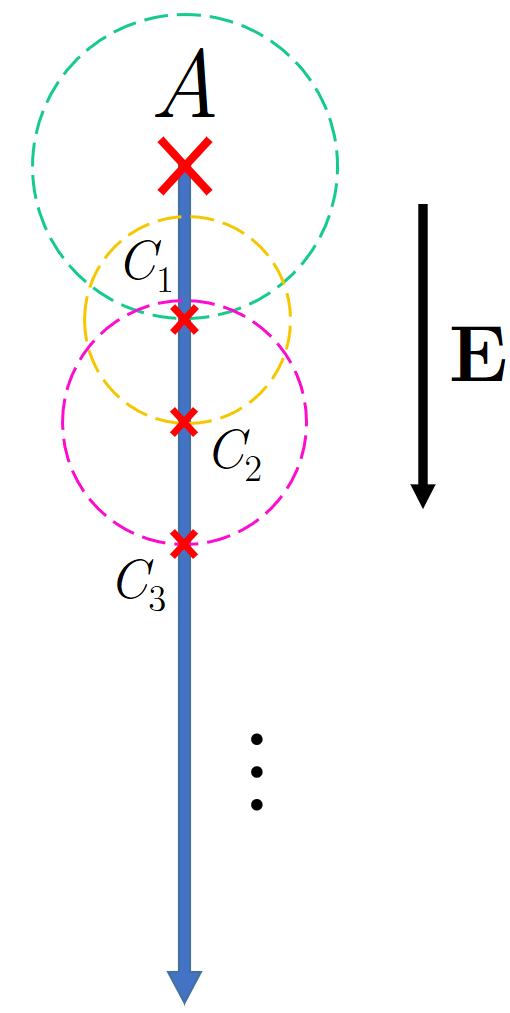} \caption{Schematic view of the \enquote{Drop} method.} \label{ctwo_93} 
\end{figure}

\begin{figure}
\centering 
\includegraphics[width=1.0\linewidth]{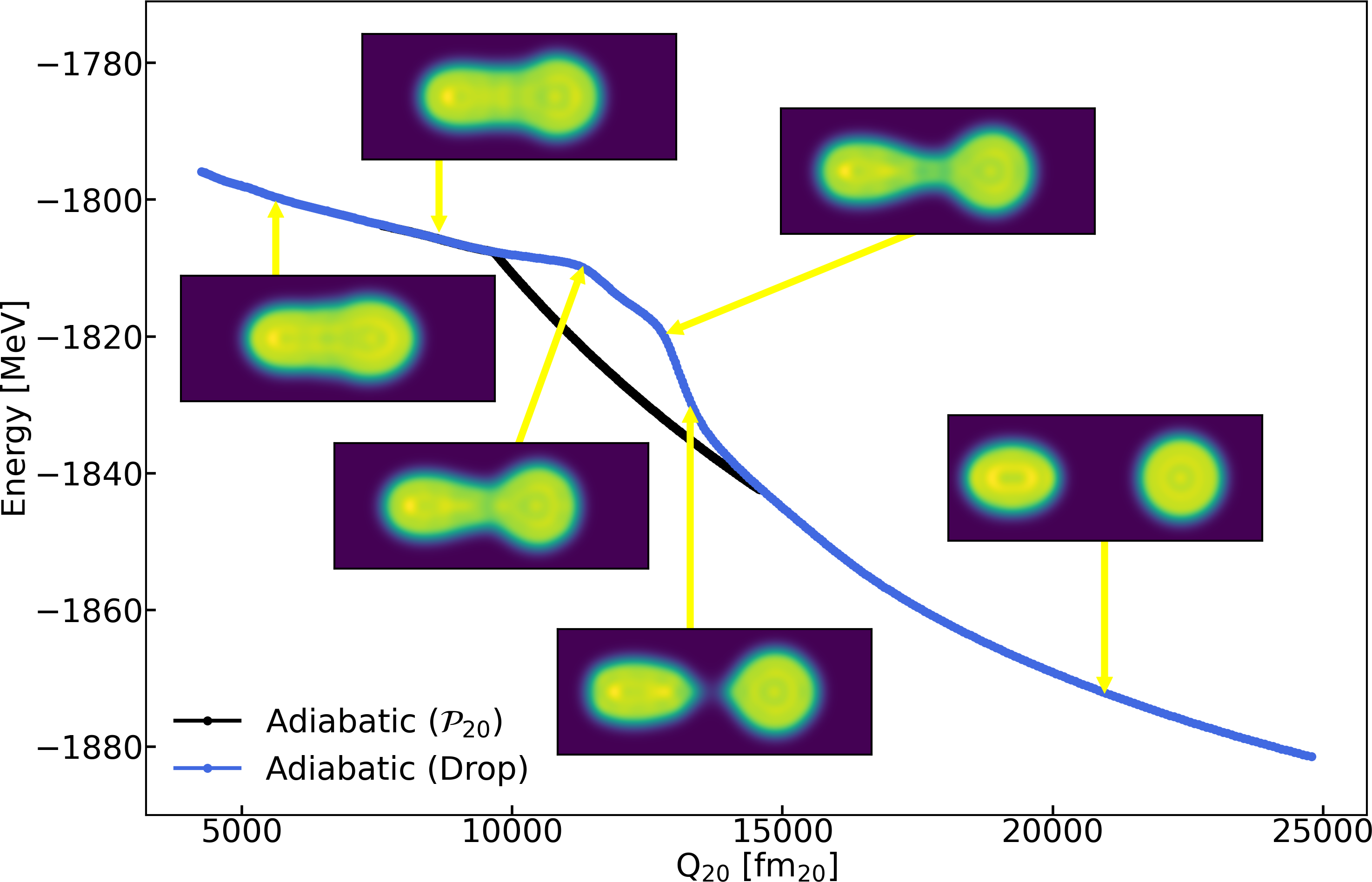} \caption{PES produced by the \enquote{Drop} method (in blue) in the $^{240}$Pu nucleus in the descent towards scission. Relevant total nucleon densities have been added as well as the PES from the $\mathcal{P}_{20}$ procedure (in black).} \label{ctwo_95} 
\end{figure}

Although the \enquote{Link} method efficiently connects two known HFB configurations, it cannot be directly applied in situations where the final state is not known in advance, such as the descent toward scission.
To address this limitation, a second overlap-based method, referred to as the \enquote{Drop} method, has been introduced.
Unlike the \enquote{Link} method, the \enquote{Drop} method does not require a target state.
Starting from a given HFB configuration, each new state is generated by minimizing the HFB energy under a fixed overlap constraint with the previous state, denoted as $x_0$.
The procedure thus continuously follows the local energy descent in the HFB space.
A schematic representation of the method is shown in FIG. \ref{ctwo_93}.
The method proves particularly efficient for continuously describing the evolution from the saddle region toward scission and beyond.
An example for the asymmetric fission path of $^{240}$Pu is displayed in FIG. \ref{ctwo_95}.
The corresponding trajectory provides a continuous description of the entire scission process, including the progressive neck rupture and the relaxation of the fragments after separation.
Such configurations are extremely challenging to obtain using standard constrained multipole methods, as the topology of the potential energy surface (PES) changes drastically near scission.

\begin{figure}
\centering 
\includegraphics[width=1.0\linewidth]{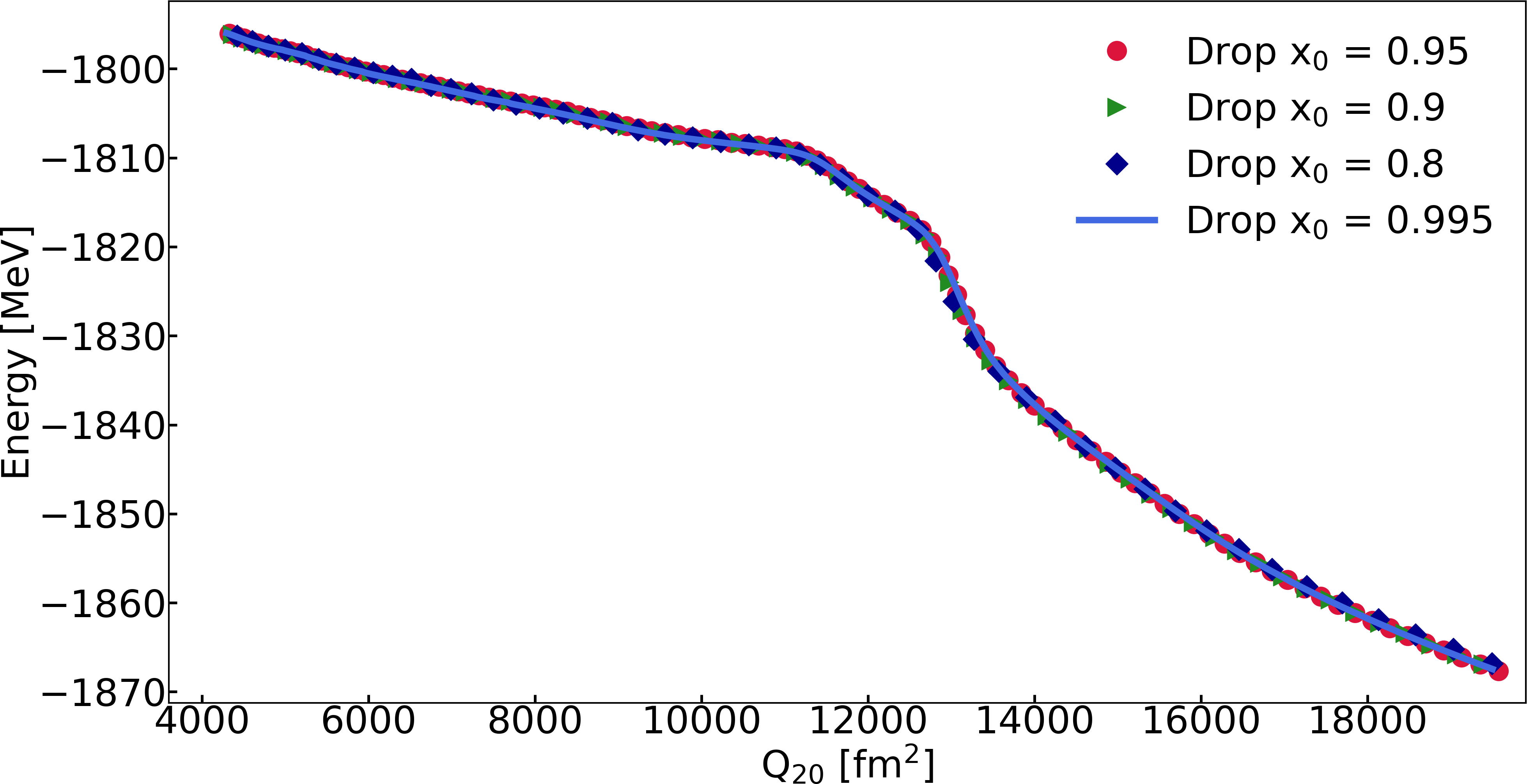} \caption{Illustration of the impact of the value of the parameter $x_0$ in the \enquote{Drop} method applied from the saddle point of the $^{240}$Pu with respect to the quadrupole deformation $Q_{20}$.} \label{ctwo_94} 
\end{figure}

Moreover, an important property of the Drop method is that its trajectory depends only weakly on the chosen overlap parameter $x_0$.
Changing the overlap spacing primarily affects the resolution of the path rather than its physical content, as illustrated in FIG. \ref{ctwo_94}.
This property makes the method particularly well-suited for both exploratory calculations using coarse overlap spacings and for constructing refined paths for use in SCIM dynamics.

\subsection{The $\mathcal{\tilde P}_{20}$ procedure}\label{proc}

The \enquote{Link} and \enquote{Drop} methods introduced in the previous sections are two complementary components of a broader strategy aimed at constructing adiabatic paths compatible with the SCIM formalism.
Together, they define a new procedure, denoted as $\mathcal{\tilde P}_{20}$, which replaces the standard $\mathcal{P}_{20}$ one with an overlap-based construction of the collective trajectory.

\subsubsection{Link part of the $\mathcal{\tilde P}_{20}$ procedure}

The $\mathcal{\tilde P}_{20}$ procedure begins with a standard adiabatic path generated using the $\mathcal{P}_{20}$ one.
This initial set of HFB states is ordered according to increasing quadrupole deformation $Q_{20}$ and truncated after the saddle region, retaining only the configurations that satisfy
$Q_{20} \leq Q_{20}^{\mathrm{saddle}} + 1000 \ \mathrm{fm}^2$.
The resulting sequence is denoted $\left\{ \ket{\Phi_i} \right\}$. The next step consists in constructing a reduced set of reference configurations, referred to as the \enquote{attractors}. 
Starting from the initial state $\ket{\Phi_0}$, one searches for the first state $\ket{\Phi_{i_0}}$ satisfying $\langle \Phi_0 | \Phi_{i_0} \rangle <0.5$.
 The procedure is then iterated by searching for the first state $\ket{\Phi_{i_1}}$ such that
$\langle \Phi_{i_0}|\Phi_{i_1} \rangle < 0.5$
and so on along the adiabatic path. At the end of the process, one obtains a sparse sequence of attractor states $\left\{ \ket{\Phi_{i_j}} \right\}$, which provides a coarse description of the global adiabatic evolution. The \enquote{Link} method is then applied successively between neighboring attractors. 
Starting from the state $\ket{\Phi_0}$, a first \enquote{Link} trajectory is generated toward the target state $\ket{\Phi_{i_0}}$ using a fixed overlap parameter $x_0$. 
In the present work, the value $x_0 = 0.995$ has been adopted. 
The iterative process is stopped once the overlap between the last generated state and the target state becomes larger than $0.9$. 
This first reconstruction produces a sequence of states $\left\{ \ket{L_0}, \ldots , \ket{L_{j_0}} \right\}$, with $\ket{L_0} = \ket{\Phi_0}$. The procedure is then repeated by taking the last generated state as the new starting point and the next attractor as the target state. 
By iterating this construction over the whole set of attractors, one finally obtains a continuous overlap-based trajectory
$ \left\{ \ket{L_0}, \ldots , \ket{L_{j_1}}, \ldots , \ket{L_{j_f}} \right\}$
which continuously reconstructs the adiabatic evolution up to and slightly beyond the saddle region. 
\begin{figure}
\centering
\includegraphics[width=1.0\linewidth]{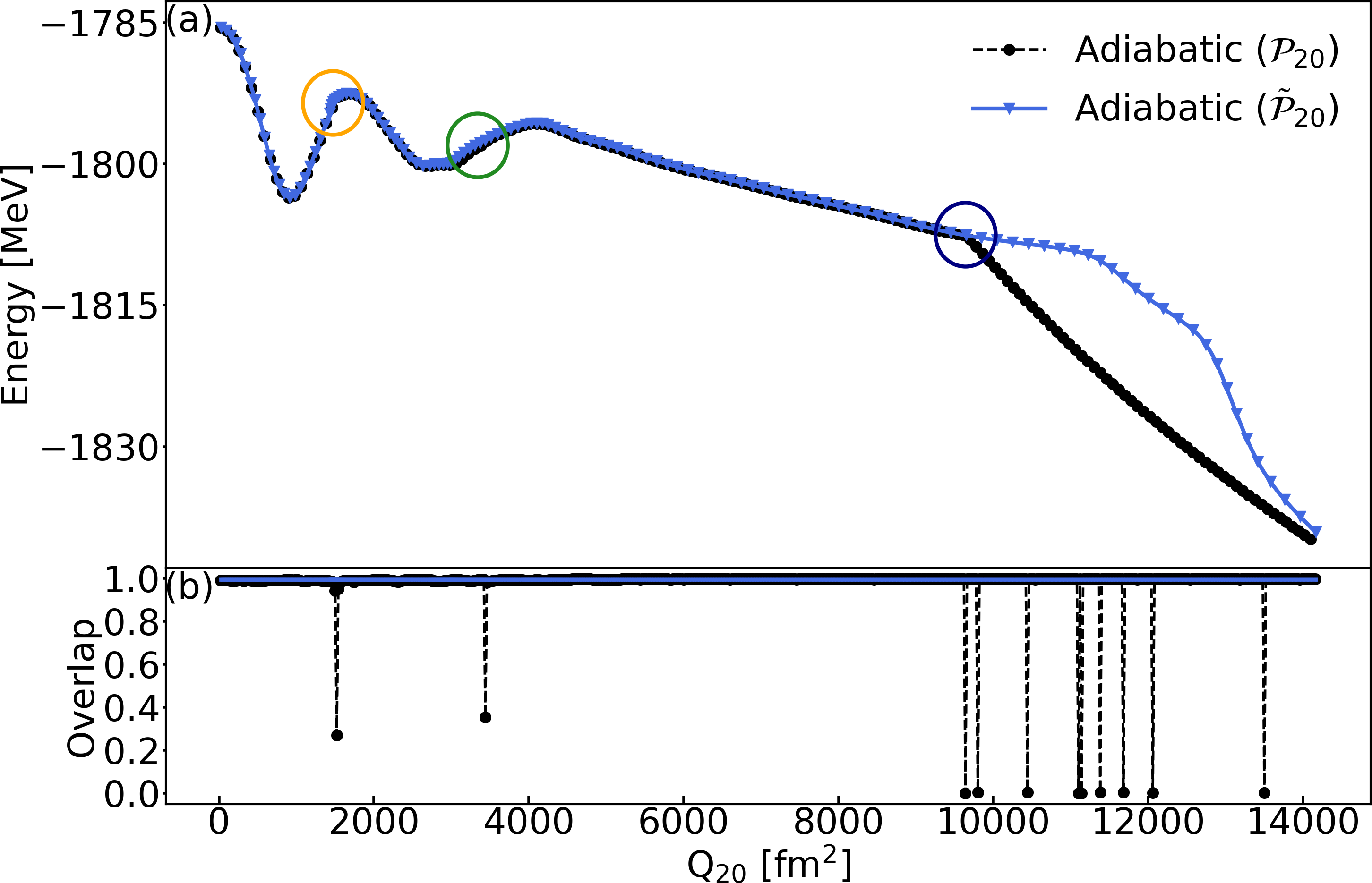}
\caption{Panel (a): PES obtained with the procedures 
$\mathcal{P}_{20}$ (in black) and $\mathcal{\tilde P}_{20}$ (in blue). 
Colored circles highlight discontinuities appearing in the $\mathcal{P}_{20}$ path. 
Panel (b): Overlap between neighboring HFB states along the $\mathcal{P}_{20}$ and $\mathcal{\tilde P}_{20}$ paths.}
\label{ctwo_???}
\end{figure}

\subsubsection{Drop part of the $\mathcal{\tilde P}_{20}$ procedure}

Beyond the saddle region, the adiabatic evolution progressively enters the scission region, where the standard $\mathcal{P}_{20}$ procedure no longer provides a reliable sequence of adiabatic states.
Thus, the path is extended using the \enquote{Drop} method which continuously follows the local energy descent in the HFB space while preserving the overlap regularity of the trajectory.

Starting from the last state $\ket{L_{j_f}}$ generated by the \enquote{Link} method, the \enquote{Drop} evolution is performed using the same overlap parameter $x_0 = 0.995$.
The method then continuously follows the local energy descent without requiring any predefined target configuration.
This second stage produces an additional sequence of states
$\left\{ \ket{D_1}, \ldots , \ket{D_f} \right\}$
which naturally extends the reconstructed trajectory through scission and beyond. 
The complete $\mathcal{\tilde P}_{20}$ path is therefore given by
$\left\{
\ket{L_0}, \ldots , \ket{L_{j_1}}, \ldots , \ket{L_{j_f}},
\ket{D_1}, \ldots , \ket{D_f}
\right\}$.
The resulting trajectory obtained for the $^{240}$Pu nucleus is displayed in FIG. \ref{ctwo_???}. 
Compared with the standard $\mathcal{P}_{20}$ procedure, the reconstructed $\mathcal{\tilde P}_{20}$ path provides a continuous and regular description of the full asymmetric fission evolution, from the ground-state region up to scission and beyond. 

\subsubsection{The $c_\#$ new collective coordinate}

An important consequence of the $\mathcal{\tilde P}_{20}$ procedure is that the collective trajectory is no longer naturally parametrized by the constrained multipole moment $Q_{20}$.
Instead, the path is now constructed from constant overlap distances (up to a given accuracy) between neighboring HFB states.
It is therefore natural to introduce a new collective coordinate directly associated with the intrinsic geometry of the HFB space, denoted as $c_\#$.
\begin{figure}
\centering
\includegraphics[width=0.9\linewidth]{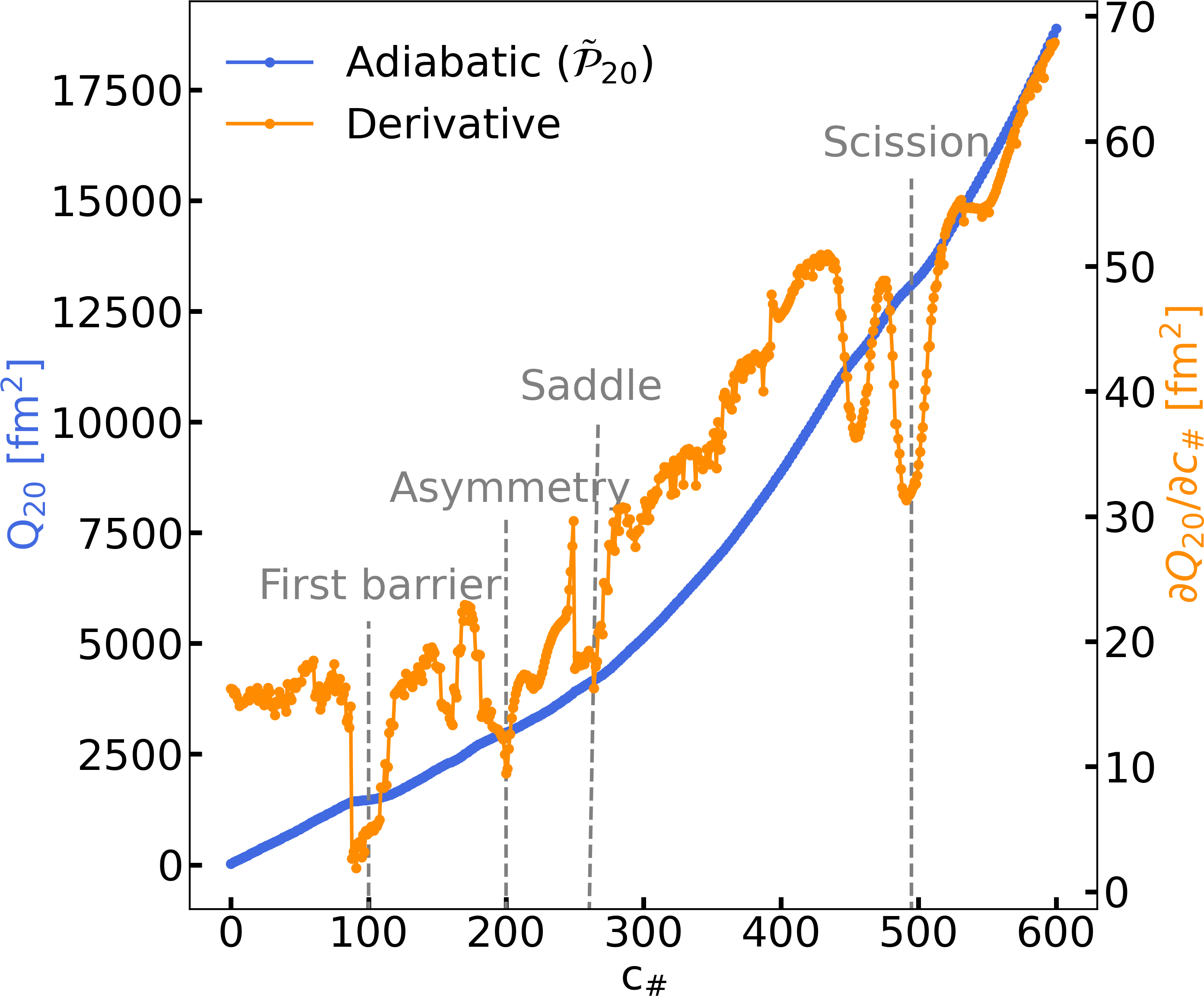}
\caption{Evolution of the quadrupole deformation $Q_{20}$ as a function of the collective coordinate $c_\#$ along the asymmetric fission path of $^{240}$Pu.}
\label{ctwo_47}
\end{figure}
In practice, the collective coordinate $c_\#$ is defined directly from the overlap spacing parameter $x_0$ used during the $\mathcal{\tilde P}_{20}$ procedure. 
Thus, one unit of $c_\#$ corresponds to an elementary step between neighboring HFB states along the trajectory. 
The collective coordinate $c_\#$ therefore provides a parametrization of the path directly connected to the overlap geometry of the HFB space. 
The relation between the usual quadrupole deformation $Q_{20}$ and the coordinate $c_\#$ is illustrated in FIG. \ref{ctwo_47}, with the blue curve. 
One clearly observes that equal variations of $Q_{20}$ do not correspond to uniform displacements along the collective trajectory in overlap space. 
Some regions of the path are strongly compressed while others are dilated, reflecting the fact that the quadrupole deformation alone does not provide a faithful measure of the actual distance between many-body states. To illustrate this non-linearity effect, the derivative $\frac{\partial Q_{20}}{\partial c_\#}$ has 
been indicated with the orange curve.
\begin{figure}
\centering
\includegraphics[width=0.95\linewidth]{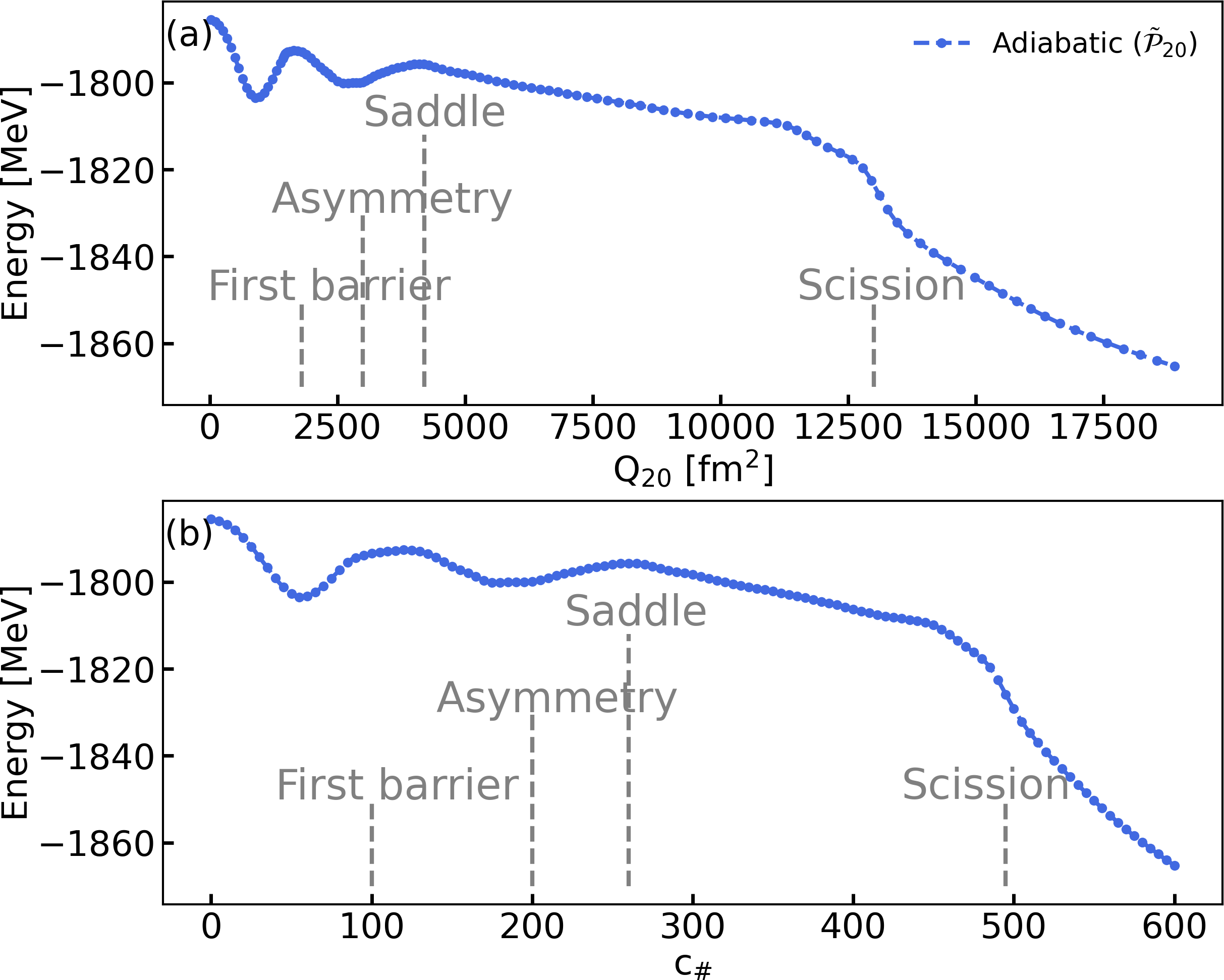}
\caption{Evolution of the total energy of $^{240}$Pu along the asymmetric path according to (a) the quadrupole deformation $Q_{20}$ and (b) the new collective variable $c_\#$.
The curves have been obtained with the procedure $\mathcal{\tilde P}_{20}$.}
\label{ctwo_48}
\end{figure}

The impact of this new parametrization on the representation of the adiabatic path is illustrated in FIG. \ref{ctwo_48}, where the same trajectory is displayed as a function of either $Q_{20}$ or $c_\#$. The value
$c_\# = 0$ is associated with $Q_{20} = 28$ fm$^2$ and  $c_\# = 600$ with $Q_{20} = 18885$ fm$^2$.
This change of perspective suggests that standard intuitive notions derived from PES plots, such as the barrier width or the steepness of the descent, should be reinterpreted in light of the underlying overlap metric.\\

\begin{figure}
\centering
\includegraphics[width=1.0\linewidth]{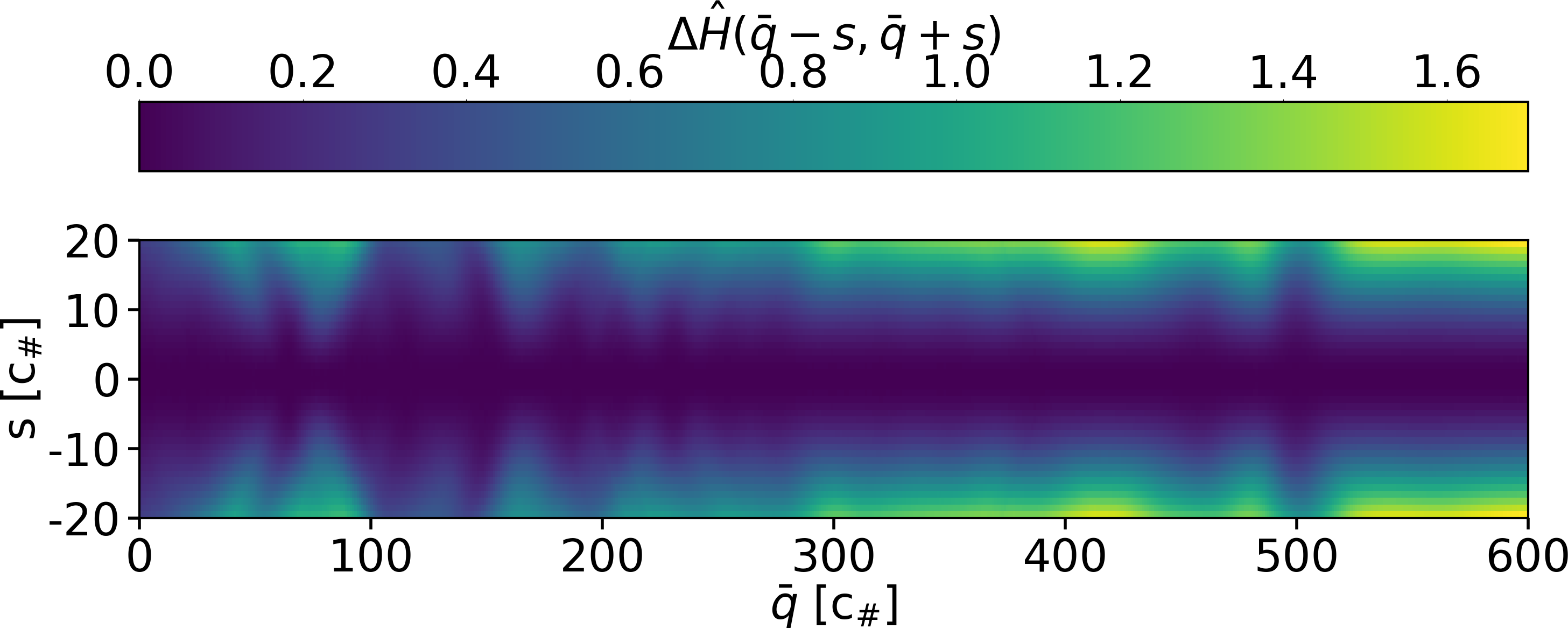}
\caption{Absolute error on the Hamiltonian kernel with respect to $\bar q$ and $s$ for the $^{240}$Pu nucleus.}
\label{ctwo_44}
\end{figure}

We conclude this section by discussing the variation of the Hamiltonian kernel in relation to the overlap kernel, as suggested by Eq. \eqref{ctwo_55_short}.
To illustrate this point, we analyze the absolute error in the Hamiltonian kernel, $\Delta \hat H(\bar q - s,\bar q + s)$, as a function of the center-of-mass coordinate $\bar q$ and relative coordinate $s$: 
\begin{eqnarray}\label{ctwo_45}
\Delta \hat H(\bar q-s,\bar q+s) = 100 \times \quad \quad \quad \quad \quad \quad
\quad  \\ \left| \frac{\bra{\Phi(\bar q-s)} \hat H \ket{\Phi(\bar q+s)} - \bra{\Phi(\bar q-s)}\ket{\Phi(\bar q +s)}E(\bar q)}{\bra{\Phi(\bar q -s)} \hat H \ket{\Phi(\bar q +s)}} \right|. \nonumber 
\end{eqnarray}
The evaluation of $\bra{\Phi(\bar q-s)} \hat H \ket{\Phi(\bar q+s)}$ can be found in Appendix \ref{hamiltonkernel}.
The results are shown in FIG.~\ref{ctwo_44} and cover the entire adiabatic PES produced by the $\mathcal{\tilde P}_{20}$ procedure. The quantities $\bar q$ and $s$ appearing in Eq.~\eqref{ctwo_45} and Fig.~\ref{ctwo_44} are expressed in units of the new collective coordinate $c_\#$.

One observes that the closer to $s=0$, the better the approximation.
Indeed, in the range $s\in$ [$-10,10$], the relative error peaks at 0.46\%, whereas in the range [$-5,5$], the maximum relative error is only 0.15\%.
Moreover, since the Hamiltonian kernel decreases rapidly with respect to $s$, the part of the kernel best described by the approximation is also the most relevant for practical applications.
For our purposes, this result is significant because it demonstrates that focusing on the regularity of the overlap kernel ensures the regularity of both the Hamiltonian and overlap kernels.
Additionally, this result fully justifies the well-known local approximation used, for example, in the GOA formalism \cite{rsbook}.

\section{Analysis of adiabatic HFB states near scission}\label{sectrue4}

In this final section, we discuss the properties of the new adiabatic path generated using the $\mathcal{\tilde P}_{20}$ procedure.
We focus specifically on the scission region, aiming to characterize it.
We will then analyze the proton and neutron distributions of the fragments at scission and extract key information for the static energy balance, such as the deformation energy of the fragments and the interaction energy between them.

\subsection{Scission area characterization}\label{scission}

The identification of the scission region remains a subtle issue in microscopic descriptions of nuclear fission. Scission does not correspond to a sharply defined geometrical configuration; rather, it emerges as a rapid reorganization of the nuclear system accompanied by the progressive decoupling of the two nascent fragments. To characterize this transition more accurately, we examine here two complementary observables along the adiabatic path obtained with the $\mathcal{\tilde P}_{20}$ procedure in $^{240}$Pu: the chemical potentials and the nucleon density in the neck region.

\subsubsection{Chemical potentials near scission}

\begin{figure}
\centering
\includegraphics[width=1.0\linewidth]{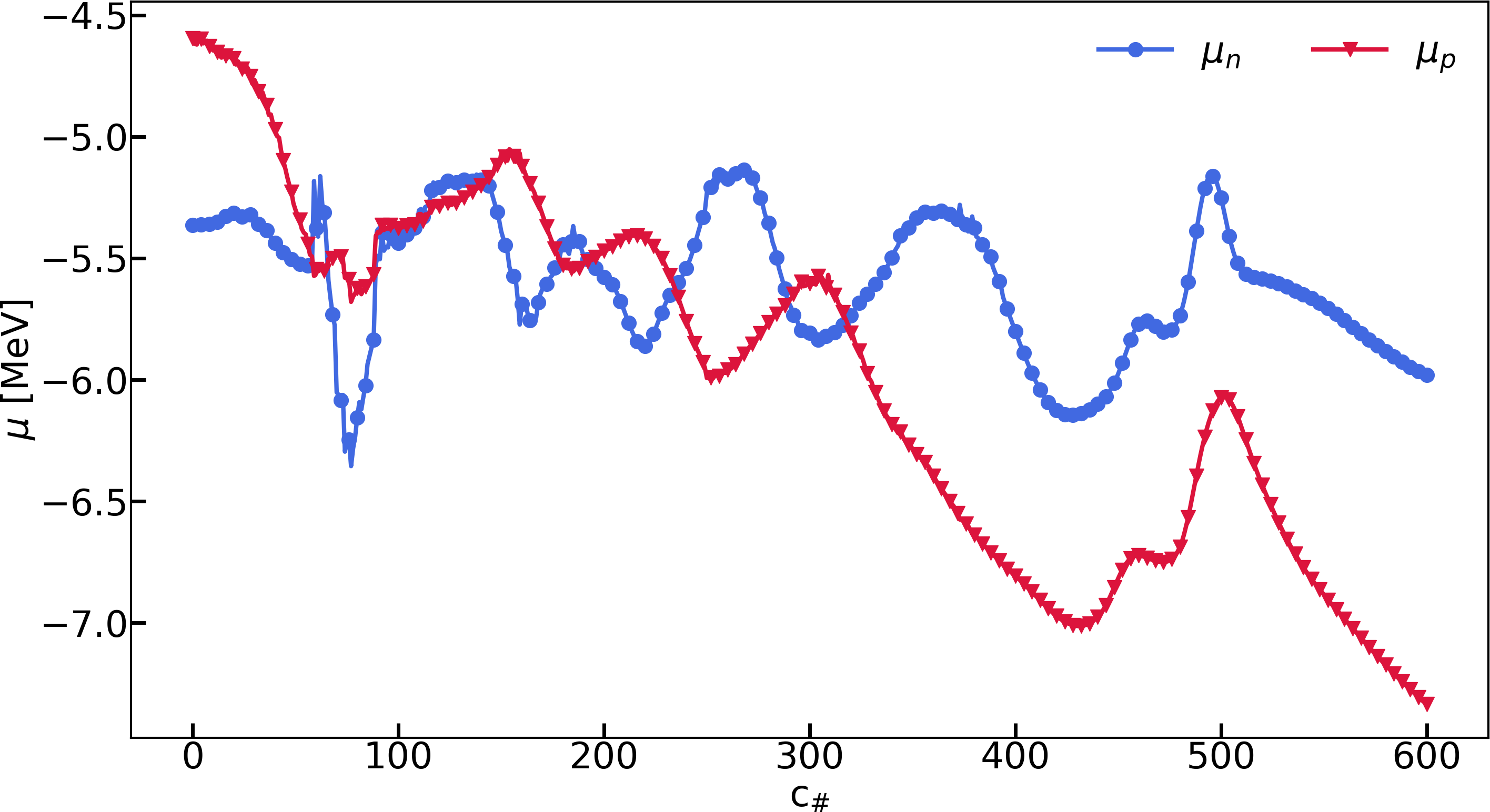}
\caption{Neutron and proton chemical potentials obtained along the adiabatic path generated with the $\mathcal{\tilde P}_{20}$ procedure in $^{240}$Pu as functions of the quadrupole deformation $Q_{20}$.}
\label{ctwo_103}
\end{figure}

In the constrained HFB theory, the neutron and proton chemical potentials arise naturally as the Lagrange multipliers associated with average particle-number conservation. 
As such, they quantify the energy required to add or remove nucleons and can provide qualitative insight into the stability of the system against particle emission in the vicinity of scission.
FIG. \ref{ctwo_103} shows the evolution of the neutron and proton chemical potentials along the $\mathcal{\tilde P}_{20}$ adiabatic path as a function of $c_\#$.
Overall, the proton chemical potential $\mu_p$ decreases with increasing deformation, reflecting the gradual reduction of Coulomb repulsion as the nuclear density distribution becomes more elongated. The neutron chemical potential $\mu_n$ also exhibits structures that correlate with the topology of the PES, particularly in the vicinity of the fission barriers. Additional localized variations may further indicate shell effects associated with the gradual emergence of the pre-fragments.

The most striking feature, however, appears near $Q_{20}\approx13000$ fm$^2$ ($c_\# \approx 495$), where pronounced peaks are observed simultaneously in both the neutron and proton chemical potentials. These peaks may be interpreted as signatures of weakly bound nucleons that belong to neither fragment and form a dilute neutron-rich nucleon medium acting as the ultimate glue before scission. This interpretation is further supported by the extremely low nucleon density in the neck region with $Q_{\mathrm{neck}}=0.38$ nucleons at $c_\#=495$, which strongly suggest that configurations around $c_\#=495$ correspond to the actual rupture of the neck between the pre-fragments. In the remainder of this work, these structures will be referred to simply as the \enquote{chemical potential peaks}.
Such features are of particular interest in the context of neutron emission at scission. Although the present analysis remains qualitative, it suggests that conditions favorable to neutron emission may indeed arise during the final stages of the scission process.

\subsubsection{Nucleon density in the neck region}

A complementary perspective can be gained by examining the local neutron-to-proton density ratio $r_\rho(\vec r)$:
\begin{eqnarray}
r_\rho(\vec r)=
\begin{cases}
\displaystyle
\frac{\rho^{\tau_n}(\vec r)}
{\rho^{\tau_p}(\vec r)}
& \text{if } \rho(\vec r)>5\times10^{-3}
\\
0
& \text{otherwise}.
\end{cases}
\end{eqnarray}
Due to the repulsive Coulomb interaction, protons are expected to localize more rapidly within the nascent pre-fragments, while neutrons may continue to sustain the residual binding between them during the final stage leading up to rupture. If this interpretation is correct, the neck region should become progressively neutron-rich in the vicinity of scission.

\begin{figure}
\includegraphics[width=1.0\linewidth]{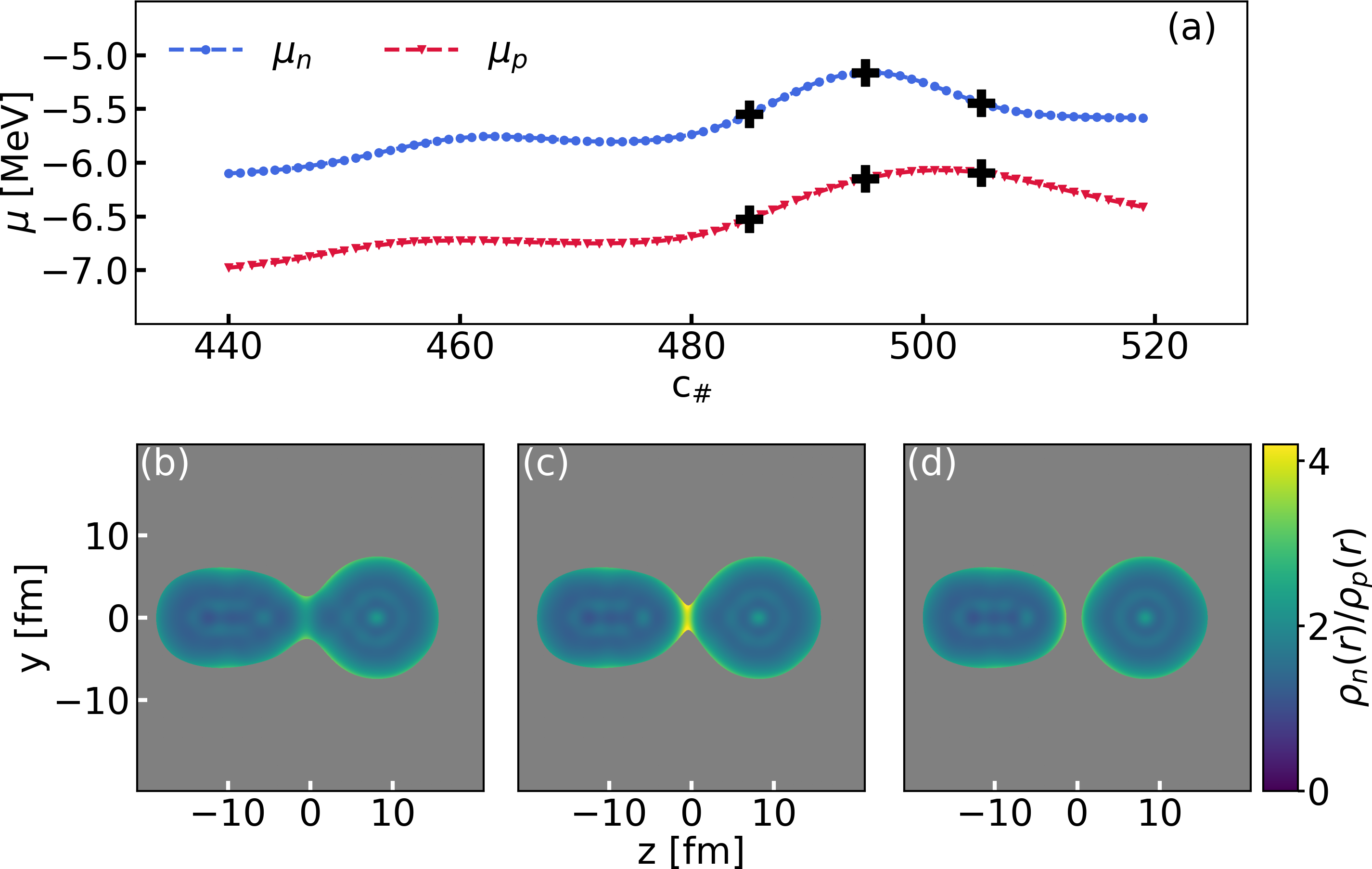}
\caption{Study of the local neutron-to-proton density ratio $r_\rho$ in the scission region. Panel (a): neutron and proton chemical potentials along the adiabatic path. Panels (b-d): spatial distributions of $r_\rho$ for three configurations located before, at, and after the chemical potential peaks.}
\label{ctwo_neck}
\end{figure}

This hypothesis is investigated in FIG. \ref{ctwo_neck} in which panel (a) shows the neutron and proton chemical potentials in the scission region together with three selected configurations. Panels (b-d) display the corresponding local neutron-to-proton ratio $r_\rho$ for configurations located before, at, and after the chemical potential peaks.
A striking enhancement of the neutron-to-proton ratio $r_\rho(\vec r)$ is observed in panel (c), corresponding to the configuration located at the chemical-potential peaks ($Q_{20}=13110$ fm$^2$, $c_\#=495$). In the neck region, the local ratio reaches values as high as 4.5, whereas the average neutron-to-proton ratio of $^{240}$Pu is only $\simeq$1.55. Such a pronounced neutron enrichment is absent both before and after scission, suggesting that this phenomenon may itself constitute a genuine microscopic signature of the scission process.
These results strongly support the interpretation according to which neutrons act as the last glue maintaining the connection between the pre-fragments immediately before scission. Similar conclusions have been reported in TDHFB-type dynamical studies
\cite{abdu}.

\subsection{Proton and neutron fragment distributions at scission}

Charge and mass yields are among the most important observables in nuclear fission, as they directly reflect the partitioning of the compound nucleus into its final fragments. 

\begin{figure}
\includegraphics[width=1.0\linewidth]{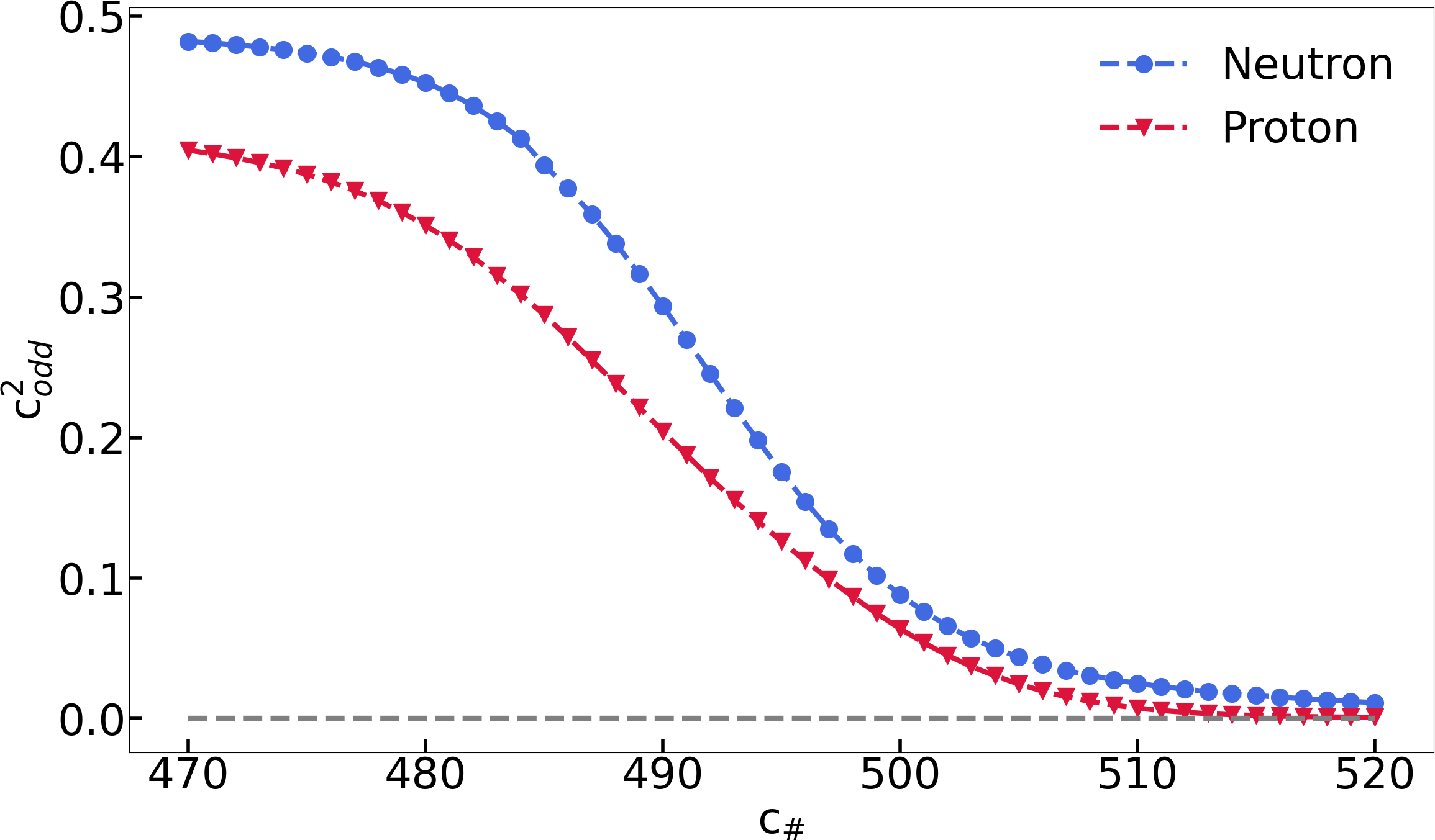}
\caption{Evolution of the odd components in the light fragment neutron and proton particle number distributions of the adiabatic states with respect to $c_\#$. }
\label{ctwo_205}
\end{figure}

In the present work, fragment properties are extracted from the full microscopic particle-number content of the HFB states using the method introduced in Refs.~\cite{zSep,zSepVer} and adapted to the two-center representation (see Appendix \ref{z-separation}). This approach goes beyond estimates based solely on average densities and provides access to the complete particle-number distributions within each fragment through the particle number projection techniques.
\begin{figure}
\includegraphics[width=1.0\linewidth]{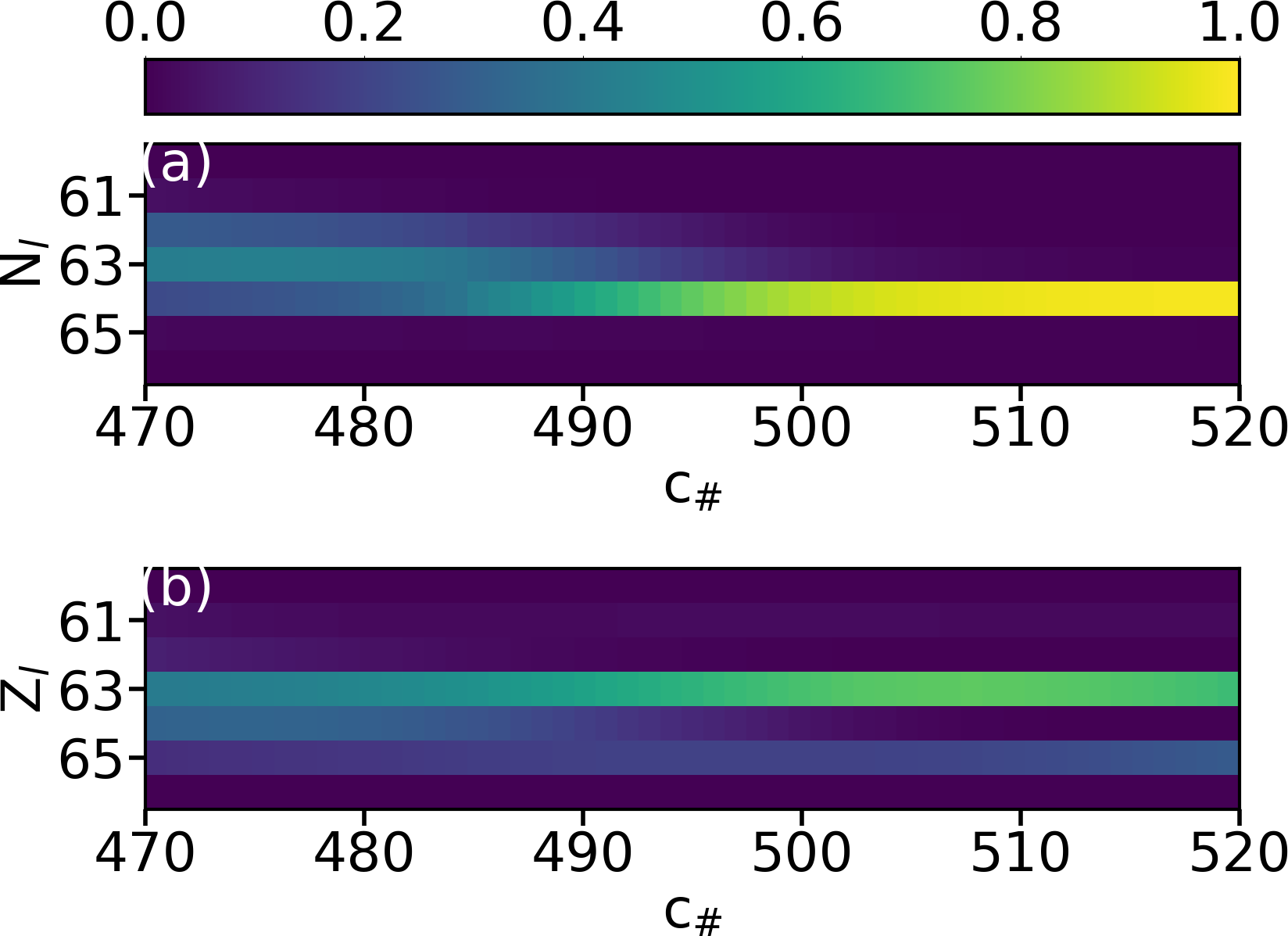}
\caption{Panel (a): Neutron particle-number distributions Y(N$_i$) of the light fragment in the scission area, as a function of $c_\#$. Panel (b): Same as in panel (a) but for protons.}
\label{cfin_19}
\end{figure}
FIG. \ref{cfin_19} illustrates the evolution of the neutron (panel (a)) and proton (panel (b)) fragment distributions along the adiabatic path. A progressive localization of the distributions is observed as the system approaches scission ($c_\# \approx 495$), reflecting the gradual formation of well-defined fragments from an initially strongly coupled configuration. One notes in particular the appearance and the
disappearance of odd-components $c^2_{odd}$ in both proton and neutron sectors whose contributions rapidly tend to zero, as illustrated in FIG. \ref{ctwo_205} for the light fragment.

\begin{figure}
\includegraphics[width=1.0\linewidth]{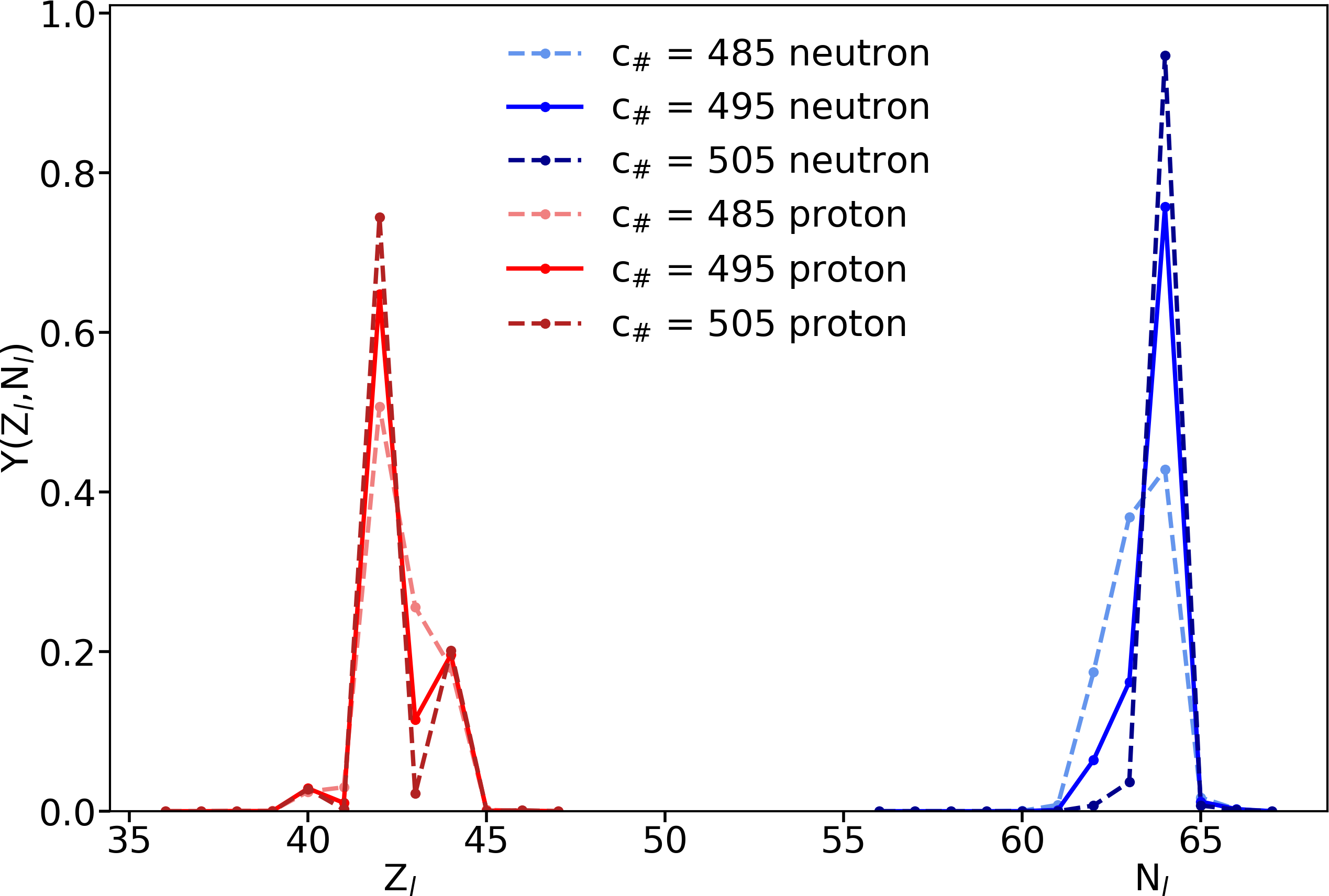}
\caption{Light fragment proton ($Z_l$) and neutron ($N_l$) particle-number distributions for selected configurations in the scission region.}
\label{ctwo_205b}
\end{figure}

Selected distributions in the vicinity of the scission region are presented in FIG. \ref{ctwo_205b} for three representative values of $c_\#$, namely $c_\#=$ 485, 495 and 505.
The results reveal a particularly striking feature: a pronounced odd–even staggering in the proton fragment distributions near scission. By contrast, the effect is absent in the neutron sector. Microscopically, this behavior can be attributed to the role of pairing correlations in the quantum partitioning of nucleons during the formation of the fragments.
This result arises directly from the intrinsic structure of the HFB states and does not rely on any phenomenological assumptions. Its relevance is further strengthened by the well-established presence of proton odd–even staggering in experimental fission charge yields \cite{morfou}. The fact that this pattern appears specifically in the region identified above as the scission domain therefore constitutes an additional microscopic signature supporting the physical characterization of that region.
\begin{figure}
\includegraphics[width=1.0\linewidth]{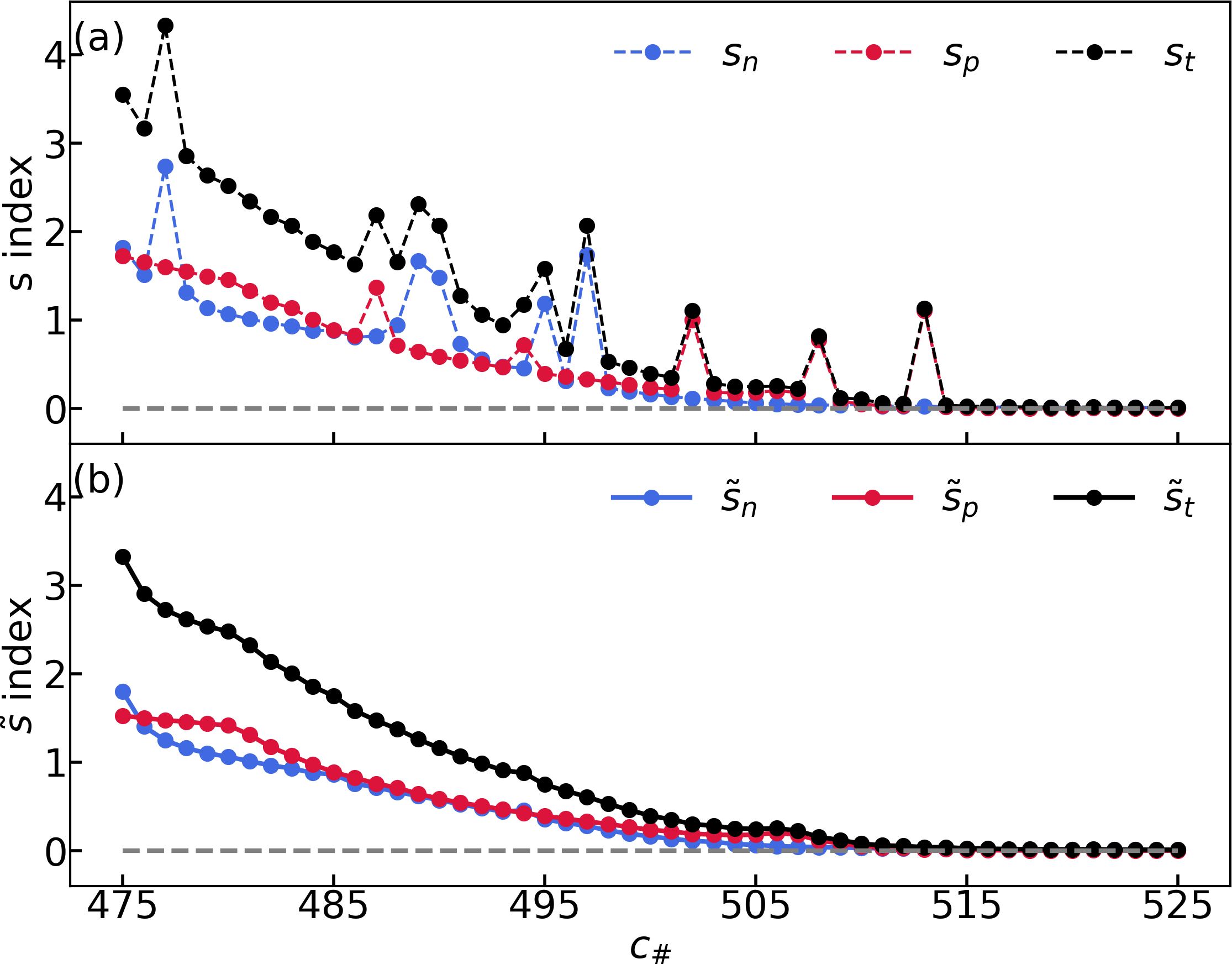}
\caption{Panel (a): Evolution of the indices $s_n$, $s_p$ and $s_t$ with respect to $c_\#$.
Panel (b): Same as for panel (a) but for the $\tilde s_n$, $\tilde s_p$ and $\tilde s_t$ indices.}
\label{ctwo_188}
\end{figure}

\subsection{Static energy balance at scission}

In this section, we analyze the static energy balance in the scission region. For this purpose, we introduce a microscopic procedure that separates each HFB state into left- and right-fragment sub-spaces. This decomposition makes it possible to evaluate both the fragment binding energies and the interaction energy between the fragments for the adiabatic path. The method builds upon the work of Ref. \cite{Rota}, which aims to minimize the interaction energy between fragments while searching for the appropriate unitary transformation. The key difference here lies in reformulating the method within the canonical representation of the HFB states.

\subsubsection{Quantum separation method in the scission area}

The starting point of the method is the decomposition of the QP content of a given HFB state into left and right subsets in the canonical basis. This decomposition, in turn, induces a corresponding partitioning of the nucleon density $\rho$ and pairing tensor $\kappa$:
\begin{eqnarray}
\rho = \rho^{(l)} + \rho^{(r)}, \qquad 
\kappa = \kappa^{(l)} + \kappa^{(r)},
\end{eqnarray}
where $\rho^{(l)}$ and $\rho^{(r)}$ ($\kappa^{(l)}$ and $\kappa^{(r)}$) represent the left and right nucleon densities (pairing tensor).

\begin{figure}
\includegraphics[width=0.8\linewidth]{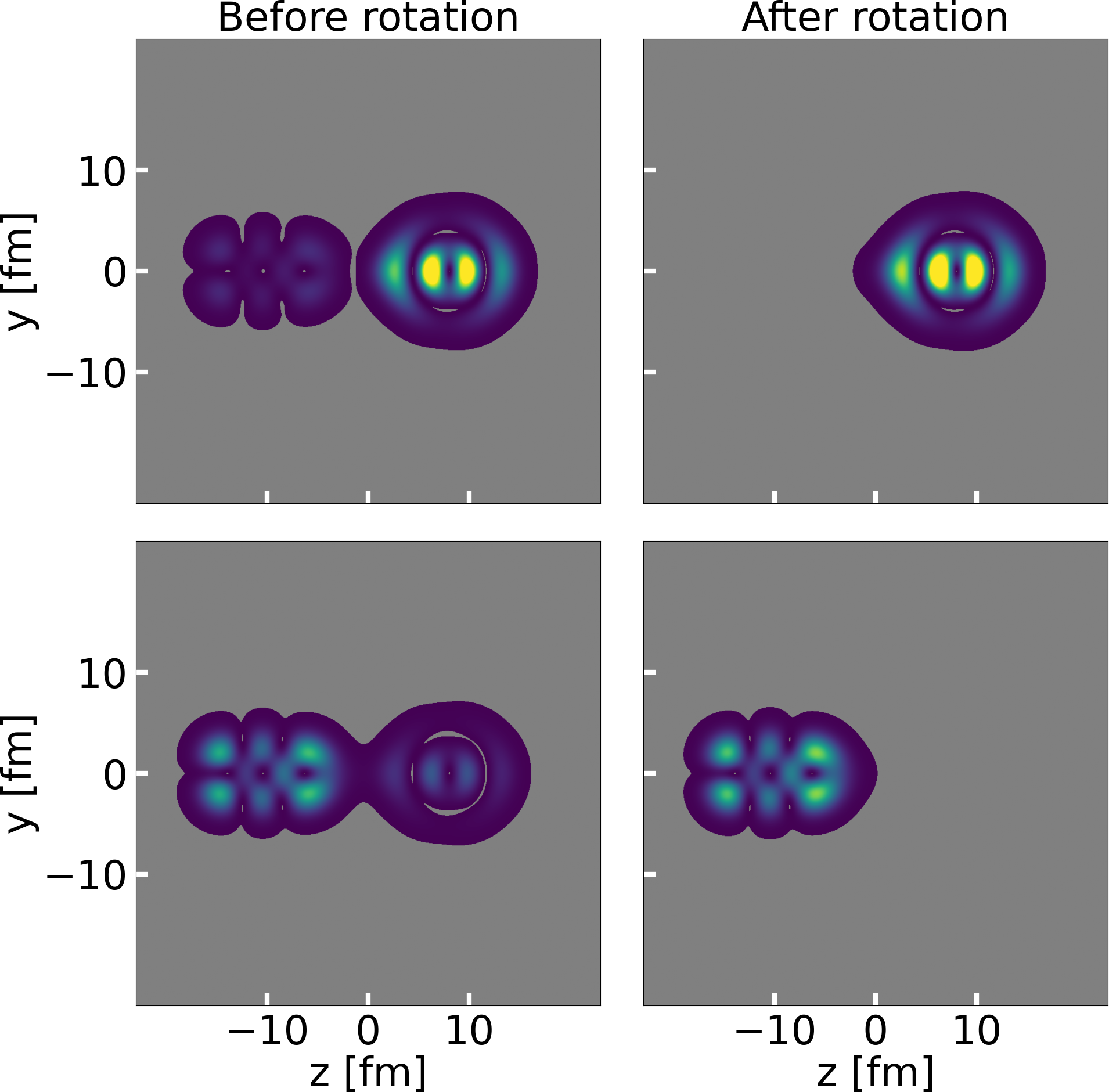}
\caption{Example of QP rotation improving the spatial localization of two nearly degenerate canonical states associated with the adiabatic HFB state at $c_\#=487$.}
\label{ctwo_189}
\end{figure}

The assignment of canonical QP states to the left or to the right fragment is based on their spatial localization with respect to the neck position $z_{\mathrm{neck}}$.
To determine which QP subset a given annihilation operator $\eta_k$ belongs to, we consider its contribution to the local density $\rho(\vec{r})$ (isospin indices are voluntarily omitted for simplicity purposes):
\begin{equation}\label{ctwo_184}
\begin{array}{lcl}
\displaystyle \rho(\vec{r}) &=&  \displaystyle \sum_k \rho^{k}(\vec{r}),
\end{array}
\end{equation}
where $\rho^{k}(\vec{r})$ represents the contribution of the QP $k$ to the local nucleon density.
Then, we introduce the two quantities $(v^{(l)}_{k})^2$ and $(v^{(r)}_{k})^2$:
\begin{eqnarray}\begin{cases}
\displaystyle (v^{(l)}_{k})^2= \int d\vec{r}_\perp \int_{-\infty}^{z_{neck}}\rho^{k}(\vec{r})\\
\displaystyle (v^{(r)}_{k})^2 = \int d\vec{r}_\perp \int_{z_{neck}}^{+\infty}\rho^{k}(\vec{r})
\end{cases},
\end{eqnarray}
which evaluate the left and right contributions to the QP $\mu_k$ to the local density. 

As mentioned above, in the present work we propose to perform the separation in the canonical basis. This approach is particularly convenient, as canonical quasiparticle states are naturally localized in space near scission, thereby reducing mixing between the fragments. Additionally, the canonical basis provides a clear and intuitive interpretation of the separation process in terms of single-particle orbitals.

The quality of the separation is quantified through the separation index $s$:
\begin{equation}
s = 2 \sum_k 
\min\left(
\left(v_k^{(l)}\right)^2,
\left(v_k^{(r)}\right)^2
\right),
\end{equation}

\noindent where $k$ labels half of the canonical basis states, while the factor two accounts for the time-reversed states.

This index measures the amount of particle density spatially localized in one fragment while formally attributed to the other one.
The evolution of the neutron $s_n$, proton $s_p$ and total $s_t$ separation indices along the adiabatic path is shown in FIG. \ref{ctwo_188}, panel (a), in the scission area. A clear decrease of the indices is observed as the system approaches scission, reflecting the progressive spatial decoupling of the fragments. 

\begin{figure}
\includegraphics[width=1.0\linewidth]{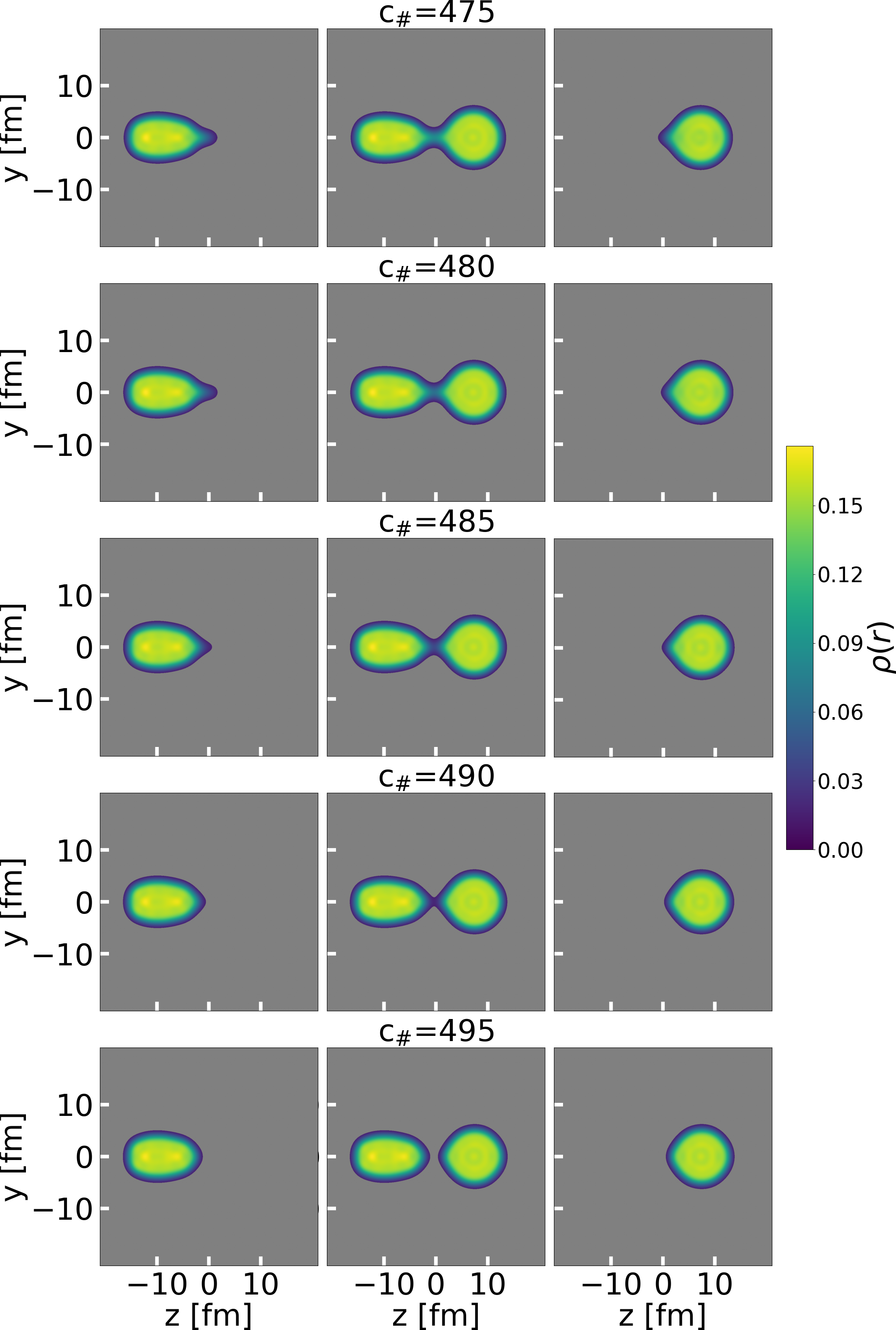}
\caption{Left fragment (left column) and right fragment (right column) total densities for selected adiabatic configurations labeled by different $c_\#$ values, in the scission area. The corresponding density of the full fissioning system is also represented in the middle column.}
\label{ctwo_191}
\end{figure}

However, a residual source of fragment mixing persists for nearly degenerate canonical states, resulting in spurious peaks in the separation indices, as observed in FIG. \ref{ctwo_188}. This issue can be significantly mitigated by performing additional pairwise rotations between QP states within the same $(\Omega,\tau)$ subspace (see Appendix \ref{rotation} for more details). Physically, these rotations disentangle orbitals that still contain small \enquote{mirror} components localized in the opposite fragment. The corresponding separation indices, obtained after rotations, will be denoted as $\tilde s$ in the following.
The effect of rotations is illustrated in FIG. \ref{ctwo_189} where we have represented the squared modulus of the wave functions $\varphi_1$ and $\varphi_2$ associated with the anomaly in the state labeled by $c_\# = 487$ before and after rotation. Before rotation, both states exhibit mixed left-right structures. After rotation, the orbitals become almost fully localized within a single fragment, substantially improving the separation quality. Indeed, as shown in FIG. \ref{ctwo_188}, panel (b), the total separation index $\tilde s_t$ exhibits smoother behavior and remains below one particle in the vicinity of the scission region ($c_\# \approx 495$). The associated neutron $\tilde s_n$ and proton $\tilde s_p$ separation indices are also presented. Both display similar behavior.

To conclude this methodological discussion, FIG. \ref{ctwo_191} displays the densities associated with the left and right sub-spaces for several adiabatic configurations for various $c_\#$ around the scission area. The progressive retraction of the fragment tails clearly demonstrates the gradual spatial decoupling of the two fragments as scission is approached. The resulting densities remain smooth and physically consistent throughout the separation process.

\subsubsection{Fragment deformation energies}

Once the fragment sub-states are constructed, the total HFB energy $E^{(tot)}$ can naturally be decomposed into a left and right contributions plus an interaction energy $E^{int}$:
\begin{equation}\label{balance}
E^{(tot)} = E^{(l)} + E^{(r)} + E^{{int}},
\end{equation}
where $E^{(l)}$ and $E^{(r)}$ denote the binding energies of the left and right fragments respectively. We can easily show that:
\begin{equation}\begin{cases}
\begin{array}{lcl}
\displaystyle E^{(l)} &=& \displaystyle \sum_{\alpha \beta} t_{\alpha \beta} \rho_{\alpha \beta}^{(l)} + \frac{1}{2} \sum_{\alpha \beta \gamma \delta}v_{\alpha \beta \gamma \delta}^{(a)}\rho_{\gamma \alpha}^{(l)} \rho_{\delta \beta}^{(l)} \\ &+& \displaystyle \frac{1}{4} \sum_{\alpha \beta \gamma \delta}(-1)^{s_\beta - s_\delta} v_{\alpha \beta \gamma \delta}^{(a)}\kappa_{\alpha \beta}^{(l)} \kappa_{\gamma \delta}^{(l)} 
\end{array}
\\
\begin{array}{lcl}
\displaystyle E^{(r)} &=& \displaystyle \sum_{\alpha \beta} t_{\alpha \beta} \rho_{\alpha \beta}^{(r)} + \frac{1}{2} \sum_{\alpha \beta \gamma \delta}v_{\alpha \beta \gamma \delta}^{(a)}\rho_{\gamma \alpha}^{(r)} \rho_{\delta \beta}^{(r)} \\ &+& \displaystyle \frac{1}{4} \sum_{\alpha \beta \gamma \delta}(-1)^{s_\beta - s_\delta} v_{\alpha \beta \gamma \delta}^{(a)}\kappa_{\alpha \beta}^{(r)} \kappa_{\gamma \delta}^{(r)}
\end{array}
\\
\begin{array}{lcl}
\displaystyle E^{int} &=& \sum_{\alpha \beta \gamma \delta}v_{\alpha \beta \gamma \delta}^{(a)}\rho_{\gamma \alpha}^{(l)} \rho_{\delta \beta}^{(r)} \\ &+& \displaystyle \frac{1}{2} \sum_{\alpha \beta \gamma \delta}(-1)^{s_\beta - s_\delta} v_{\alpha \beta \gamma \delta}^{(a)}\kappa_{\alpha \beta}^{(l)} \kappa_{\gamma \delta}^{(r)}
\end{array}
\end{cases},
\end{equation}
 where  $v^{(a)}_{\alpha \beta \gamma \delta}$ denotes the antisymmetrized two-body matrix elements of the D1S Gogny interaction, and $t_{\alpha \beta}$ denotes the matrix elements of the kinetic-energy operator.
 
 To evaluate the contribution of the density-dependent term to $E^{(l)}$, $E^{(r)}$ and $E_{int}$, we have used the prescription given in \cite{Rota}. This prescription involves using the one-body densities $\rho^{(l)}$ and $\rho^{(r)}$ to evaluate the binding energies $E^{(l)}$ and $E^{(r)}$ respectively. The interaction energy $E_{int}$ is then deduced from the difference between the total binding energy $E^{(tot)}$ and the left and right binding energies $E^{(l)}$ and $E^{(r)}$.

\begin{figure}
\includegraphics[width=1.0\linewidth]{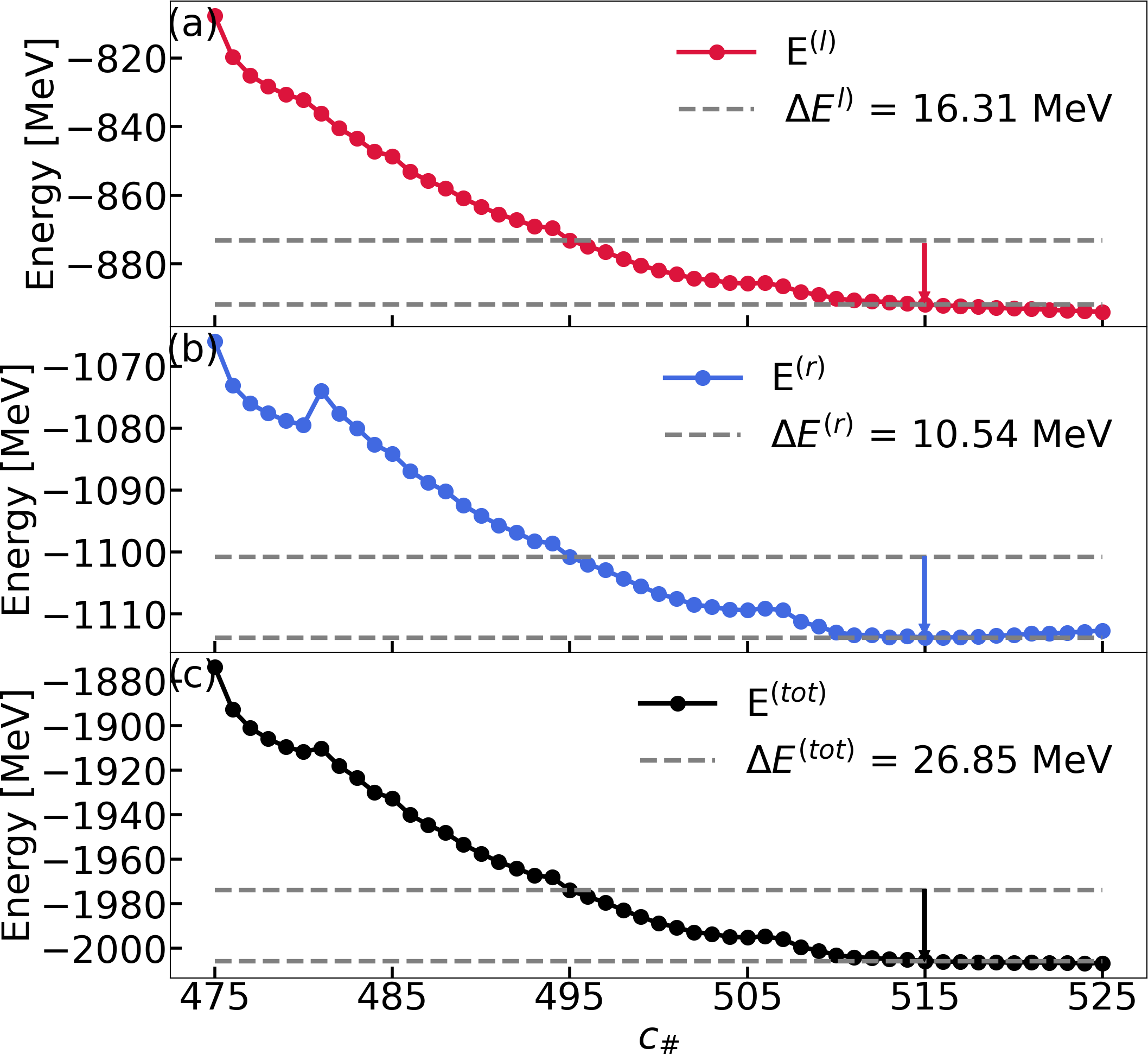}
\caption{Evolution of the fragment binding energies . Panel (a): Light fragment (left). Panel (b): Heavy fragment (right). Panel (c): Full fissioning nucleus.}
\label{ctwo_193}
\end{figure}

In FIG. \ref{ctwo_193}, the evolution of the fragment binding energies is shown, along with the total binding energy of the fissioning system. Colored arrows indicate the left, right, and total deformation energies, $\Delta E^{(l)}$, $\Delta E^{(r)}$ and $\Delta E^{(tot)}$, respectively. These are calculated as the difference between the scission point at $c_\#=495$ and $c_\# = 515$.
Indeed, as seen in panel (c), after scission, around $c_\# = 515$, $E^{(tot)}$ tends to stabilize, indicating that both fragments have reached their ground-state deformations.
The difference between the fragment energies at scission and after relaxation thus provides an estimate of the deformation energy stored in the fragments at the moment of scission.
Thus, we obtain a total deformation energy of approximately
$\Delta E^{(\mathrm{tot})} \approx 26.85 \mathrm{MeV}$. Depending on the precise scission configuration, moderate variations of a few MeV are observed, which serve as an estimate of the associated uncertainty.
These findings suggest that a substantial amount of energy remains stored in fragment deformation at scission and may later contribute to fragment excitation and post-scission relaxation.
We recall that this estimate depends on the separation method used to isolate the fragments. Further investigation is needed.

\subsubsection{Interaction energy at scission}

With the previous separation method, we have also a direct access to the interaction energy $E^{int}$ between the fragments (see Eq. \ref{balance}). It contains both a Coulomb and a nuclear terms:
\begin{equation}
E^{int} = E^{int}_{Coulomb} + E^{int}_{nuclear},
\end{equation}
where $E^{int}_{nuclear}$ is built from all the terms of
the D1S Gogny interaction.

\begin{figure}
\includegraphics[width=1.0\linewidth]{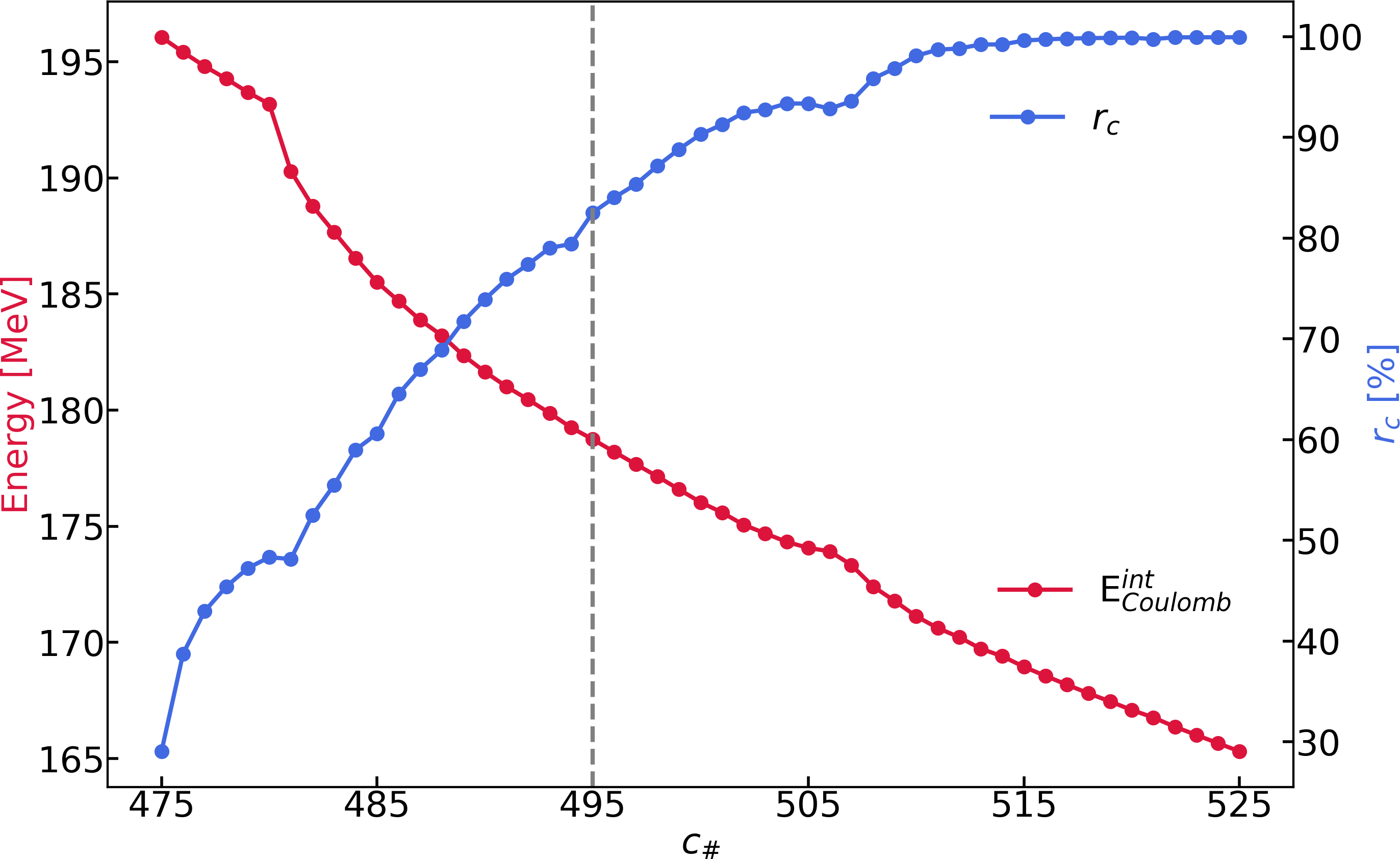}
\caption{Coulomb interaction energy between the fragments (red curve) and ratio $r_c$ (blue curve).}
\label{ctwo_192}
\end{figure}

\begin{figure}
\includegraphics[width=1.0\linewidth]{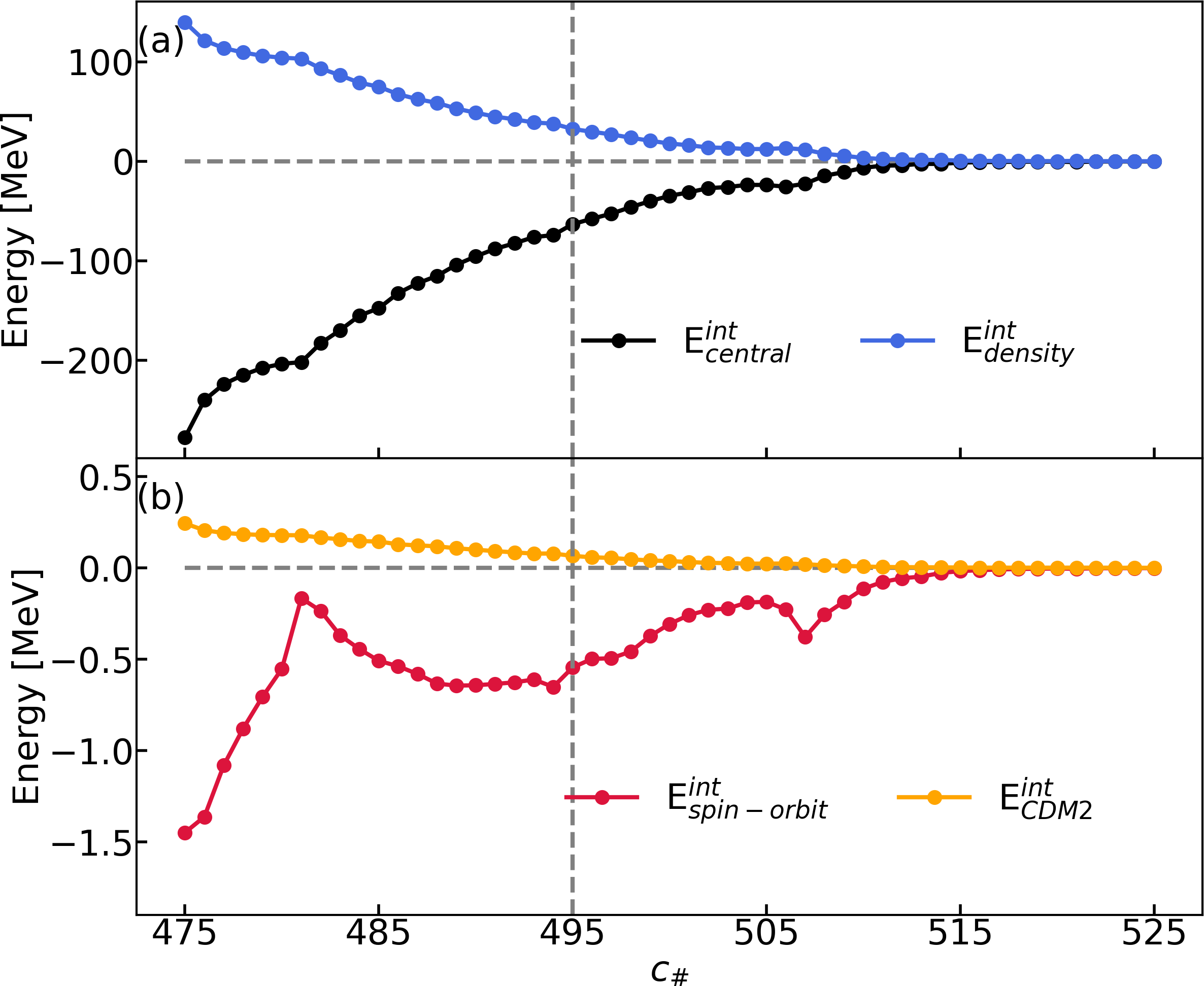}
\caption{Panel (a): Evolution of the central and density components of the interaction energy $E^{int}$ with respect to $c_\#$. Panel (b): Same as panel (a) but for the spin-orbit and two-body center of mass contribution.}
\label{ctwo_194}
\end{figure}

The evolution of the Coulomb interaction energy is shown in FIG. \ref{ctwo_192} (in red) in the scission area. At scission ($c_\# = 495$, grey dashed line), a value of
$E^{int}_{{Coulomb}} \approx 178.74$ MeV
is found. This quantity represents the Coulomb contribution to the final total kinetic energy (TKE) of the fragments.
Furthermore, this analysis reveals that a noticeable nuclear interaction between the fragments persists in this region. At scission, the total interaction energy still reaches approximately $E_{\mathrm{int}} \approx 153 \ \mathrm{MeV}$, which remains comparable to the Coulomb interaction itself.
In FIG. \ref{ctwo_194}, the evolution of the central, density, spin-orbit and two-body center-of-mass (CDM2) contributions to the nuclear interaction energy is shown. It is clear that, while the spin-orbit and two-body center-of-mass contributions (panel (b)) are negligible, most of the nuclear interaction energy arises from the central and density-dependent terms (panel (a)) with opposite signs.
This result may suggest that the fragments are far from being fully independent at scission. Instead, the system remains strongly coupled through residual nuclear interactions even when the neck has already ruptured.
This is revealed by the ratio $r_c = 100 \times  \frac{E^{int}_{Coulomb}}{E^{int}}$ shown in FIG. \ref{ctwo_192}, blue curve. Its value reaches around $\simeq$83\% at $c_\# = 495$.
This observation may have important consequences for the physical interpretation of the scission and for the evaluation of the post-scission kinetic energy. In particular, it suggests that approximating the fragment interaction solely through Coulomb repulsion may miss an important part of the scission dynamics. This point
will be adressed in the third article of the trilogy \cite{trilogy3}.

Overall, the separation method provides a coherent microscopic picture of the energy balance at scission. Using this approach, the analysis simultaneously reveals the gradual spatial decoupling of the fragments, the persistence of significant residual nuclear interaction and the presence of substantial deformation energy stored in the fragments at scission.
These results further support the interpretation of scission as an extended and highly collective reorganization process, rather than an abrupt geometric rupture. However, further investigations into the separation method are warranted to improve the energy balance.

\section{Conclusions and perspectives}\label{sec4}

In this first article of the trilogy, we have discussed the necessary conditions for a concrete implementation of the SCIM formalism. This paper focuses on the adiabatic path and the importance of ensuring both high-level continuity and  regularity properties for applying the SCIM.
In particular, we have presented two newly proposed protocols from Ref. \cite{carpentier2024}, the \enquote{Link} and \enquote{Drop} methods, to design a 1D path that enables the description of a fissioning compound nucleus, from its ground state deformation up to scission and beyond. A comparison between the exact Gaussian Overlap Approximation and the SCIM for the adiabatic states has also been performed and has revealed close properties between both approaches.

The application of these two new protocols to $^{240}$Pu, using the D1S Gogny interaction and the two-center HFB3 solver, has enabled us to establish key properties at scission. These include chemical potential peaks, the composition of the neck, where neutrons dominate, the emergence of odd components in the HFB wave function leading to an odd-even staggering effect on the proton side, and a static energy balance. This balance includes the evaluation of the deformation energy of the fragments as well as the interaction energy, which comprises both Coulomb and nuclear contributions.

This work opens several perspectives for future developments. Some of these will be explored in the second and third papers of this trilogy \cite{trilogy2,trilogy3}, which will focus on the construction of excited states and on dynamical calculations. More specifically, regarding adiabatic states, the present study lays the groundwork for a two-dimensional extension of the SCIM, which should enable a more meaningful comparison with experiment. At this stage, we envision two possible approaches: a relatively simple one and a more ambitious one.

The simplest one is related to the implementation of the \enquote{Link} and \enquote{Drop} methods introducing additional multipole constraints. During this work, we have tested the compatibility of both the \enquote{Link} and the \enquote{Drop} protocols with a constraint on the octupole moment $Q_{30}$ \cite{TPaul}. This clearly opens the possibility of restoring continuity and smoothness at low computational cost, at least locally, starting from two-dimensional PES obtained with $\mathcal{P}_{20}$-type procedures.

\begin{figure}
\centering
\includegraphics[width=0.9\linewidth]{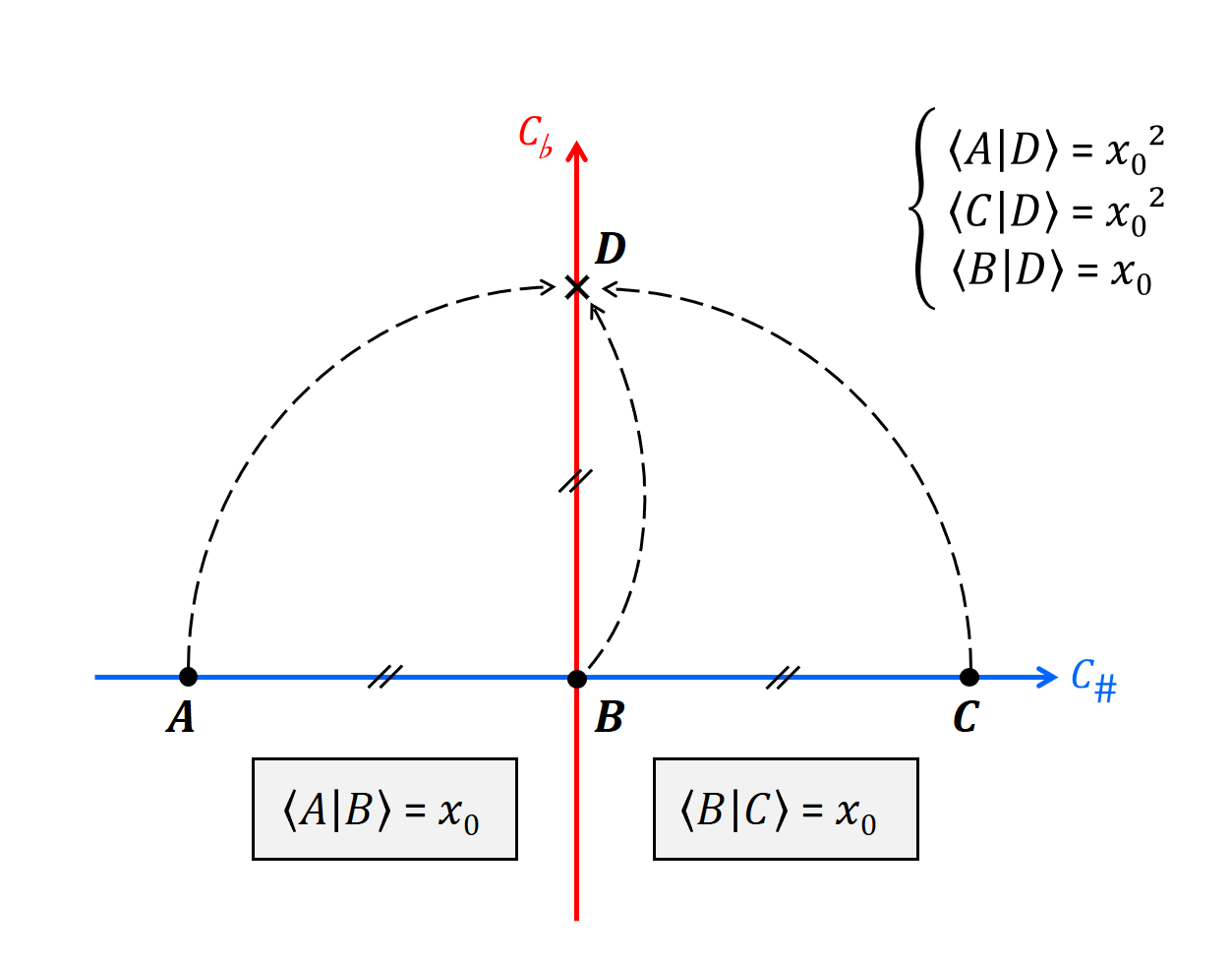}
\caption{Schematic view of the definition of a direction ``orthogonal'' to another.}
\label{conclu_1}
\end{figure}

Developing a new overlap constraint method to construct two-dimensional, continuous, and regular PES in terms of collective coordinates is critical. To this end, we have tested a novel approach called the \enquote{Nuclear Paving} (NP) method, based on a few simple yet powerful ideas.
The method introduces a new collective coordinate, $c_\flat$, defined as geometrically orthogonal to an existing coordinate $c_\#$ using the GOA predictions. As illustrated in Fig. \ref{conclu_1}, for three states (A, B, C) characterized by a fixed overlap $x_0$ along $c_\#$, the NP method constructs a fourth state D with overlap $x_0$ with state B and $x_0^2$ with states A and C.
Indeed, within the GOA paradigm, simultaneously imposing the three constraints $\langle A|D\rangle = x_0^2$, $\langle C|D\rangle = x_0^2$, and $\langle B|D\rangle = x_0$ defines a state D adjacent to state B, but along the direction associated with $c_\flat$, which is geometrically orthogonal to that associated with $c_\#$.

\begin{figure}
\centering
\includegraphics[width=0.9\linewidth]{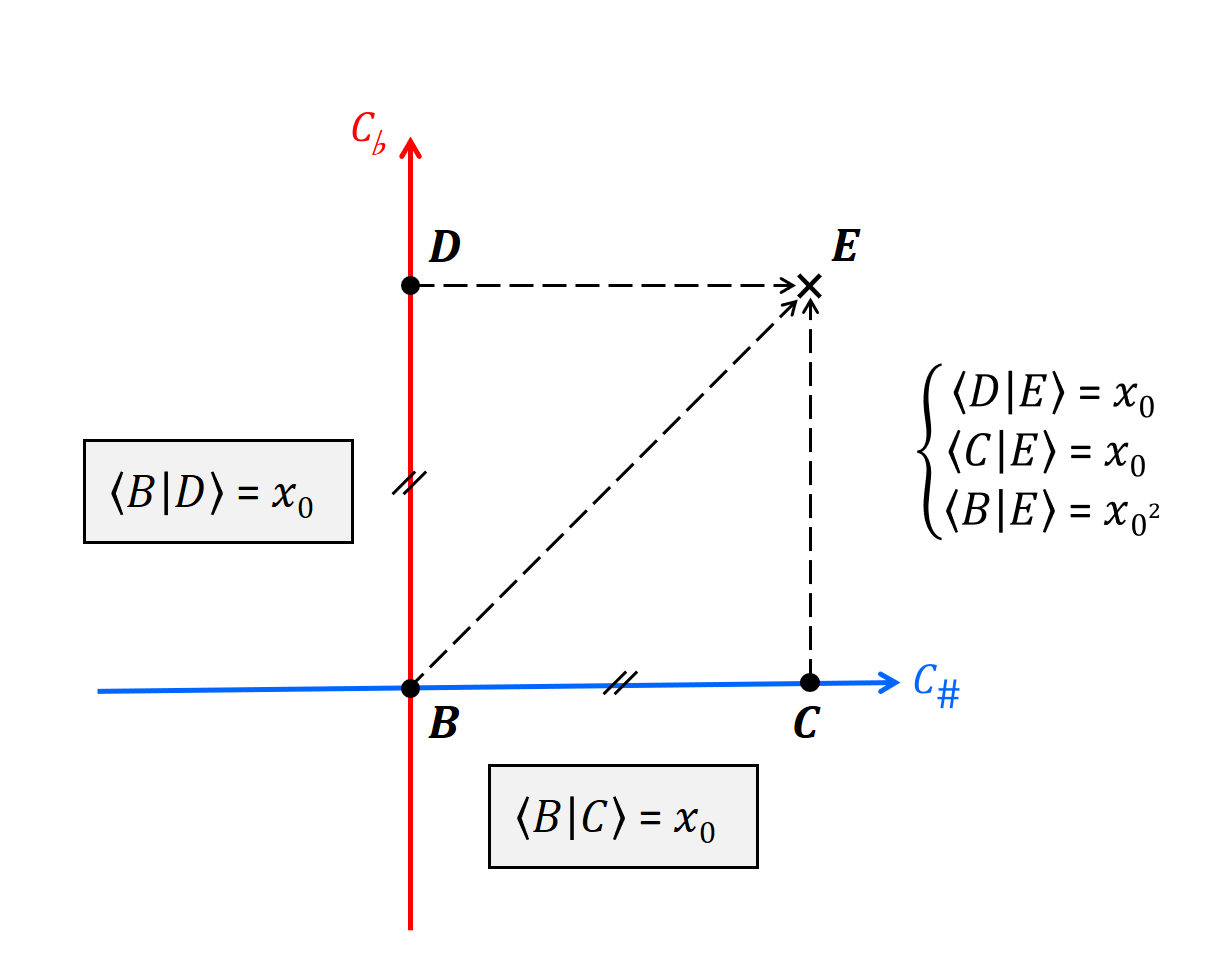}
\caption{Schematic view of a $c_\flat - c_\#$ grid construction using overlap constraints plus the GOA predictions.}
\label{conclu_2}
\end{figure}

Once such an orthogonal direction is defined, it becomes straightforward to construct a two-dimensional grid, guided by GOA predictions. As illustrated in FIG. \ref{conclu_2}, a state E can be defined from three states, D, B, and C, forming an isosceles triangle in the $c_\flat-c_\#$ plane.
In this case also, three different overlap constraints are imposed at the same time, namely $\bra{D}\ket{E} = x_0$, $\bra{C}\ket{E} = x_0$, and $\bra{B}\ket{E} = x^2_0$. 
\begin{figure}
\centering
\includegraphics[width=0.9\linewidth]{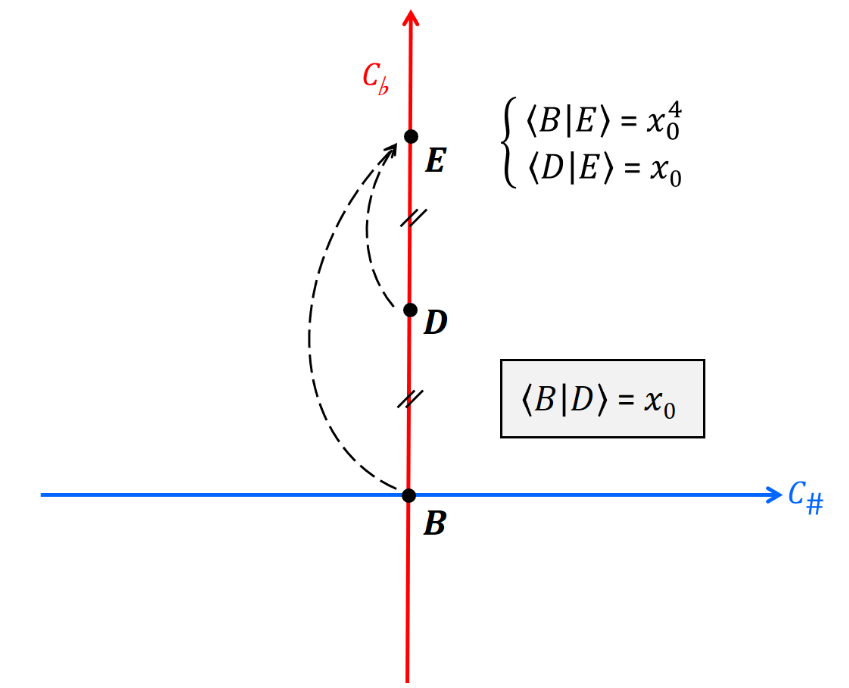}
\caption{Schematic view of $c_\flat$ climbing using the GOA predictions.}
\label{conclu_3}
\end{figure}
Finally, it is possible to jump from a $c_\flat$ level to another using the GOA predictions once again. In FIG. \ref{conclu_3}, we have represented this latter idea showing how a state E($c_\#$,$c_\flat$) is defined from two other states D ($c_\#$,$c_\flat - 1$) and B ($c_\#$,$c_\flat - 2$).

\begin{figure}
\centering
\includegraphics[width=1.0\linewidth]{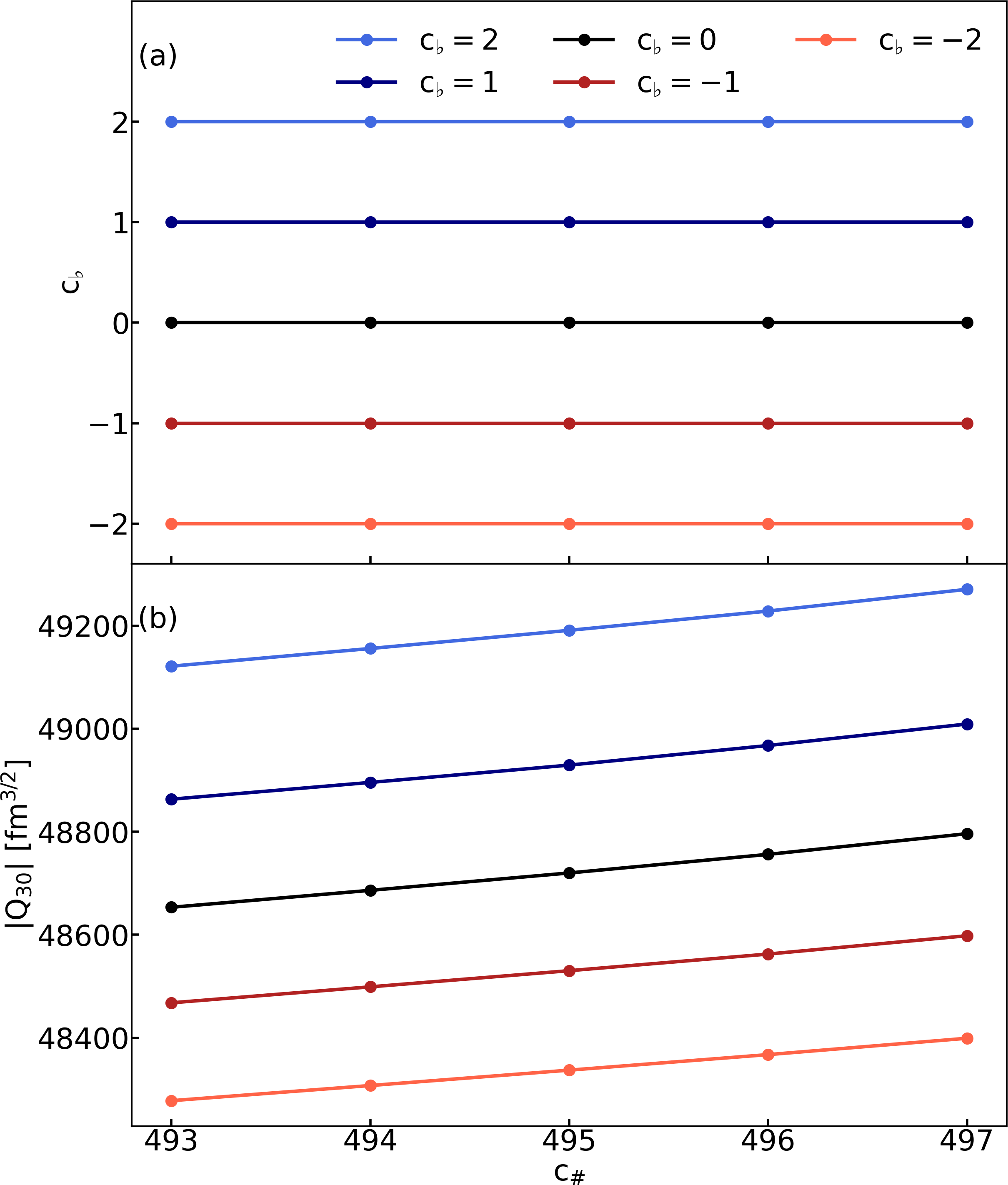}
\caption{Panel (a): 5 $\times$ 5 grid produced by the NP method according to $c_\#$ and $c_\flat$
Panel (b) 5 $\times$ 5 grid produced by the NP method according to $c_\#$ and $|Q_{30}|$.}
\label{conclu_4}
\end{figure}

\begin{figure}
\centering
\includegraphics[width=1.0\linewidth]{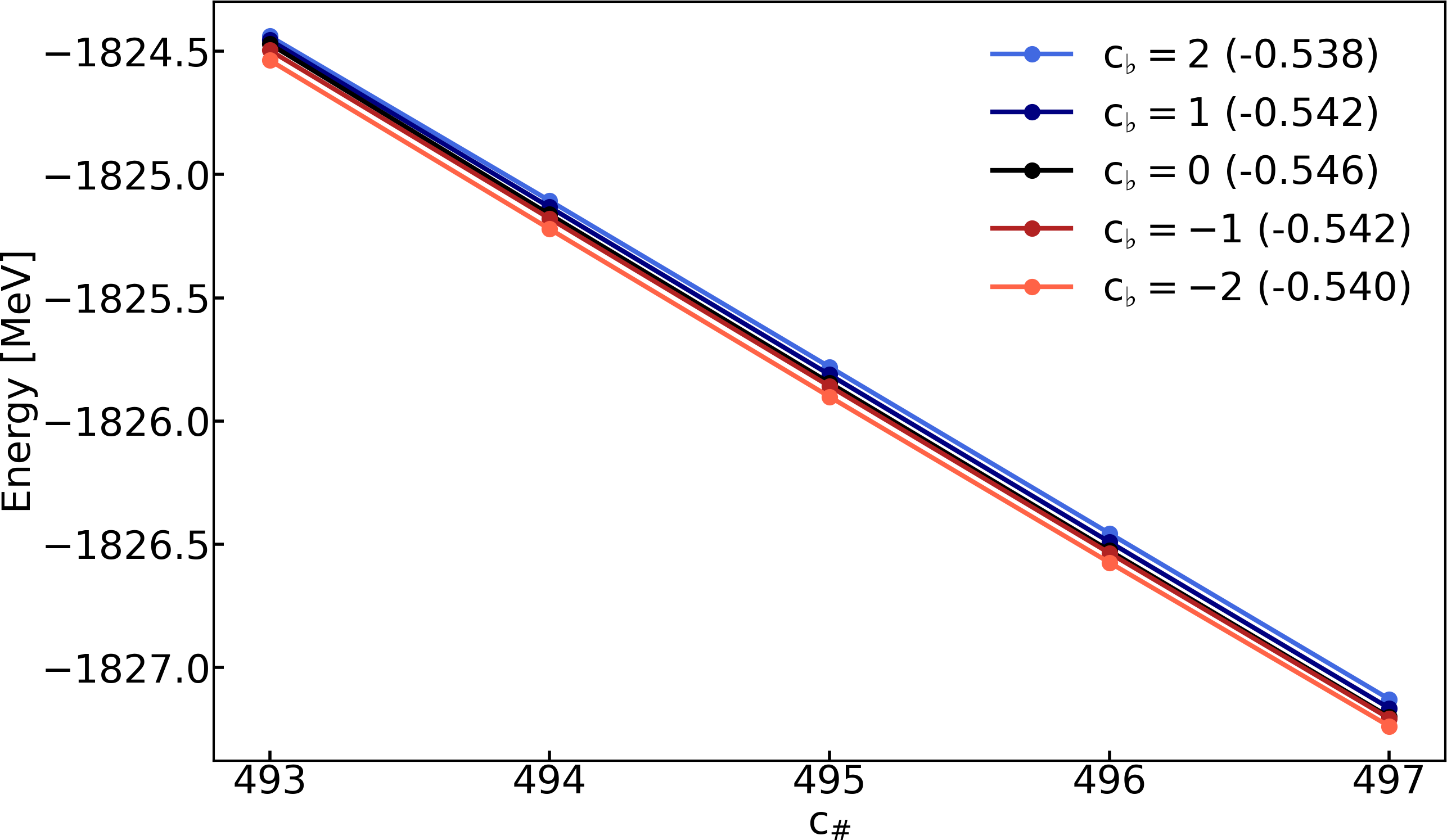}
\caption{Energy of the states belonging to the 5 $\times$ 5 grid obtained from the NP method. The energy gradient is indicated in parenthesis.}
\label{conclu_6}
\end{figure}

We have tested these ideas locally around the scission point ($c_\# = 495$). In FIG. \ref{conclu_4}, panels (a) and (b), we have displayed a 5 $\times$ 5 grid created with the NP method with respect to $c_\#$ and $c_\flat$ and $c_\#$ and $|Q_{30}|$, respectively. The black line labeled with $c_\flat = 0$ corresponds to the Drop results.
The new coordinate $c_\flat$ captures an asymmetry gradient, as seen in panel (b). In addition, FIG. \ref{conclu_6} shows the energy of all states in the grid, with the strongest energy gradient (in parentheses) observed for $c_\flat = 0$ states, as expected from the "Drop" method from which they were generated.

While the NP method shows promise, it may require refinement for global applications, particularly to address topological challenges. Relaxing certain constraints could mitigate these issues. Nevertheless, we believe this approach offers a viable path for extending overlap-based methods to 2D PES construction. Work in this direction is ongoing.\\ \\

\noindent \textbf{Acknowledgment}: N.P and P.C. would like to thank J.F. Berger
for his kindness throughout this work. N.P. dedicates this first application 
of the SCIM approach to the memory of D. Gogny. The work of W.Y. was supported by the U.S. Department of Energy, Office of Science, Office of Nuclear Physics under the contract No.
DE-AC02- 05CH11231 (LBNL). The work of L.M.R. is supported by Spanish Agencia Estatal de Investigacion 
(AEI) of the Ministry of Science and Innovation under Grant No. 
PID2024-159559NB-C21.

\appendix

\section{Symmetric Ordered Products of Operators}\label{appendixa}

The Symmetric Ordered Product of Operators (SOPO), hereafter referred to as SOPO, provides a convenient way to express the product of operators in a compact form. It is defined as follows \cite{rsbook}:
\begin{eqnarray}
\displaystyle \left[AB\right]^{(n)} = \sum_{k=0}^{n}\begin{pmatrix}n \\ k \end{pmatrix}
 B^{k}AB^{n-k}\qquad  \forall n \in \mathbb{N}.
\end{eqnarray}
\noindent It is clear that a SOPO is $\mathbb{C}$-linear on the left such that:
\begin{eqnarray}
\displaystyle \left[\lambda (A+C)B\right]^{(n)} = \lambda  ([AB]^{(n)} + [CB]^{(n)}) 
\end{eqnarray}
\noindent for all $\lambda \in \mathbb{C}$ and $n \in \mathbb{N}$.\\

In the SCIM formalism, the SOPOs is always present with a derivative operator on the right:
\begin{eqnarray}\label{sopo_0}\left[A(q)\frac{\partial}{\partial q}\right]^{(n)} = \sum_{k=0}^{n}\begin{pmatrix}n \\ k \end{pmatrix}
 \frac{\partial^k}{\partial q^k}A(q) \frac{\partial^{n-k}}{\partial q^{n-k}},
\end{eqnarray}
where the derivative operator $\partial /\ \partial q$ acts not only on $A$ but also on everything to its right.

Products of SOPOs frequently appear in the SCIM framework. It is therefore useful to derive a closed-form expression for handling them. These products explicitly read as:
\begin{eqnarray}\label{Product_1}
\left[A(q)\frac{\partial}{\partial q}\right]^{(n)} \left[B(q)\frac{\partial}{\partial q}\right]^{(p)} =
\sum_{k=0}^{n}\sum_{l=0}^{p} \begin{pmatrix}n \\ k \end{pmatrix}\begin{pmatrix}p \\ l \end{pmatrix} \times \notag \\
 \frac{\partial^k}{\partial q^k}A(q) \frac{\partial^{n-k}}{\partial q^{n-k}}
 \frac{\partial^l}{\partial q^l}B(q) \frac{\partial^{p-l}}{\partial q^{p-l}}. \quad \quad \quad
\end{eqnarray}
The commutation of the derivatives in the middle of Eq.(\ref{Product_1}) leads to the following expression:
\begin{eqnarray}
\displaystyle \left[A(q)\frac{\partial}{\partial q}\right]^{(n)} \left[B(q)\frac{\partial}{\partial q}\right]^{(p)} =
\sum_{k=0}^{n}\sum_{l=0}^{p}\sum_{\alpha=0}^{l}\sum_{s=0}^{n-k} \notag \\ \times \begin{pmatrix}n \\ k \end{pmatrix}\begin{pmatrix}p \\ l \end{pmatrix} \notag \begin{pmatrix}l \\ \alpha \end{pmatrix}\begin{pmatrix}n-k \\ \displaystyle  s\end{pmatrix} \\
 \times \frac{\partial^{(k+l)-\alpha}}{\partial q^{(k+l)-\alpha}}(-1)^{\alpha}A^{(\alpha)}(q) B^{(s)}(q)  \notag \displaystyle \frac{\partial^{(p + n) - (k+l) -s}}{\partial q^{(p + n) - (k+l) -s}}. 
\end{eqnarray}
\noindent Changing the order of the sums provides:
\begin{eqnarray}
\left[A(q)\frac{\partial}{\partial q}\right]^{(n)} \left[B(q)\frac{\partial}{\partial q}\right]^{(p)} =
\sum_{s=0}^{n}\sum_{\alpha=0}^{p} \sum_{l=\alpha}^{p}\sum_{k=0}^{n-s} \notag \\ \times \begin{pmatrix}n \\ k \end{pmatrix} \begin{pmatrix}p \\ l \end{pmatrix}   \begin{pmatrix}l \\ \alpha \end{pmatrix}\begin{pmatrix}n-k \\ s\end{pmatrix} \\
 \times \frac{\partial^{(k+l)-\alpha}}{\partial q^{(k+l)-\alpha}}(-1)^{\alpha}A^{(\alpha)}(q) B^{(s)}(q)  \notag  \frac{\partial^{(p + n) - (k+l) -s}}{\partial q^{(p + n) - (k+l) -s}}.
\end{eqnarray}
Then, setting $l'= l-\alpha$:
\begin{eqnarray}
\displaystyle \left[A(q)\frac{\partial}{\partial q}\right]^{(n)} \left[B(q)\frac{\partial}{\partial q}\right]^{(p)} =
\sum_{s=0}^{n}\sum_{\alpha=0}^{p} \sum_{l=0}^{p-\alpha}\sum_{k=0}^{n-s} \notag \\ \times \begin{pmatrix}n \\ k \end{pmatrix}\begin{pmatrix}p \\ l+\alpha \end{pmatrix}  \begin{pmatrix}l+\alpha \\ \alpha \end{pmatrix}\begin{pmatrix}n-k \\ s\end{pmatrix} \\
 \times \frac{\partial^{(k+l)}}{\partial q^{(k+l)}}(-1)^{\alpha}A^{(\alpha)}(q) B^{(s)}(q) \notag  \displaystyle \frac{\partial^{(p + n) - (k+l) -(s+\alpha)}}{\partial q^{(p + n) - (k+l) -(s+\alpha)}}, \notag 
\end{eqnarray}
\noindent and $z = k+l$:
\begin{eqnarray}
\displaystyle \left[A(q)\frac{\partial}{\partial q}\right]^{(n)} \left[B(q)\frac{\partial}{\partial q}\right]^{(p)} =
\sum_{s=0}^{n}\sum_{\alpha=0}^{p} \sum_{z=0}^{p+n-(s+\alpha)} \notag\\
\times \frac{\partial^{z}}{\partial q^{z}}(-1)^{\alpha}A^{(\alpha)}(q) B^{(s)}(q) \notag \\ \times \frac{\partial^{(p + n) -(s+\alpha)-z}}{\partial q^{(p + n) -(s+\alpha)-z}}  \sum_{k=\text{max}(0,z-(p-\alpha))}^{\text{min}(z,n-s)} \\ \times \begin{pmatrix}n \\ k \end{pmatrix}\begin{pmatrix}p \\ z-k +\alpha \end{pmatrix}\begin{pmatrix}z- k +\alpha \\ \alpha \end{pmatrix}\begin{pmatrix}n-k \\ s\end{pmatrix}. \notag
\end{eqnarray}
The binomial coefficients are then reorganized in such a way that:
\begin{eqnarray}\label{prod_2}
\displaystyle \left[A(q)\frac{\partial}{\partial q}\right]^{(n)} \left[B(q)\frac{\partial}{\partial q}\right]^{(p)} =
\sum_{s=0}^{n}\sum_{\alpha=0}^{p} \sum_{z=0}^{p+n-(s+\alpha)} \notag \\
\times \frac{\partial^{z}}{\partial q^{z}}(-1)^{\alpha}A^{(\alpha)}(q) B^{(s)}(q) \frac{\partial^{(p + n) -(s+\alpha)-z}}{\partial q^{(p + n) -(s+\alpha)-z}}  \\ \times\begin{pmatrix}n \\ s \end{pmatrix}\begin{pmatrix} p \\ \alpha \end{pmatrix}
\sum_{k=\text{max}(0,z-(p-\alpha))}^{\text{min}(z,n-s)}
\begin{pmatrix}n-s \\ k \end{pmatrix}
\begin{pmatrix}p- \alpha \\ z-k \end{pmatrix}. \notag
\end{eqnarray}
Using the following identity in Eq.(\ref{prod_2}):
\begin{eqnarray}\label{prod_3}
\sum_{k=\text{max}(0,z-(p-\alpha))}^{\text{min}(z,n-s)}
\begin{pmatrix}n-s \\ k \end{pmatrix}
\begin{pmatrix}p- \alpha \\ z-k \end{pmatrix} = \begin{pmatrix}n+p - (s+\alpha) \\ k \end{pmatrix}, \notag
\end{eqnarray}
leads to:
\begin{eqnarray}\label{sopo_1}
\displaystyle \left[A(q)\frac{\partial}{\partial q}\right]^{(n)} \left[B(q)\frac{\partial}{\partial q}\right]^{(p)} =
\sum_{s=0}^{n}\sum_{\alpha=0}^{p} \begin{pmatrix}n \\ s \end{pmatrix}\begin{pmatrix} p \\ \alpha \end{pmatrix}  \notag \\ \times \sum_{z=0}^{p+n-(s+\alpha)}\begin{pmatrix}n+p - (s+\alpha) \\ k \end{pmatrix} \quad \quad \quad 
\\ \displaystyle  \times \nonumber \frac{\partial^{z}}{\partial q^{z}}(-1)^{\alpha}A^{(\alpha)}(q) B^{(s)}(q) \frac{\partial^{(p + n) -(s+\alpha)-z}}{\partial q^{(p + n) -(s+\alpha)-z}}.
\end{eqnarray}
We clearly identify the expression of a SOPO on the right hand side of Eq.(\ref{sopo_1}). Thus,
\begin{eqnarray}
\displaystyle \left[A(q)\frac{\partial}{\partial q}\right]^{(n)} \left[B(q)\frac{\partial}{\partial q}\right]^{(p)} =
\sum_{s=0}^{n}\sum_{\alpha=0}^{p} \begin{pmatrix}n \\ s \end{pmatrix}\begin{pmatrix} p \\ \alpha \end{pmatrix} \notag \\ \times (-1)^{\alpha}\left[A^{(\alpha)}(q)B^{(s)}(q)\frac{\partial}{\partial q}\right]^{(n+p-(s+\alpha))}. \quad \quad \quad
\end{eqnarray}
Setting $i = \alpha + s$, we finally obtain for the product of two SOPOs:
\begin{widetext}
\begin{eqnarray}\label{final_prod}
\displaystyle \left[A(q)\frac{\partial}{\partial q}\right]^{(n)} \left[B(q)\frac{\partial}{\partial q}\right]^{(p)} =
\sum_{i=0}^{n+p}\sum_{s=\text{max}(0,i-p)}^{\text{min}(i,n)} \begin{pmatrix}n \\ s \end{pmatrix}  \begin{pmatrix} p \\ i-s \end{pmatrix} (-1)^{i-s}\left[A^{(i-s)}(q)B^{(s)}(q)\frac{\partial}{\partial q}\right]^{(n+p-i)}.
\end{eqnarray}    
The previous result can be generalized to the product of three SOPOs. The derivation is tedious. We only give here the final result:
\begin{eqnarray}
\left[A(q)\frac{\partial}{\partial q}\right]^{(n)} \left[B(q)\frac{\partial}{\partial q}\right]^{(p)} \left[C(q)\frac{\partial}{\partial q}\right]^{(r)}=
\sum_{i=0}^{n+p}   \sum_{s=\text{max}(0,i-p)}^{\text{min}(i,n)}   \begin{pmatrix}n \\ s \end{pmatrix} \begin{pmatrix} p \\ i-s \end{pmatrix} (-1)^{i-s} 
 \sum_{j=0}^{n+p+r-i} \sum_{z = \text{max}(0,j-r)}^{\text{min}(j,n+p-i)}   \quad \notag \\  \begin{pmatrix}n+p-i \\ z \end{pmatrix}  \begin{pmatrix} r \\ j-z \end{pmatrix} (-1)^{j-z}\left[(A^{(i-s)}(q)B^{(s)}(q))^{(j-z)}  C^{(z)}(q)\frac{\partial}{\partial q}\right]^{(n+p+r-i-j)}. \quad \quad \quad \quad \quad \quad     
\end{eqnarray}
\end{widetext}

\section{Inversion of the square root of the norm kernel $\mathcal{\bar N}$}\label{appendixb}

In this Appendix, the method we have used to calculate the inverse of 
the square root of the operator $\mathcal{\bar N}$ is explained. This inversion is not straightforward and follows the technics detailed in Refs.\cite{bernard1,TBernard}. 

We first recall that the expression of the special norm kernel $\mathcal{\bar N}$ at second order 
in SOPO is (see Eq. (\ref{eq:SCIM14})):
\begin{eqnarray}\label{appendixc1_1}
\mathcal{\bar N}(\bar q)  = \mathcal{N}^{(0)}(\bar q) + \left[\mathcal{N}^{(1)}(\bar q)\frac{\partial}{\partial q}\right]^{(1)} + \frac{1}{2}\left[\mathcal{N}^{(2)}(\bar q)\frac{\partial}{\partial q}\right]^{(2)}.
\end{eqnarray}
In Eq.\eqref{appendixc1_1}, the quantities $\mathcal{N}^{(p)}(\bar q)$ are matrices whose elements are noted $\mathcal{N}_{ji}^{(p)}(\bar q)$:
\begin{eqnarray}
\mathcal{N}^{(p)}(\bar q) = \begin{pmatrix}
      \mathcal{N}_{00}^{(p)}(\bar q) & \ldots & \mathcal{N}_{0n}^{(p)}(\bar q) \\
\vdots & \ddots & \vdots \\
\mathcal{N}_{n0}^{(p)}(\bar q) & \ldots & \mathcal{N}_{nn}^{(p)}(\bar q)\end{pmatrix}.
\end{eqnarray}
\noindent The definition of the moments implies that the matrices $\mathcal{N}^{(p)}(\bar q)$ are symmetric for even values of $p$ and skew-symmetric for odd ones. As $\mathcal{N}^{(0)}(\bar q)$ is a symmetric positive definite matrix for all $\bar q$, it is possible to define its square root:
\begin{eqnarray}
\mathcal{N}^{(0)1/2}(\bar q)\mathcal{N}^{(0)1/2}(\bar q) = \mathcal{N}^{(0)}(\bar q), \quad \quad \quad \forall \bar{q}.
\end{eqnarray}

\noindent The full operator $\mathcal{\bar N}(\bar q)$ is factorized with this square root. In the following, the $\bar q$ dependence is implicit in order to make the 
notations clearer. Thus,
\begin{eqnarray}\label{cone_11}
\mathcal{\bar N}  = \mathcal{N}^{(0)1/2}\left(I + \mathcal{N}^{(0)-1/2}\left[\mathcal{N}^{(1)}\frac{\partial}{\partial q}\right]^{(1)}\mathcal{N}^{(0)-1/2} + \notag \right.\\ \left. 
\frac{1}{2}\mathcal{ N}^{(0)-1/2}\left[\mathcal{N}^{(2)}\frac{\partial}{\partial q}\right]^{(2)}\mathcal{ N}^{(0)-1/2}\right)\mathcal{N}^{(0)1/2}. \quad \quad
\end{eqnarray}
\noindent Using the SOPO product formula derived in appendix \ref{appendixa}, one obtains:
\begin{eqnarray}
 \mathcal{N}^{(0)-1/2}\left[\mathcal{N}^{(1)}\frac{\partial}{\partial q}\right]^{(1)}\mathcal{N}^{(0)-1/2} = \qquad \qquad \qquad \\ \mathcal{N}^{(0)-1/2}\mathcal{N}^{(1)}\left(\mathcal{N}^{(0)-1/2}\right)'  - \left(\mathcal{N}^{(0)-1/2}\right)' \mathcal{N}^{(1)} \mathcal{N}^{(0)-1/2} \notag \\ +  \left[\mathcal{N}^{(0)-1/2} \mathcal{N}^{(1)}\mathcal{N}^{(0)-1/2}\frac{\partial}{\partial q}\right]^{(1)}, \notag
\end{eqnarray}
\noindent and
\begin{widetext}
\begin{eqnarray}
 \frac{1}{2}\mathcal{ N}^{(0)-1/2}\left[\mathcal{N}^{(2)}\frac{\partial}{\partial q}\right]^{(2)}\mathcal{ N}^{(0)-1/2} = 
 \frac{1}{2}\left((\mathcal{N}^{(0)-1/2})''\mathcal{N}^{(2)}\mathcal{N}^{(0)-1/2} + \mathcal{N}^{(0)-1/2}\mathcal{N}^{(2)}(\mathcal{N}^{(0)-1/2})'' \right. \notag\\ \left.  
 - 2 (\mathcal{N}^{(0)-1/2})'\mathcal{N}^{(2)}(\mathcal{N}^{(0)-1/2})'\right) \notag \\ + \left[(\mathcal{N}^{(0)-1/2}\mathcal{N}^{(2)}(\mathcal{N}^{(0)-1/2})'  - (\mathcal{N}^{(0)-1/2})'\mathcal{N}^{(2)}\mathcal{N}^{(0)-1/2}) \frac{\partial}{\partial q}\right]^{(1)} + \left[\frac{1}{2}\mathcal{N}^{(0)-1/2}\mathcal{N}^{(2)}\mathcal{N}^{(0)-1/2}\frac{\partial}{\partial q}\right]^{(2)}.
\end{eqnarray}
We define the following notations:
\begin{eqnarray}
\alpha_0 = \mathcal{N}^{(0)-1/2}\mathcal{N}^{(1)}\left(\mathcal{N}^{(0)-1/2}\right)' 
- \left(\mathcal{N}^{(0)-1/2}\right)' \mathcal{N}^{(1)} \mathcal{N}^{(0)-1/2} 
 + \frac{1}{2}\left( \left(\mathcal{N}^{(0)-1/2}\right)''\mathcal{N}^{(2)}\mathcal{N}^{(0)-1/2} \right.  \\ \left. + \mathcal{N}^{(0)-1/2}\mathcal{N}^{(2)} \left(\mathcal{N}^{(0)-1/2}\right)'' 
- 2 \left(\mathcal{N}^{(0)-1/2}\right)'\mathcal{N}^{(2)}\left(\mathcal{N}^{(0)-1/2}\right)'\right), \notag
\end{eqnarray}
\begin{eqnarray}
\mathcal{N}^{(1)}_{R_0} = \mathcal{N}^{(0)-1/2}\mathcal{N}^{(1)}\mathcal{N}^{(0)-1/2} + \mathcal{N}^{(0)-1/2}\mathcal{N}^{(2)}(\mathcal{N}^{(0)-1/2})'  - (\mathcal{N}^{(0)-1/2})'\mathcal{N}^{(2)}\mathcal{N}^{(0)-1/2},
\end{eqnarray}
\noindent and
\begin{eqnarray}
\mathcal{N}^{(2)}_{R_0} = \frac{1}{2}\mathcal{N}^{(0)-1/2}\mathcal{N}^{(2)}\mathcal{N}^{(0)-1/2}.
\end{eqnarray}
Thus, Eq.($\ref{cone_11}$) transforms into:
\begin{eqnarray}\label{cone_12}
\mathcal{\bar N}  = \mathcal{N}^{(0)1/2}\left(I + \alpha_0 + \left[\mathcal{N}_{R_0}^{(1)}\frac{\partial}{\partial q}\right]^{(1)} + \left[\mathcal{N}_{R_0}^{(2)}\frac{\partial}{\partial q}\right]^{(2)}\right)\mathcal{N}^{(0)1/2}.
\end{eqnarray}
\end{widetext}
Eq. (\ref{cone_12}) represents the first iteration of an iterative process where the Frobenius norm $||\alpha_0||$ is assumed to be small and decreases with each factorization step. In practice, the process is stopped at the i-th iteration when $||\alpha_i||<10^{-10}$. At the end of this process, we obtains the following expression:
\begin{eqnarray}
\mathcal{\bar N}(\bar q)  = \mathcal{F}(\bar q)\left(I + \left[\mathcal{N}_R^{(1)}(\bar q)\frac{\partial}{\partial q}\right]^{(1)} \notag \quad \quad \quad \qquad \right.\\ \left. + \left[\mathcal{N}_R^{(2)}(\bar q)\frac{\partial}{\partial q}\right]^{(2)}\right)\mathcal{F}^T(\bar q),
\end{eqnarray}
where $\mathcal{F}(\bar q)$ is defined by :
\begin{eqnarray}
\mathcal{F}(\bar q) = \mathcal{N}^{(0)1/2}(\bar q)\left(I+\alpha_0(\bar q)\right)^{1/2}... \notag \quad \quad \quad \\ \times \left(I+\alpha_{i-1}(\bar q)\right)^{1/2}.
\end{eqnarray}
Then, we define the quantity $\mathcal{J}(\bar q)$:
\begin{equation}\label{ji}
\begin{array}{lcl}
\mathcal{J}(\bar q) &=& \displaystyle I + \left[\mathcal{N}_R^{(1)}(\bar q)\frac{\partial}{\partial q}\right]^{(1)} + \left[\mathcal{N}_R^{(2)}(\bar q)\frac{\partial}{\partial q}\right]^{(2)} \\
 & = & \displaystyle I + \mathcal{U}(\bar q).
\end{array}
\end{equation}
We see that the inverse of the square root of $\mathcal{J}$ can be written as a series. The convergence of this series is likely the strongest approximation of the model, as it assumes that the derivatives of the overlap kernel's moments rapidly tend toward zero with increasing derivation order:
\begin{eqnarray}
\mathcal{J}^{-1/2}(\bar q) = I + \sum_{k=1}^{+\infty} \frac{(-1/2)...(-1/2 -k +1)}{k!}\mathcal{U}^{k}(\bar q). \notag
\end{eqnarray}
\noindent Keeping this series only up to the second order in SOPO, the following expression can be deduced for the inverse square root of $\mathcal{\bar N}$:
\begin{eqnarray}\label{cone_13}
\mathcal{\bar N}^{-1/2}(\bar q) = \mathcal{J}^{-1/2}(\bar q) \mathcal{F}^{-1}(\bar q).
\end{eqnarray}

\section{Explicit expression of the SCIM Hamiltonian $\mathcal{H}_{SCIM}$}\label{appendixc}

This appendix aims to provide explicit formulas of the SCIM potential $V$, dissipation tensor $D$, and inertia tensor $B$ that appear in the SCIM Hamiltonian $\mathcal{H}_{SCIM}$:
\begin{eqnarray}\label{scimH}
\mathcal{H}_{SCIM}(\bar q) = V(\bar q) + \left[D(\bar q)\frac{\partial}{\partial q}\right]^{(1)} + \left[B(\bar q)\frac{\partial}{\partial q}\right]^{(2)}.
\end{eqnarray}
In Eq.\eqref{eq:SCIM18}, we have defined the SCIM Hamiltonian $\mathcal{H}_{SCIM}$ as 
the result of the following product:
\begin{eqnarray}\label{am_0}
\mathcal{H}_{SCIM}(\bar q) = \mathcal{\bar N}^{-1/2}(\bar q) \mathcal{\bar H}(\bar q) \mathcal{\bar N}^{-1/2^T}(\bar q).
\end{eqnarray}
The operator $\mathcal{\bar N}^{-1/2}$ is defined in Appendix~\ref{appendixb}. In particular, it incorporates the operator $\mathcal{J}^{-1/2}$, for which we also provide an explicit expression.

\subsection{Expression of $\mathcal{J}^{-1/2}$}

In this section, we give the explicit expression of the operator $\mathcal{J}^{-1/2}$, which reads as:
\begin{eqnarray}
\mathcal{J}^{-1/2}(\bar q) = j_0(\bar q) + \left[j_1(\bar q) \frac{\partial}{\partial q}\right]^{(1)} + \left[j_2(\bar q) \frac{\partial}{\partial q}\right]^{(2)}.
\end{eqnarray}
From Eq.\eqref{ji}, the inverse square root of the operator $\mathcal{J}$ is calculated thanks to a Taylor series expansion truncated at the second order:
\begin{eqnarray}
\mathcal{J}^{-1/2}(\bar q) = I - \frac{1}{2} \mathcal{U}(\bar q) + \frac{3}{8} \mathcal{U}^2(\bar q).
\end{eqnarray}
To simplify the notation, the dependence of the operators on $\bar q$ is omitted. The quantity $\mathcal{U}$ can then be expressed as
\begin{equation}
\begin{array}{lcl}
\displaystyle \mathcal{U} &=& \displaystyle \left[\mathcal{N}_R^{(1)}\frac{\partial}{\partial q}\right]^{(1)} + \left[\mathcal{N}_R^{(2)}\frac{\partial}{\partial q}\right]^{(2)} \\ &=& \displaystyle \left[u_1 \frac{\partial}{\partial q}\right]^{(1)} + \left[u_2\frac{\partial}{\partial q}\right]^{(2)}.
\end{array}
\end{equation}
The expression of $j_0$ is:
\begin{eqnarray}\label{lab1}
j_0 = I - \frac{3}{8} u_1^{(1)} u_1^{(1)} + \frac{3}{8} \left(u_1^{(2)} u_2^{(1)} -  u_2^{(1)}u_1^{(2)}\right) \nonumber \\ + \frac{3}{8}u_2^{(2)}u_2^{(2)}. \qquad \qquad \quad 
\end{eqnarray}
For $j_1$, one has:
\begin{eqnarray}\label{lab2}
j_1 = \frac{3}{8}\left[ u_1 u_1^{(1)} - u_1^{(1)}u_1 + u_1^{(2)}u_2 + u_2u_1^{(2)} \quad \nonumber \right. \\ \left. - 2 \left(u_1^{(1)}u_2^{(1)} + u_2^{(1)}u_1^{(1)}\right) +2\left(u_2^{(2)}u_2^{(1)} - u_2^{(1)}u_2^{(2)}\right) \right]
\end{eqnarray}
while $j_2$ is given by:
\begin{eqnarray}\label{lab3}
j_2 = \frac{3}{8}\left[ u_1 u_1 + u_1u_2^{(1)} - u_2^{(1)}u_1 -2\left(u_1^{(1)}u_2 - u_2u_1^{(1)}\right) \right. \notag \\ \left. + u_2^{(2)} u_2 + u_2 u_2^{(2)} - 4( u_2^{(1)} u_2^{(1)}\right]. \qquad
\end{eqnarray}
In Eqs.(\ref{lab1}) to (\ref{lab3}), the expression of $\mathcal{J}^{-1/2}$ is truncated at second order in SOPO, in coherence with the SCIM formalism.

\subsection{Expression of $\mathcal{H}_{SCIM}$}

We start by rewriting the expression (\ref{scimH}) of $\mathcal{H}_{SCIM}$ in terms of the quantities $\mathcal{F}$ and $\mathcal{J}^{-1/2}$
\begin{eqnarray}
\mathcal{H}_{SCIM}(\bar q) = \mathcal{J}^{-1/2}(\bar q)\mathcal{F}^{-1}(\bar q) \quad \quad \quad \nonumber \\ \times \mathcal{\bar H}(\bar q)\mathcal{F}^{-1^T}(\bar q)\mathcal{J}^{-1/2}(\bar q),
\end{eqnarray}
where the quantity $\mathcal{\bar H}$ has for expression:
\begin{eqnarray}
 \mathcal{\bar H}(\bar q) = \mathcal{H}^{(0)}(\bar q) + \left[\mathcal{H}^{(1)}(\bar q)\frac{\partial}{\partial q}\right]^{(1)} + \frac{1}{2} \left[\mathcal{H}^{(2)}(\bar q)\frac{\partial}{\partial q}\right]^{(2)}
\end{eqnarray}
\noindent Now, one defines the operator $h$ such that:
\begin{eqnarray}\label{am_2}
 h(\bar q) = h_0(\bar q) + \left[h_1(\bar q)\frac{\partial}{\partial q}\right]^{(1)} +  \left[h_2(\bar q)\frac{\partial}{\partial q}\right]^{(2)},
\end{eqnarray}
with
\begin{eqnarray}
 h(\bar q) = \mathcal{F}^{-1}(\bar q)\left(\mathcal{H}^{(0)}(\bar q) + \left[\mathcal{H}^{(1)}(\bar q)\frac{\partial}{\partial q}\right]^{(1)} \quad \quad \nonumber \right. \\ \left. + \frac{1}{2} \left[\mathcal{H}^{(2)}(\bar q)\frac{\partial}{\partial q}\right]^{(2)} \right) \mathcal{F}^{-1T}(\bar q).
\end{eqnarray}
In the following, the $\bar q$ dependency of the operators is omitted for simplification purposes. The quantity $h_0$ has for expression:
\begin{equation}
\begin{array}{lcl}
h_0 &=& \displaystyle \mathcal{F}^{-1} \mathcal{H}^{(0)} \mathcal{F}^{-1^T} + \mathcal{F}^{-1} \mathcal{H}^{(1)} (\mathcal{F}^{-1^T})^{(1)} \nonumber \\ &-& \displaystyle (\mathcal{F}^{-1})^{(1)} \mathcal{H}^{(1)} \mathcal{F}^{-1^T} + \mathcal{F}^{-1} \mathcal{H}^{(2)} (\mathcal{F}^{-1^T})^{(2)} \nonumber \\ &+& \displaystyle (\mathcal{F}^{-1})^{(2)} \mathcal{H}^{(2)} \mathcal{F}^{-1^T} - 2  (\mathcal{F}^{-1})^{(1)}\mathcal{H}^{(2)} (\mathcal{F}^{-1^T})^{(1)}.
\end{array}
\end{equation}
The quantity $h_1$ is equal to:
\begin{eqnarray}
h_1 = \mathcal{F}^{-1} \mathcal{H}^{(1)} \mathcal{F}^{-1T} + 2 [\mathcal{F}^{-1} \mathcal{H}^{(2)} (\mathcal{F}^{-1T})^{(1)} \nonumber \\ - (\mathcal{F}^{-1})^{(1)} \mathcal{H}^{(2)} \mathcal{F}^{-1T}]. \quad \quad 
\end{eqnarray}
Finally, one obtains for $h_2$:
\begin{eqnarray}
h_2 = \mathcal{F}^{-1} \mathcal{H}^{(2)} \mathcal{F}^{-1T}.
\end{eqnarray}
The definition of $h$ in Eq.(\ref{am_2}) leads to the following expression for $\mathcal{H}_{SCIM}$:
\begin{widetext}
\begin{eqnarray}
\displaystyle \mathcal{H}_{SCIM} = \displaystyle \left(j_0 + \left[j_1 \frac{\partial}{\partial q}\right]^{(1)} + \left[j_2 \frac{\partial}{\partial q}\right]^{(2)} \right) \displaystyle  \left(h_0 + \left[h_1 \frac{\partial}{\partial q}\right]^{(1)} + \left[h_2 \frac{\partial}{\partial q}\right]^{(2)}\right) \displaystyle \left( j_0 + \left[ j_1 \frac{\partial}{\partial q}\right]^{(1)} + \left[j_2 \frac{\partial}{\partial q}\right]^{(2)}\right).  \nonumber
\end{eqnarray}
We start by giving the expression of the inertia tensor $B_{SCIM}$, which is the simplest one:
\begin{equation}
\begin{array}{lcl}
B_{SCIM} &=& j_0 h_2 j_0
 + j_0 h_0 j_2 + j_2 h_0 j_0
 + j_1 h_0 j_1
 + j_0 h_1 j_1 + j_1 h_1 j_0 +j_2 h_0^{(1)} j_1 - j_1 h_0^{(1)} j_2 \nonumber
\\ &-& j^{(1)}_2 h_0 j_1 + j_1 h_0 j^{(1)}_2
+2(j_2 h_0 j^{(1)}_1 - j^{(1)}_1 h_0 j_2) \nonumber
\\ &+& 3( - j_0^{(1)} h_1 j_2 + j_2 h_1 j_0^{(1)})
 + j_0 h_1 j^{(1)}_2 - j^{(1)}_2 h_1 j_0
 +2( - j_0 h^{(1)}_1 j_2 + j_2 h^{(1)}_1 j_0) \nonumber
\\ &+& 3( - j_0^{(1)} h_2 j_1 + j_1 h_2 j_0^{(1)})
 + 2( j_0 h_2 j^{(1)}_1 - j^{(1)}_1 h_2 j_0)
 - j_0 h^{(1)}_2 j_1 + j_1 h^{(1)}_2 j_0 \nonumber
\\ &+& j_2^{(2)} h_0 j_2 + j_2 h_0 j_2^{(2)}
 +2(-j_2 h^{(1)}_0 j^{(1)}_2 - j^{(1)}_2 h^{(1)}_0 j_2)
 -4 j_2^{(1)} h_0 j_2^{(1)}
 -2 j_2 h_0^{(2)} j_2 \nonumber
\\ &+& 6 (j_0^{(2)} h_2 j_2 + j_2 h_2 j_0^{(2)})
 + j_0 h_2 j_2^{(2)} + j_2^{(2)}h_2 j_0
 + 6 (-j_0^{(1)} h_2 j^{(1)}_2 - j^{(1)}_2 h_2 j_0^{(1)}) \nonumber
\\ &+& 6 (j_0^{(1)} h^{(1)}_2 j_2 + j_2 h^{(1)}_2 j_0^{(1)})
 + 4 (-j_0 h^{(1)}_2 j^{(1)}_2 - j^{(1)}_2 h^{(1)}_2 j_0)
 + j_0 h_2^{(2)} j_2 + j_2 h_2^{(2)} j_0 \nonumber
\\ &+& 2(j_1 h_1 j_1^{(1)} - j_1^{(1)} h_1 j_1)
 +3(j_1 h_2 j_1^{(2)} + j_1^{(2)} h_2 j_1)
 - j_1 h^{(1)}_2 j_1^{(1)} - j_1^{(1)} h^{(1)}_2 j_1
 - 7 j_1^{(1)}h_2 j_1^{(1)} \nonumber
 \\ &-& j_1 h_2^{(2)} j_1
  + j_2 h_1^{(1)} j_1^{(1)} + j_1^{(1)}h_1^{(1)} j_2
  + 5(- j^{(1)}_2 h_1 j_1^{(1)} - j_1^{(1)}h_1 j^{(1)}_2)
  + 3(j_2 h_1 j_1^{(2)} + j_1^{(2)}h_1 j_2) \nonumber
 \\ &+& 3( - j^{(1)}_2 h_1^{(1)} j_1 - j_1 h_1^{(1)} j^{(1)}_2)
  - j_2 h_1^{(2)} j_1 - j_1 h_1^{(2)} j_2
  + j^{(2)}_2 h_1 j_1 + j_1 h_1 j^{(2)}_2 \nonumber
  \\ &+& 19 j_2^{(2)} h_2 j^{(2)}_2
   + 16 j_2^{(1)} h^{(2)}_2 j^{(1)}_2
   + j_2 h_2^{(4)} j_2
   + j_2 h_2 j_2^{(4)} + j_2^{(4)}h_2 j_2 \nonumber
  \\ &+& 10(- j_2^{(1)} h_2 j_2^{(3)} - j_2^{(3)}h_2 j_2^{(1)})
    +8(j_2^{(1)} h^{(1)}_2 j_2^{(2)} + j_2^{(2)}h^{(1)}_2 j_2^{(1)}) \nonumber
  \\ &+& 6(- j_2 h^{(1)}_2 j_2^{(3)} - j_2^{(3)}h^{(1)}_2 j_2)
   +5(- j_2 h^{(2)}_2 j_2^{(2)} - j_2^{(2)}h^{(2)}_2 j_2) \nonumber
  \\ &+& 2( j_2 h^{(3)}_2 j_2^{(1)} + j_2^{(1)}h^{(3)}_2 j_2)
   +7( j_2^{(2)} h_1 j_2^{(1)} - j_2^{(1)} h_1j_2^{(2)})
   + j_2 h_1 j_2^{(3)} - j_2^{(3)} h_1j_2 \nonumber
  \\ &+& 4( - j_2 h^{(1)}_1 j_2^{(2)} + j_2^{(2)} h^{(1)}_1 j_2)
   +4( - j_2 h^{(2)}_1 j_2^{(1)} + j_2^{(1)} h^{(2)}_1 j_2) \nonumber
  \\ &+& 4(j_2 h_2 j_1^{(3)} - j_1^{(3)}h_2 j_2 )
  - j^{(3)}_2 h_2 j_1 + j_1 h_2 j^{(3)}_2
   + 12(- j^{(1)}_2 h_2 j_1^{(2)} + j_1^{(2)}h_2 j^{(1)}_2 ) \nonumber
 \\ &+& 8( j^{(2)}_2 h_2 j_1^{(1)} - j_1^{(1)}h_2 j^{(2)}_2 )
  + 5( j^{(2)}_2 h^{(1)}_2 j_1 - j_1 h^{(1)}_2 j^{(2)}_2 )
  + 8( - j^{(1)}_2 h^{(1)}_2 j_1^{(1)} + j_1^{(1)}h^{(1)}_2 j^{(1)}_2 ) \nonumber
\\ &+& 4( - j_2 h^{(2)}_2 j_1^{(1)} + j_1^{(1)}h^{(2)}_2 j_2 )
  + j^{(1)}_2 h^{(2)}_2 j_1 - j_1h^{(2)}_2 j^{(1)}_2
 - j_2 h^{(3)}_2 j_1 + j_1 h^{(3)}_2 j_2 
 \end{array}
\end{equation}
The explicit expression of the potential $V_{SCIM}$ is very complex. It is given by:
\begin{equation}
\begin{array}{lcl}
V_{SCIM} &=& j_0 h_0 j_0 + j_0 h_1 j_0^{(1)} - j_0^{(1)} h_1 j_0
 -j_0 h_0^{(1)} j_1 + j_1 h_0^{(1)} j_0 -j_0^{(1)} h_0 j_1 + j_1 h_0 j_0^{(1)} \nonumber
\\ &+& j_0 h_2 j_0^{(2)} + j_0^{(2)} h_2 j_0  -2 j_0^{(1)} h_2 j_0^{(1)} \nonumber
\\ &+& j_0^{(2)} h_0 j_2 + j_2 h_0 j_0^{(2)}
 + j_0 h_0^{(2)} j_2 + j_2 h_0^{(2)} j_0
 + 2 ( j_0^{(1)} h_0^{(1)} j_2 + j_2 h_0^{(1)} j_0^{(1)}) \nonumber
\\ &-& j_1 h^{(1)}_0 j_1^{(1)} - j_1^{(1)} h_0^{(1)} j_1
-j_1^{(1)} h_0 j_1^{(1)}
-j_1 h^{(2)}_0 j_1 \nonumber
\\ &-& j_0^{(1)} h_1 j_1^{(1)} - j_1^{(1)} h_1 j_0^{(1)}
 + j_0^{(2)} h_1 j_1 + j_1 h_1 j_0^{(2)}
 - j_0 h_1^{(1)} j_1^{(1)} - j_1^{(1)} h_1^{(1)} j_0 \nonumber
\\ &+& j_0^{(1)} h_1^{(1)} j_1 + j_1  h_1^{(1)} j_0^{(1)}
 - j^{(1)}_2 h_0 j_1^{(2)} + j_1^{(2)} h_0 j_2^{(1)}
 + 2(-j_2^{(1)}h_0^{(1)}j_1^{(1)} + j_1^{(1)}h_0^{(1)}j_2^{(1)}) \nonumber
\\ &-& j_2 h_0^{(1)}j_1^{(2)} + j_1^{(2)} h_0^{(1)}j_2
 + 2(-j_2 h_0^{(2)}j_1^{(1)} + j_1^{(1)}h_0^{(2)}j_2 ) \nonumber
\\ &-& j_2^{(1)} h_0^{(2)} j_1 + j_1 h_0^{(2)}j_2^{(1)}
 -j_2 h_0^{(3)} j_1 + j_1 h_0^{(3)} j_2 \nonumber
\\ & +& j_0^{(2)} h_1 j_2^{(1)} - j_2^{(1)}h_1 j_0^{(2)}
  - j_0^{(3)} h_1 j_2 + j_2h_1j_0^{(3)}
  + 2(j_0^{(1)} h_1^{(1)} j_2^{(1)} - j_2^{(1)} h_1^{(1)} j_0^{(1)}) \nonumber
\\ &+& 2(- j_0^{(2)} h_1^{(1)} j_2 + j_2 h_1^{(1)} j_0^{(2)})
  + j_0 h_1^{(2)} j^{(1)}_2 -  j^{(1)}_2  h_1^{(2)}j_0
  - j_0^{(1)}h_1^{(2)}j_2 + j_2 h_1^{(2)}j_0^{(1)} \nonumber
\\ &-& j_0^{(1)} h_2 j_1^{(2)} + j_1^{(2)}h_2 j_0^{(1)}
   + 2(j_0^{(2)} h_2 j_1^{(1)} - j_1^{(1)} h_2 j_0^{(2)})
   - j_0^{(3)} h_2 j_1 + j_1 h_2 j_0^{(3)} \nonumber
\\ &+& 2(j_0^{(1)} h_2^{(1)} j_1^{(1)} - j_1^{(1)}h_2^{(1)}j_0^{(1)})
   - j_0 h_2^{(1)}j_1^{(2)} + j_1^{(2)}h_2^{(1)}j_0
   - j_0^{(2)}h_2^{(1)}j_1 + j_1 h_2^{(1)} j_0^{(2)} \nonumber
\\ &+& 2(j_2^{(1)} h_0^{(1)} j_2^{(2)} +  j_2^{(2)}h_0^{(1)}j_2^{(1)})
 + j_2 h_0^{(2)} j_2^{(2)} +  j_2^{(2)}h_0^{(2)}j_2
 + 2(j_2 h_2^{(3)}j_2^{(1)} + j_2^{(1)}h_2^{(3)}j_2 ) \nonumber
\\ &+& j_2^{(2)} h_2 j_2^{(2)}
 + 4 j_2^{(1)} h_2^{(2)} j_2^{(1)}
 + j_2 h_2^{(4)} j_2 \nonumber
 \\ &+& j_0^{(2)} h_2 j_2^{(2)} + j_2^{(2)} h_2 j_0^{(2)}
  + 2( - j_0^{(3)} h_2 j_2^{(1)} - j_2^{(1)}h_2 j_0^{(3)})
  + j_0^{(4)} h_2 j_2 + j_2 h_2 j_0^{(4)} \nonumber
 \\ &+& 2(j_0^{(1)}h_2^{(1)}j_2^{(2)} +j_2^{(2)} h_2^{(1)}j_0^{(1)})
  + 4( - j_0^{(2)} h_2^{(1)} j_2^{(1)} - j_2^{(1)}h_2^{(1)}j_0^{(2)})
  + 2(j_0^{(3)}h_2^{(1)}j_2 + j_2h_2^{(1)}j_0^{(3)}) \nonumber
 \\ &+& j_0 h_2^{(2)}j_2^{(2)} + j_2^{(2)}h_2^{(2)}j_0
  + 2 ( - j_0^{(1)}h_2^{(2)}j_2^{(1)} - j_2^{(1)}h_2^{(2)}j_0^{(1)})
  + j_0^{(2)} h_2^{(2)} j_2 + j_2 h_2^{(2)}j_0^{(2)} \nonumber
  \\ &-& j_1^{(1)} h_1 j_1^{(2)} + j_1^{(2)} h_1 j_1^{(1)}
   - j_1 h_1^{(1)} j_1^{(2)} + j_1^{(2)} h_1^{(1)}j_1
   - j_1 h_1^{(2)} j_1^{(1)} + j_1^{(1)} h_1^{(2)}j_1 \nonumber
   \\ &-& j_1^{(1)} h_2 j_1^{(3)} - j_1^{(3)}h_2 j_1^{(1)}
    + 2 j_1^{(2)}h_2j_1^{(2)}
   - j_1 h_1^{(1)} j_1^{(3)} - j_1^{(3)}h_1^{(1)}j_1 \nonumber
   \\ &+& j_1^{(1)}h_2^{(1)} j_1^{(2)} + j_1^{(2)}h_2^{(1)}j_1^{(1)}
    - j_1 h_2^{(2)} j_1^{(2)} - j_1^{(2)}h_2^{(2)}j_1
    + 2 j_1^{(1)} h_2^{(2)} j_1^{(1)} \nonumber
\\ &-& j_2 h_1^{(1)} j_1^{(3)} - j_1^{(3)}h_1^{(1)} j_2
 + 2( - j_2 h_1^{(2)} j_1^{(2)} -j_1^{(2)} h_1^{(2)}j_2)
 - j_2 h_1^{(3)} j_1^{(1)} -j_1^{(1)} h_1^{(3)}j_2 \nonumber
\\ &- & j_2^{(1)} h_1 j_1^{(3)} -j_1^{(3)} h_1 j_2^{(1)}
 + j_2^{(2)} h_1^{(2)} j_1 + j_1 h_1^{(2)} j_2^{(2)}
 + j_2^{(1)} h_1^{(3)} j_1 + j_1 h_1^{(3)} j_2^{(1)} \nonumber
\\ &-& j_2^{(1)} h_1^{(1)} j^{(2)}_1 - j^{(2)}_1 h_1^{(1)} j_2^{(1)}
 + j_2^{(2)} h_1 j^{(2)}_1 + j^{(2)}_1 h_1 j_2^{(2)}
 + j_2^{(1)} h_1^{(2)} j^{(1)}_1 + j^{(1)}_1 h^{(2)}_1 j_2^{(1)} \nonumber
\\ &+& 2(j_2^{(2)} h_1^{(1)} j^{(1)}_1 + j^{(1)}_1 h^{(1)}_1 j_2^{(2)})
 +j_2^{(2)} h_2 j_2^{(4)} + j_2^{(4)}h_2j_2^{(2)}
 +2(-j_2^{(2)} h_2^{(1)} j_2^{(3)} - j_2^{(3)}h_2^{(1)}j_2^{(2)}) \nonumber
\\ &+& 2(j_2 h_2^{(3)}j_2^{(3)} + j_2^{(3)} h_2^{(3)}j_2)
 + 2(j_2^{(1)}h_2^{(1)} j_2^{(4)} + j_2^{(4)}h_2^{(1)}j_2^{(1)})
 + j_2 h_2^{(2)} j_2^{(4)} + j_2^{(4)}h_2^{(2)}j_2 \nonumber
\\ &+& 2( - j_2^{(1)} h_2^{(3)} j_2^{(2)} - j_2^{(2)} h_2^{(3)} j_2^{(1)})
 + j_2 h_2^{(4)} j_2^{(2)} + j_2^{(2)}h_2^{(4)}j_2
 + 2(j_2^{(1)} h_2^{(2)}j_2^{(3)} + j_2^{(3)}h_2^{(2)}j_2^{(1)}) \nonumber
\\ &+& 2(-j_2^{(1)} h_2^{(4)} j_2^{(1)} - j_2^{(3)} h_2 j_2^{(3)})
 - 6j_2^{(2)} h_2^{(2)} j_2^{(2)} \nonumber
\\ &+& j_2^{(2)} h_1 j_2^{(3)} - j_2^{(3)}h_1j_2^{(2)}
 + 2(j_2^{(1)} h^{(1)}_1 j_2^{(3)} - j_2^{(3)}h_1^{(1)}j_2^{(1)})
 + 3(j_2^{(1)}h_1^{(2)}j_2^{(2)} - j_2^{(2)}h_1^{(2)}j_2^{(1)}) \nonumber
\\ &+& j_2 h_1^{(2)} j_2^{(3)} - j_2^{(3)}h_1^{(2)}j_2
 + 2( j_2 h_1^{(3)} j_2^{(2)} - j_2^{(2)}h_1^{(3)}j_2)
 + j_2 h_1^{(4)}j_2^{(1)} - j_2^{(1)} h_1^{(4)}j_2 \nonumber
\\ &-& j_2^{(1)}h_2 j_1^{(4)} + j_1^{(4)}h_2 j_2^{(1)}
 + 2(j_2^{(2)} h_2 j_1^{(3)} - j_1^{(3)} h_2 j_2^{(2)})
 - j_2^{(3)}h_2j_1^{(2)} + j_1^{(2)}h_2j_2^{(3)} \nonumber
\\ &-& j_2 h_2^{(1)} j_2^{(4)} + j_2^{(4)}h_2^{(1)}j_2
 + 3(j_2^{(2)} h_2^{(1)} j_1^{(2)} - j_1^{(2)}h_2^{(1)}j_2^{(2)})
 +2(-j_2^{(3)} h_2^{(1)} j_1^{(1)} + j_1^{(1)}h_2^{(1)}j_2^{(3)}) \nonumber
\\ &+& 2(-j_2 h_2^{(2)} j_1^{(3)} + j_1^{(3)}h_2^{(2)}j_2)
 -j_2^{(3)} h_2^{(2)} j_1 + j_1h_2^{(2)}j_2^{(3)}
 + 3(j_2^{(1)} h_2^{(2)} j_1^{(2)} - j_1^{(2)}h_2^{(2)}j_2^{(1)}) \nonumber
\\ &+& 2(j_2^{(1)} h_2^{(3)} j_1^{(1)} - j_1^{(1)}h_2^{(3)}j_2^{(1)})
 -j_2 h_2^{(3)} j^{(2)}_1 + j^{(2)}_1h_2^{(3)}j_2
 -j_2^{(2)} h_2^{(3)} j_1 + j_1h_2^{(3)}j^{(2)}_2
 \end{array}
\end{equation}
Finally, the expression of the dissipation tensor $D_{SCIM}$ is:
\begin{equation}
\begin{array}{lcl}
D_{SCIM} & = & j_0 h_1 j_0
 + j_0 h_0 j_1 + j_1 h_0 j_0
 +2(j_0 h_2 j_0^{(1)} - j_0^{(1)} h_2 j_0)
 +2(-j_0 h_0^{(1)} j_2 - j_2h_0^{(1)} j_0) \nonumber
  \\ &+& 2(-j^{(1)}_0 h_0 j_2 - j_2h_0 j^{(1)}_0)
  + j_1 h_0 j_1^{(1)} - j_1^{(1)}h_0j_1 \nonumber
  \\ &-& j_0 h_1^{(1)} j_1 + j_1h_1^{(1)}j_0
   + 2( - j_0^{(1)} h_1 j_1 +j_1 h_1j_0^{(1)})
  + j_0 h_1 j_1^{(1)} - j_1^{(1)} h_1 j_0 \nonumber
\\ &+& 2(-j_2^{(1)} h_0 j_1^{(1)} - j_1^{(1)}h_0j_2^{(1)})
+ j_2 h_0 j_1^{(2)} + j_1^{(2)} h_0 j_2 \nonumber
\\ &+& 2(-j_2^{(1)} h_0^{(1)} j_1 - j_1h_0^{(1)}j_2^{(1)})
- j_2 h_0^{(2)} j_1 - j_1 h_0^{(2)} j_2 \nonumber
\\ &+& 3( j_0^{(2)} h_1 j_2 + j_2 h_1 j_0^{(2)})
+ 2( - j_0^{(1)} h_1 j_2^{(1)} - j_2^{(1)}h_1j_0^{(1)})
+4(j_0^{(1)} h^{(1)}_1 j_2 + j_2 h^{(1)}_1 j_0^{(1)}) \nonumber
\\ &+& 2( - j_0 h^{(1)}_1 j_2^{(1)} - j_2^{(1)}h^{(1)}_1j_0)
+ j_0 h_1^{(2)} j_2 + j_2 h_1^{(2)} j_0 \nonumber
\\ &+& 4( - j_0^{(1)} h_2 j_1^{(1)} - j_1^{(1)} h_2j_0^{(1)})
+ j_0 h_2 j_1^{(2)} + j_1^{(2)}  h_2 j_0
+ 3 (j_0^{(2)} h_2 j_1 + j_1 h_2 j_0^{(2)}) \nonumber
\\ &+& 2(-j_0 h^{(1)}_2 j^{(1)}_1 - j^{(1)}_1 h^{(1)}_2 j_0)
+ 2(j^{(1)}_0 h^{(1)}_2 j_1 + j_1 h^{(1)}_2 j^{(1)}_0) \nonumber
\\ &+& 2(j_2^{(2)} h_0 j_2^{(1)} - j_2^{(1)} h_0 j_2^{(2)})
+2 (-j_2 h^{(1)}_0 j_2^{(2)} + j_2^{(2)} h^{(1)}_0 j_2)
+2 (-j_2 h^{(2)}_0 j_2^{(1)} + j_2^{(1)} h^{(2)}_0 j_2) \nonumber
\\ &+ & 2( - j_0^{(1)} h_2 j_2^{(2)} + j_2^{(2)} h_2 j_0^{(1)} )
 +4 ( - j_0^{(3)} h_2 j_2 + j_2 h_2 j_0^{(3)})
 + 6 (j_0^{(2)} h_2 j_2^{(1)} - j_2^{(1)} h_2 j_0^{(2)}) \nonumber
\\ &+& 2( - j_0 h_2^{(1)} j_2^{(2)} + j_2^{(2)}h_2^{(1)}j_0)
 + 8 (j_0^{(1)} h^{(1)}_2 j_2^{(1)} - j_2^{(1)} h^{(1)}_2 j_0^{(1)})
 + 6 (- j_0^{(2)} h_2^{(1)} j_2 + j_2 h_2^{(1)}j_0^{(2)}) \nonumber
\\ &+& 2 (j_0 h^{(2)}_2 j_2^{(1)} - j_2^{(1)} h^{(2)}_2 j_0)
 + 2(- j_0^{(1)}h_2^{(2)} j_2 + j_2 h_2^{(2)}j_0^{(1)}) \nonumber
\\ &+& j_1 h_1 j_1^{(2)} + j_1^{(2)} h_1 j_1
 -j_1 h_1^{(1)}j_1^{(1)} - j_1^{(1)} h_1^{(1)}j_1
 -j_1 h_1^{(2)}j_1
 - 3 j_1^{(1)}h_1 j_1^{(1)} \nonumber
\\ &+& 5(-j_1^{(1)} h_2 j_1^{(2)} +  j_1^{(2)}h_2j_1^{(1)})
 +2(-j_1 h^{(1)}_2 j_1^{(2)} +  j_1^{(2)}h^{(1)}_2j_1) \nonumber
\\ &+& 2(-j_1 h^{(2)}_2 j_1^{(1)} +  j_1^{(1)}h^{(2)}_2 j_1)
 + j_1 h_2 j_1^{(3)} - j_1^{(3)}h_2j_1 \nonumber
 \\ &-& j_2 h_1^{(1)} j_1^{(2)} + j_1^{(2)} h_1^{(1)} j_2
  + 4(-j_2^{(1)} h_1 j_1^{(2)} + j_1^{(2)} h_1 j_2^{(1)})
  + j_2 h_1 j_1^{(3)} - j_1^{(3)} h_1 j_2 \nonumber
 \\ &+& 2(j_2^{(2)} h_1 j_1^{(1)} - j_1^{(1)} h_1 j_2^{(2)})
 -j_2 h_1^{(3)} j_1 + j_1 h_1^{(3)} j_2
 + 4( -j_2^{(1)} h^{(1)}_1 j_1^{(1)} + j_1^{(1)} h^{(1)}_1 j_2^{(1)}) \nonumber
\\ &+& 3( -j_2 h^{(2)}_1 j_1^{(1)} + j_1^{(1)} h^{(2)}_1 j_2)
 + 2( j^{(2)}_2 h^{(1)}_1 j_1 - j_1 h_1^{(1)} j^{(2)}_2) \nonumber
\\ &+& 8 ( - j_2^{(3)} h_2 j_2^{(2)} +  j_2^{(2)}h_2 j_2^{(3)})
 + 2 ( - j_2^{(1)} h_2 j_2^{(4)} +  j_2^{(4)}h_2 j_2^{(1)})
 + 8 (j_2^{(1)} h^{(1)}_2 j_2^{(3)} -  j_2^{(3)}h^{(1)}_2 j_2^{(1)}) \nonumber
\\ &+& 2 ( - j_2 h^{(1)}_2 j_2^{(4)} +  j_2^{(4)}h^{(1)}_2 j_2)
 + 12 ( j_2^{(1)} h^{(2)}_2 j_2^{(2)} -  j_2^{(2)}h^{(2)}_2 j_2^{(1)}) \nonumber
\\ &+& 4 ( j_2 h^{(3)}_2 j_2^{(2)} -  j_2^{(2)}h^{(3)}_2 j_2)
 + 2 ( j_2 h^{(4)}_2 j_2^{(1)} -  j_2^{(1)}h^{(4)}_2 j_2) \nonumber
\\ &+& 5 j_2^{(2)} h_1 j_2^{(2)}
 + 8 j_2^{(1)} h^{(2)}_1 j_2^{(1)}
 + j_2 h^{(4)}_1 j_2
 + 2(-j_2^{(1)}h_1 j_2^{(3)} - j_2^{(3)}h_1 j_2^{(1)}) \nonumber
\\ &+& 4(j_2^{(1)}h^{(1)}_1 j_2^{(2)} + j_2^{(2)}h^{(1)}_1 j_2^{(1)})
 + 2(-j_2 h^{(1)}_1 j_2^{(3)} - j_2^{(3)}h^{(1)}_1 j_2) \nonumber
\\ &-& j_2 h^{(2)}_1 j_2^{(2)} - j_2^{(2)}h^{(2)}_1 j_2
 + 2(j_2 h^{(3)}_1 j_2^{(1)} + j_2^{(1)}h^{(3)}_1 j_2) \nonumber
\\ &+& j_2 h_2 j_1^{(4)} + j_1^{(4)} h_2 j_2
 + 6(-j_2^{(1)} h_2 j_1^{(3)} - j_1^{(3)} h_2 j_2^{(1)})
 + 7(j_2^{(2)} h_2 j_1^{(2)} + j_1^{(2)} h_2 j_2^{(2)}) \nonumber
\\ &+& 2(- j_2^{(3)} h_2 j_1^{(1)} - j_1^{(1)} h_2 j_2^{(3)})
 + 2(- j_2 h^{(1)}_2 j_1^{(3)} - j_1^{(3)} h^{(1)}_2 j_2)
 + 2(- j_2^{(3)} h^{(1)}_2 j_1 - j_1 h^{(1)}_2 j_2^{(3)}) \nonumber
\\ &+& 8(j_2^{(2)} h^{(1)}_2 j_1^{(1)} + j_1^{(1)} h^{(1)}_2 j_2^{(2)})
 + 4(- j_2^{(1)} h^{(1)}_2 j^{(2)}_1 - j^{(2)}_1 h^{(1)}_2 j_2^{(1)}) \nonumber
 \\ &+& 4( j_2^{(1)} h^{(2)}_2 j^{(1)}_1 + j^{(1)}_1 h^{(2)}_2 j_2^{(1)})
 + 5(- j_2 h^{(2)}_2 j^{(2)}_1 - j^{(2)}_1 h^{(2)}_2 j_2) \nonumber
 \\ &+& j_2^{(2)} h^{(2)}_2 j_1 + j_1 h^{(2)}_2 j_2^{(2)}
 + 2(- j_2 h^{(3)}_2 j^{(1)}_1 - j^{(1)}_1 h^{(3)}_2 j_2) \nonumber
\\ &+& 2(j^{(1)}_2 h^{(3)}_2 j_1 + j_1 h^{(3)}_2 j^{(1)}_2)
\end{array}
\end{equation}
The components $V_{SCIM}$ and $B_{SCIM}$ entering the SCIM Hamiltonian $\mathcal{H}_{SCIM}$ are truncated at second order in SOPO, in accordance with the assumptions of the formalism.
\end{widetext}

\section{Evaluation of overlap kernels between two HFB vacua}\label{overlap}

One of the major challenges of the SCIM approach lies in the evaluation of the overlap and Hamiltonian kernels. In this section, we have compiled all the formulas used in the present SCIM application. The primary difficulty was likely handling kernels between states constructed with different harmonic-oscillator representations. In some cases, existing solutions were available in the literature. In others, we derived our own new formulas, particularly for the two-center representation, which is specific to the present work.

This appendix is devoted to presenting numerous results related to overlap kernels.
First, for pedagogical purposes, we review the evaluation of the overlap between two HFB states constructed using the same harmonic-oscillator representations, following the derivations in \cite{PfaRob}. We then focus on the axial and time-reversal case. Finally, we establish the connection between our derived formulas and the well-known Onishi-Yoshida formula \cite{OniYo}.
Second, we present the new formulas we have developed to address the overlap between two HFB states built with different two-center harmonic-oscillator representations. This work was inspired by Ref. \cite{2BrBz}. We then examine the axial and time-reversal case. Finally, we demonstrate the relation between our new formulas and those obtained in 
Ref. \cite{2BrBz}, as well as their connection to the Haider-Gogny formulas \cite{HaiGo}.
We conclude this section by addressing the critical \enquote{V Phasis} issue we discovered during PES calculations. We present both our analysis of the nature of this phenomenon and the practical solution we implemented to resolve it.
Finally, we note that, throughout this section, all the $\{ u_k \}$ coefficients associated with the Bogoliubov matrices $U$ considered are non-zero. In practice, this has always been the case when using the D1S Gogny force.

\subsection{HFB states normalization}

We begin by presenting the expression for the normalization factor $\mathcal{N}$ of a given HFB state, which will play a crucial role in the subsequent discussion. Let $\ket{\Phi}$ denote an HFB state defined by the set of QP annihilation operators $\{ \xi_i \}$:
\begin{eqnarray}
\ket{\Phi} = \mathcal{N} \prod_i \xi_i \ket{0}.
\end{eqnarray}
We transform the set of QP annihilation operators $\{ \xi_i \}$ into the set $\{ \eta_k \}$, which corresponds to the canonical representation of $\ket{\Phi}$:
\begin{eqnarray}
\ket{\Phi} = \mathcal{N} \text{Det}\left(C^*\right)\left(\prod_{k>}  \eta_{k} \bar \eta_k\right)\ket{0}.
\end{eqnarray}
Then, we remark the following property:
\begin{equation}\label{ctro_61}
\begin{array}{lcl}
\eta_k  \bar \eta_{k}\ket{0} &=& v_k \left( u_k + v_k a^+_{k} \bar a_{ k}^+ \right)\ket{0}.
\end{array}
\end{equation}
Thus, we can write:
\begin{eqnarray}\label{ctro_63}
\ket{\Phi} = \mathcal{N}\text{Det}\left(C^*\right)\prod_{k'>}v_{k'}\prod_{k>}\left(u_k + v_ka^+_{k}\bar a_{k}^+\right)\ket{0}.
\end{eqnarray}
The evaluation of the norm of the state \\ \noindent  $\displaystyle \prod_{k>}(u_k + v_ka^+_{k}\bar a_{k}^+)\ket{0}$ leads to:
\begin{equation}\label{ctro_62}
\begin{array}{lcl}
\displaystyle \left| \left( \prod_{k>}u_k + v_ka_{k}\bar a_{k}^+ \right)\ket{0} \right|^2 &=& \displaystyle 
\bra{0} \prod_{k>}\left(u_k^2 + v_k^2 \right)\ket{0} = 1.
\end{array}
\end{equation}
Inserting Eq.(\ref{ctro_62}) into Eq.(\ref{ctro_63}) gives:
\begin{eqnarray}
\displaystyle \bra{\Phi}\ket{\Phi} = \frac{\left|\mathcal{N}\right|^2}{\displaystyle \prod_{k>} u_k^2} = 1.
\end{eqnarray}
Among the different possibilities, we choose:
\begin{eqnarray}
\mathcal{N} = \text{Det}\left(C\right) \prod_{k>} u_k.
\end{eqnarray}
Indeed, this definition enables us to write the state $\ket{\Phi}$ in the canonical representation as:
\begin{eqnarray}\label{ctro_77}
\ket{\Phi} = \prod_{k>}\left(u_k + v_k a_k^+ \bar a_k^+\right)\ket{0}.
\end{eqnarray}

\subsection{Overlap between HFB states built with the same harmonic-oscillator representations}\label{ctro_82}

We consider two distinct HFB states, $\ket{\Phi_0}$ and $\ket{\Phi_1}$, associated with the QP annihilation operators $\{ \xi_{0,i} \}$ and $\{ \xi_{1,i} \}$, respectively. By applying the Thouless theorem and denoting $U^{(0)}$, $V^{(0)}$ and $U^{(1)}$, $V^{(1)}$ as the Bogoliubov matrices corresponding to the states $\ket{\Phi_0}$ and $\ket{\Phi_1}$, respectively, we can express:
\begin{eqnarray}\label{ctro_64}
\displaystyle \bra{\Phi_0}\ket{\Phi_1} = \bra{0}\ket{\Phi_1} \bra{\Phi_0}\ket{0} \bra{0}e^{\displaystyle \frac{1}{2}\sum_{kk'}\left(V^{(0)}U^{(0)-1}\right)_{kk'}c_{k'} c_k} \nonumber \\ \times e^{\displaystyle \frac{1}{2}\sum_{kk'}\left(V^{(1)}U^{(1)-1}\right)^*_{kk'}c_k^+c_{k'}^+} \ket{0}. \qquad
\end{eqnarray}
In the following, we will use the notations:
\begin{eqnarray}\label{ctro_74}
\begin{cases}
M^{(0)} = \left(V^{(0)}U^{(0)-1}\right)^* \\ M^{(1)} = \left(V^{(1)}U^{(1)-1}\right)^* 
\end{cases}.
\end{eqnarray}
Both matrices $M^{(0)}$ and $M^{(1)}$ are skew-symmetric. 
Thus, Eq.(\ref{ctro_64}) transforms into:
\begin{eqnarray}\label{ctro_65}
\bra{\Phi_0}\ket{\Phi_1} = \displaystyle \bra{0}\ket{\Phi_1} \bra{\Phi_0}\ket{0} \bra{0} e^{\displaystyle \frac{1}{2}\sum_{kk'}M^{(0)*}_{kk'}c_{k'} c_k} \nonumber \\ \displaystyle \times e^{\displaystyle \frac{1}{2}\sum_{kk'}M^{(1)}_{kk'}c_k^+c_{k'}^+} \ket{0}.
\end{eqnarray}
Then, we use the fermionic coherent states built using Grassmann algebras (see Appendix L of Ref. \cite{TPaul}). We can introduce them into Eq.(\ref{ctro_65}) thanks to their completeness relation:
\begin{eqnarray}\label{ctro_66}
\int \ket{z}\bra{z} \prod_{k=1}^{n} \left( dz_k^*dz_k \right) = \text{Id}.
\end{eqnarray}
The equation (\ref{ctro_66}) does not corresponds to the one introduced in Ref.\cite{PfaRob}. This difference comes from a scaling in the super Hilbert space dot product. Inserting Eq.(\ref{ctro_66}) into Eq.(\ref{ctro_65}) leads to:
\begin{equation}\label{ctro_68}
\begin{array}{lcl}
\displaystyle \bra{\Phi_0}\ket{\Phi_1} &=& \displaystyle \bra{0}\ket{\Phi_1} \bra{\Phi_0}\ket{0} \int  \bra{0}e^{\displaystyle \frac{1}{2}\sum_{kk'}M^{(0)*}_{kk'}c_{k'} c_k} \\ & & \displaystyle \times \ket{z}\bra{z} e^{\displaystyle \frac{1}{2}\sum_{kk'}M^{(1)}_{kk'}c_k^+c_{k'}^+} \ket{0} \prod_{k=1}^{n} \left( dz_k^*dz_k \right).
\end{array}
\end{equation}
By definition, the fermionic state $\ket{z}$ is an eigenvector of the particle annihilation operators with:
\begin{eqnarray}\label{ctro_67}
 c_k\ket{z} = z_k\ket{z}, \qquad  \bra{z}c^+_k = \bra{z}z^*_k,
\end{eqnarray}
where $z_k$ et $z_k^*$ are Grassmann numbers. Using Eq.(\ref{ctro_67}) into Eq.(\ref{ctro_68}) brings:
\begin{equation}\label{ctro_70}
\begin{array}{lcl}
\bra{\Phi_0}\ket{\Phi_1} &=& \displaystyle \bra{0}\ket{\Phi_1} \bra{\Phi_0}\ket{0} \int  \bra{0}\ket{z}\bra{z}\ket{0}  e^{\displaystyle \frac{1}{2}\sum_{kk'}M^{(0)*}_{kk'}z_{k'} z_k}  \\ & & \times \displaystyle   e^{\displaystyle \frac{1}{2}\sum_{kk'}M^{(1)}_{kk'}z_k^* z_{k'}^*} \prod_{k=1}^{n} (dz_k^*dz_k ).
\end{array}
\end{equation}
It can be shown that (see Appendix L of Ref. \cite{TPaul}):
\begin{eqnarray}\label{ctro_69}
\displaystyle \bra{0}\ket{z} = e^{-z^*.z/2}.
\end{eqnarray}
This property also differs from the one reported in Ref.\cite{PfaRob}. In fact, Eq.  (\ref{ctro_69}) explicitly provides the scaling factor between the two dot product definitions. Naturally, the effects of the differences arising from Eq. (\ref{ctro_66}) and Eq. (\ref{ctro_69}) cancel out in the equations. Substituting Eq. (\ref{ctro_69}) into Eq. (\ref{ctro_70}) yields:
\begin{equation}\label{ajout1}
\begin{array}{lcl}
\bra{\Phi_0}\ket{\Phi_1} &=& \displaystyle \bra{0}\ket{\Phi_1} \bra{\Phi_0}\ket{0} \int  e^{-z^*.z} e^{\displaystyle \frac{1}{2}\sum_{kk'}M^{(0)*}_{kk'}z_{k'} z_k} \\ & & \times \displaystyle e^{\displaystyle \frac{1}{2}\sum_{kk'}M^{(1)}_{kk'}z_k^* z_{k'}^*} \prod_{k=1}^{n} \left(dz_k^*dz_k \right).
\end{array}
\end{equation}
To simplify the expression, we define the following quantities:
\begin{eqnarray}
z = \begin{pmatrix} z^*_1 \\ . \\ z^*_n \\ z_1 \\ . \\ z_n\end{pmatrix}
\quad \text{and} \quad M =
\begin{pmatrix}
         M^{(1)}    & -I \\
             I & -M^{(0)*}
\end{pmatrix}.
\end{eqnarray}
It is clear that the matrix $M$ is a $2n \times 2n$ skew-symmetric matrix. Thanks to these notations, Eq. \eqref{ajout1} becomes:
\begin{eqnarray}\label{ctro_71}
\bra{\Phi_0}\ket{\Phi_1} =\bra{0}\ket{\Phi_1} \bra{\Phi_0}\ket{0} \int  e^{\displaystyle  \frac{1}{2}z^T M z}  \prod_{k=1}^{n} (dz_k^*dz_k )
\end{eqnarray}
To apply the results concerning Gaussian integration over a Grassmann algebra, we must reorder the volume elements on the right-hand side of Eq.(\ref{ctro_71}):
\begin{equation}
\begin{array}{lcl}
\displaystyle \bra{\Phi_0}\ket{\Phi_1} &=&(-1)^{\frac{\displaystyle n(n+1)}{\displaystyle 2}}\bra{0}\ket{\Phi_1} \bra{\Phi_0}\ket{0} \\ & & \times \displaystyle \int  e^{\displaystyle \frac{1}{2}z^T M z}  dz_n...dz_1 dz_n^*...dz_1^*.
\end{array}
\end{equation}
Thus,
\begin{eqnarray}\label{ctro_72}
\bra{\Phi_0}\ket{\Phi_1} =(-1)^{\frac{\displaystyle n(n+1)}{\displaystyle 2}}\bra{0}\ket{\Phi_1} \qquad \qquad \qquad  \notag \\ \times \bra{\Phi_0}\ket{0} \text{Pf}\begin{pmatrix}
         M^{(1)}    & -I \\
             I & -M^{(0)*}
\end{pmatrix}.
\end{eqnarray}
In Eq.(\ref{ctro_72}), the notation \enquote{Pf} stands for the Pfaffian. It is possible to factorize the matrix $M$:
\begin{eqnarray}
\begin{pmatrix}
         I    & 0 \\  -M^{(1)-1} & I
\end{pmatrix}
\begin{pmatrix}
         M^{(1)}    & -I \\      I & -M^{(0)*}
\end{pmatrix}
\begin{pmatrix}
         I    & M^{(1)-1} \\  0 & I
\end{pmatrix}= \nonumber \\
\begin{pmatrix}
         M^{(1)}    & 0 \\  0 & -M^{(0)*} + M^{(1)-1}
\end{pmatrix}. \quad 
\end{eqnarray}
Thus:
\begin{eqnarray}\nonumber
\text{Pf}\begin{pmatrix}
         M^{(1)}    & -I \\
             I & -M^{(0)*}
\end{pmatrix}
 = \text{Pf}(M^{(1)})\text{Pf}(-M^{(0)*} + M^{(1)-1}).
\end{eqnarray}
Finally:
\begin{eqnarray}\label{ctro_75}
\bra{\Phi_0}\ket{\Phi_1} =(-1)^{\frac{\displaystyle n(n+1)}{\displaystyle 2}}\bra{0}\ket{\Phi_1} \bra{\Phi_0}\ket{0} \quad \quad \qquad \nonumber \\ \times \text{Pf}(M^{(1)}) ~ \text{Pf}(-M^{(0)*} + M^{(1)-1}). \quad
\end{eqnarray}

\subsection{Application to the axial and time-reversal invariance with the same harmonic-oscillator representations}\label{ctro_84}

In the special case of axial and time-reversal invariant HFB states, the matrices $\tilde U^{(i)}$ and $\tilde V^{(i)}$ (where $i = 0$ or $1$) are real and exhibit the following structures:
\begin{eqnarray}\label{ctro_73}
\tilde U^{(i)} =
\begin{pmatrix}
       U^{(i)}    & 0 \\  0 &  U^{(i)}
\end{pmatrix}
\quad \;
\tilde V^{(i)} =
\begin{pmatrix}
       0 & - V^{(i)} \\  V^{(i)} & 0
\end{pmatrix}.
\end{eqnarray}
Here, the $U^{(i)}$ and $V^{(i)}$ are of dimension $n/2$. Eq.(\ref{ctro_73}) implies that the matrices $M^{(i)}$ defined in Eq.(\ref{ctro_74}) also have a special structure:
\begin{eqnarray}
M^{(i)} =
\begin{pmatrix}
       0    & -V^{(i)}U^{(i)-1} \\  V^{(i)}U^{(i)-1} &  0
\end{pmatrix}.
\end{eqnarray}

\noindent It is important to remark that the matrices $M^{(i)}$ are skew-symmetric, while the matrices $V^{(i)}U^{(i)-1}$ are symmetric. Using the Pfaffian properties, we write:
\begin{eqnarray}
\displaystyle \text{Pf}(M^{(1)}) = (-1)^{\frac{\displaystyle n\left(\frac{n}{2}-1\right)}{\displaystyle 4}}(-1)^{\displaystyle \frac{n}{2}} \notag \qquad \qquad \\ \times \text{Det}\left(V^{(1)}U^{(1)-1}\right),
\end{eqnarray}
and
\begin{eqnarray}
\displaystyle \text{Pf}\left(-M^{(0)*} + M^{(1)-1}\right) = (-1)^{\displaystyle \frac{n(\frac{n}{2}-1)}{\displaystyle 4}} \nonumber \qquad \quad\\ \times \displaystyle \text{Det}\left(V^{(0)}U^{(0)-1} + U^{(1)}V^{(1)-1}\right). \; \; \;
\end{eqnarray}
With these results, we can now express:
\begin{eqnarray}\label{ctro_92}
\bra{\Phi_0}\ket{\Phi_1} = \bra{0}\ket{\Phi_1} \bra{\Phi_0}\ket{0} \qquad \qquad \qquad \\ \times \text{
Det} \left( I + V^{(0)}U^{(0)-1}V^{(1)}U^{(1)-1} \right). \notag
\end{eqnarray}

\subsection{Link with the \enquote{Onishi-Yoshida} formula}

The \enquote{Onishi-Yoshida} formula \cite{OniYo} gives the absolute value of the overlap between two given HFB states such that:
\begin{eqnarray}\label{ctro_76}
\left| \bra{\Phi_0}\ket{\Phi_1} \right| = \sqrt{ \left| \text{Det}(U^{(0)+}U^{(1)}+V^{(0)+}V^{(1)})\right|}.
\end{eqnarray}
Starting from Eq.(\ref{ctro_75}), one uses the Pfaffian properties and obtains:
\begin{eqnarray}\label{ctro_78}
\left|\bra{\Phi_0}\ket{\Phi_1}\right| = \left|\bra{0}\ket{\Phi_1}\bra{\Phi_0}\ket{0}\right| \quad \quad \quad \quad \quad \quad \quad \nonumber \\ \displaystyle \times \sqrt{\left|\text{Det}(M^{(1)})\text{Det}(-M^{(0)*} + M^{(1)-1})\right|}.
\end{eqnarray}
From Eq.(\ref{ctro_77}), it is clear that: 
\begin{eqnarray}
\displaystyle \left|\bra{0}\ket{\Phi_i}\right| = \prod_{k>} u^{(i)}_k = \sqrt{\left|\text{Det}(U^{(i)})\right|}.
\end{eqnarray}
Then:
\begin{widetext}
\begin{eqnarray}
|\bra{\Phi_0}\ket{\Phi_1}| = \sqrt{|\text{Det}(U^{(1)})\text{Det}(U^{(0)+})\text{Det}(I+(U^{(0)+})^{-1}V^{(0)+}V^{(1)}U^{(1)-1})}|,
\end{eqnarray}
\end{widetext}
which allows to recover the \enquote{Onishi-Yoshida} formula back.

\subsection{Overlap between HFB states built with two different harmonic-oscillator representations}\label{ctro_87}

The derivations presented in this section are entirely novel. They allow us to consider two distinct harmonic-oscillator representations of different dimensions without the need to compute their rectangular overlap matrix $R$.
In this context, we examine two HFB states, $\ket{\Phi_0}$ and $\ket{\Phi_1}$, each associated with a set of QP annihilation operators, $\{ \xi_{0,j} \}$ and $\{ \xi_{1,j} \}$, respectively. The QP annihilation operators of these two sets are not constructed from the same harmonic-oscillator representations. More precisely, they are expressed as follows:
\begin{eqnarray}\begin{cases}
\displaystyle \xi_{0,i} = \sum_l U^{(0)*}_{li} c_{0,l} + V^{(0)*}_{li} c^+_{0,l} \\
\\ \displaystyle  \xi_{1,i} = \sum_l U^{(1)}_{li*} c_{1,l} + V^{(1)*}_{li} c^+_{1,l} \end{cases}.
\end{eqnarray}
These two harmonic-oscillator sets will be noted $\{ 0 \}$ and $\{ 1 \}$ in the following. Both are truncated subsets of the full harmonic-oscillator sets $\{ \bar 0 \}$ and $\{ \bar 1 \}$. The matrix $\bar R$ represents the full overlap matrix between them:
\begin{eqnarray}\label{ctro_79}\begin{cases}
\displaystyle  c_{0,l} = \sum_k^{+\infty} \bar R^T_{kl} c_{1,k} \\
\\ \displaystyle  c_{1,l} = \sum_k^{+\infty} \bar R_{kl} c_{0,k}\end{cases}.
\end{eqnarray}
The Thouless theorem is used to transform the overlap $\bra{\Phi_0}\ket{\Phi_1}$:
\begin{eqnarray}
\displaystyle \bra{\Phi_0}\ket{\Phi_1} = \bra{0}\ket{\Phi_1} \bra{\Phi_0}\ket{0} \quad \quad \quad \quad \quad \quad \quad \quad \quad \nonumber \\ \times \bra{0}e^{\displaystyle \frac{1}{2}\sum_{kk'}\left(V^{(0)}U^{(0)-1}\right)_{kk'}c_{0,k'} c_{0,k}} \nonumber \\ \qquad \times e^{\displaystyle \frac{1}{2}\sum_{kk'}\left(V^{(1)}U^{(1)-1}\right)^*_{kk'}c_{1,k}^+c_{1,k'}^+} \ket{0}.
\end{eqnarray}
Using the matrix $\bar R$ defined in Eq.(\ref{ctro_79}), one obtains:
\begin{eqnarray}
\displaystyle \bra{\Phi_0}\ket{\Phi_1} = \bra{0}\ket{\Phi_1} \bra{\Phi_0}\ket{0} \quad \quad \quad \quad \quad \quad \quad \quad \quad \qquad \nonumber \\ \times \bra{0}e^{\displaystyle \frac{1}{2}\sum_{kk'}\left(V^{(0)}U^{(0)-1}\right)_{kk'}c_{0,k'} c_{0,k}} \quad \quad \nonumber \\ \times e^{\displaystyle \frac{1}{2}\sum_{ll'}^{+\infty} \left(\bar RV^{(1)}U^{(1)-1}\bar R^T \right)^*_{ll'}c_{0,l}^+c_{0,l'}^+} \ket{0}. \; \; \; \; \; 
\end{eqnarray}
The summation is restricted to the indices of the truncated harmonic-oscillator representation associated with $\{ 0 \}$, whereas the sum in the exponential on the right-hand side extends over the complete infinite set $\{ \bar 0 \}$. For an arbitrary particle creation operator $c_{0,\gamma}^+ \notin \{0\}$, we can express:
\begin{eqnarray}
 e^{\displaystyle \frac{1}{2} \sum_{ll'}^{+\infty}\left(\bar RV^{(1)}U^{(1)-1}\bar R^T\right)^*_{ll'}c_{0,l}^+c_{0,l'}^+} \ket{0} = \qquad \nonumber \\ \left( 1+\frac{1}{2}\sum_{l}^{+\infty} \left( \bar RV^{(1)}U^{(1)-1}\bar R^T \right)^*_{l\gamma}c_{0,l}^+c_{0,\gamma}^+ \right) \qquad \\ \nonumber \times e^{\displaystyle \frac{1}{2}\sum_{ll' (l' \ne \gamma)}^{+\infty} \left( \bar RV^{(1)}U^{(1)-1}\bar R^T \right)^*_{ll'}c_{0,l}^+c_{0,l'}^+} \ket{0}.
\end{eqnarray}
The operator $c_{0,\gamma}^+$ commutes with $\exp\left( \frac{1}{2} {\displaystyle \sum_{kk'}(V^{(0)}U^{(0)-1})_{kk'}c_{0,k'} c_{0,k}} \right)$?, consequently, we find:
\begin{eqnarray}
\displaystyle \bra{\Phi_0}\ket{\Phi_1} = \bra{0}\ket{\Phi_1} \bra{\Phi_0}\ket{0} \qquad \qquad \qquad \qquad \nonumber \\  \times \bra{0}e^{\displaystyle \frac{1}{2}\sum_{kk'}\left(V^{(0)}U^{(0)-1}\right)_{kk'}c_{0,k'} c_{0,k}} \qquad \quad \\ \qquad \qquad \nonumber \times e^{\displaystyle \frac{1}{2}\sum_{ll'(l'\ne \gamma)}^{+\infty}\left(\bar RV^{(1)}U^{(1)-1}\bar R^T\right)^*_{ll'}c_{0,l}^+c_{0,l'}^+} \ket{0}.
\end{eqnarray}
Repeating this process for all the indices spanning $\{ \bar0 \}/\{0\}$ and for both $l$ and $l'$, we finally find:
\begin{eqnarray}\label{ctro_80}
\bra{\Phi_0}\ket{\Phi_1} = \bra{0}\ket{\Phi_1} \bra{\Phi_0}\ket{0} \qquad \qquad \qquad \qquad \nonumber \\  \times \bra{0}e^{\displaystyle \frac{1}{2}\sum_{kk'}\left(V^{(0)}U^{(0)-1}\right)_{kk'}c_{0,k'} c_{0,k}} \qquad \quad \\ \qquad \qquad \nonumber \times e^{\displaystyle \frac{1}{2}\sum_{kk'}\left(RV^{(1)}U^{(1)-1}R^T \right)^*_{kk'}c_{0,k}^+c_{0,k'}^+} \ket{0},
\end{eqnarray}
where the matrix $R$ stands for the restriction of $\bar R$ to the subsets $\{0\}$ and $\{1\}$. We can define the matrices $M^{(0)}$ and $M^{(1)}_R$ as :
\begin{eqnarray}\label{ctro_81}\begin{cases}
M^{(0)} = \left( V^{(0)}U^{(0)-1} \right)^*
\\
M^{(1)}_R = \left( RV^{(1)}U^{(1)-1}R^T \right)^* \end{cases}.
\end{eqnarray}
The insertion of Eq.(\ref{ctro_81}) into Eq.(\ref{ctro_80}) leads to:
\begin{eqnarray}
\bra{\Phi_0}\ket{\Phi_1} = \bra{0}\ket{\Phi_1} \bra{\Phi_0}\ket{0} \bra{0} e^{\displaystyle \frac{1}{2}\sum_{kk'}M^{(0)*}_{kk'}c_{0,k'} c_{0,k}} \qquad \nonumber \\ \times e^{\displaystyle  \frac{1}{2} \sum_{kk'}(M^{(1)}_R)_{kk'}c_{0,k}^+c_{0,k'}^+} \ket{0}.  \qquad
\end{eqnarray}
Both $M^{(0)}$ and $M^{(1)}_R$ are skew-symmetric matrices. Consequently, we can directly express the final form of $\bra{\Phi_0}\ket{\Phi_1}$  by analogy with the derivations presented in section \ref{ctro_82}.
\begin{eqnarray}\label{ctro_83}
\displaystyle \bra{\Phi_0}\ket{\Phi_1} =(-1)^{\frac{\displaystyle n(n+1)}{\displaystyle 2}}\bra{0}\ket{\Phi_1} \bra{\Phi_0}\ket{0} \quad \quad \quad \quad \nonumber \\ \times \text{Pf}\left( M_R^{(1)} \right) \text{Pf}\left( -M^{(0)*}+(M_R^{(1)})^{-1} \right). \; \; \;
\end{eqnarray}
Using the pfaffian properties, we directly write for the particular axial and time-reversal invariance case:
\begin{eqnarray}
\bra{\Phi_0}\ket{\Phi_1} = \bra{0}\ket{\Phi_1} \bra{\Phi_0}\ket{0} \qquad \qquad \qquad \qquad \nonumber \\ \times \text{Det}\left(I + V^{(0)}U^{(0)-1}RV^{(1)}U^{(1)-1}R^T \right).
\end{eqnarray}
This formula is very useful in practice as it does not require to find the inverse 
matrix of $R$.

\subsection{Link with Robledo's formula}

In Ref. \cite{2BrBz}, the following expression is proposed for the calculation of the norm of the overlap $\left| \bra{\Phi_0}\ket{\Phi_1} \right|$:
\begin{eqnarray}\label{ctro_85}
\left|\bra{\Phi_0}\ket{\Phi_1}\right| = \qquad \qquad \qquad \qquad \qquad \qquad \qquad \qquad \\ \sqrt{\left| \text{Det}(U^{(0)T}(R^T)^{-1}U^{(1)*} + V^{(0)T}RV^{(1)*})\text{Det}(R) \right| }. \nonumber
\end{eqnarray}
This formula has two major drawbacks. First, it requires to complete $R$ into a square matrix. Then, it does not give the phase of $\bra{\Phi_0}\ket{\Phi_1}$.
Starting from Eq.(\ref{ctro_85}), we write:
\begin{eqnarray}
\left| \bra{\Phi_0}\ket{\Phi_1} \right| =
\sqrt{ \left|\text{Det}(U^{(0)})\right|}  \sqrt{\left| \text{Det}\left(U^{(1)}\right) \right|} \times \qquad \qquad \\ \nonumber 
\sqrt{\left| \text{Det}\left(I + U^{(0)T-1}V^{(0)T} R V^{(1)*}(U^{(1)-1})^* R^T \right) \right|}. 
\end{eqnarray}
After some manipulations:
\begin{eqnarray}
\left|\bra{\Phi_0}\ket{\Phi_1}\right| =
\left|\bra{0}\ket{\Phi_1} \bra{\Phi_0}\ket{0}\right|
\sqrt{\left| \text{det}( I - M^{(0)*} M_R^{(1)})\right|}. \nonumber 
\end{eqnarray}
Thus,
\begin{eqnarray}
\left|\bra{\Phi_0}\ket{\Phi_1}\right|= \left|\bra{0}\ket{\Phi_1} \bra{\Phi_0}\ket{0}\right| \qquad \qquad \qquad \nonumber \\ \times
\sqrt{\left| \text{Det}(M_R^{(1)}) \text{Det}( - M^{(0)*} + M_R^{(1)-1})\right|}.
\end{eqnarray}
Then, we finally go back to the Pfaffians:
\begin{eqnarray}
\left| \bra{\Phi_0}\ket{\Phi_1}  \right| = \left| \bra{0}\ket{\Phi_1} \bra{\Phi_0}\ket{0}
  \right. \qquad \qquad \qquad \nonumber \\ \left. \times \text{Pf} \left(M_R^{(1)} \right) \text{Pf} \left( - M^{(0)*} + M_R^{(1)-1} \right) \right|.
\end{eqnarray}
which is exactly Eq.(\ref{ctro_83}) up to the phase.

\subsection{Link with the \enquote{Haider-Gogny} formula}

In this section, we connect the previously obtained results to the \enquote{Haider-Gogny} \cite{HaiGo}. Below, we consider two HFB states, $\ket{\Phi_0}$ and $\ket{\Phi_1}$, both constructed using the same harmonic-oscillator basis. We begin by writing the canonical transformation associated with each state:
\begin{equation}\label{ctro_86}
\begin{array}{lcl}
\begin{pmatrix}
\eta^{(k)} \\
\bar \eta^{(k)} \\
\eta^{(k)+} \\
\bar \eta^{(k)+}
\end{pmatrix}
&=&
\begin{pmatrix}
u^{(k)} & 0 & 0 & - v^{(k)} \\
0 & u^{(k)} & v^{(k)} & 0 \\
0 & -v^{(k)} & u^{(k)} & 0 \\
v^{(k)} & 0 & 0 & u^{(k)} \\
\end{pmatrix}
\begin{pmatrix}
a^{(k)} \\
\bar a^{(k)} \\
a^{(k)+} \\
\bar a^{(k)+}
\end{pmatrix}, \\ 
&= & \begin{pmatrix}
\tilde u^{(k)} & \tilde v^{(k)} \\
\tilde v^{(k)} & \tilde u^{(k)}
\end{pmatrix}
\begin{pmatrix}
a^{(k)} \\
\bar a^{(k)} \\
a^{(k)+} \\
\bar a^{(k)+}
\end{pmatrix}.
\end{array}
\end{equation}
Eq.(\ref{ctro_86}) describes a case that is very similar to the one involving two different harmonic-oscillator bases. In this context, the equivalent of the overlap matrix $R$ is the matrix $\tau^{01}$ defined as follows:
\begin{eqnarray}
a^{(1)+}_{k'} = \sum_k \tau^{01}_{kk'}a_k^{(0)+},
\end{eqnarray}
where the matrix $\tau^{01}$ can be explicitly rewritten using the $D^{(i)}$ Bloch-Messiah matrices associated with both $\ket{\Phi_0}$ and $\ket{\Phi_1}$:
\begin{equation}
\begin{array}{lcl}
\displaystyle a^{(1)+}_{k'} &=& \displaystyle \sum_l D^{(1)}_{lk'} c_l^+ = \sum_l\sum_k D^{(1)}_{lk'} D^{(0)+}_{kl}a_k^{(0)+}, \\ &=& \displaystyle \sum_k \left( D^{(0)+} D^{(1)} \right)_{kk'} a_k^{(0)+}.
\end{array}
\end{equation}
Thus,
\begin{eqnarray}
\tau^{01}_{kk'} = \left( D^{(0)+} D^{(1)} \right)_{kk'}.
\end{eqnarray}
Then, we use the analogy with section \ref{ctro_87} to write:
\begin{eqnarray}
\bra{\Phi_0}\ket{\Phi_1} = (-1)^{\frac{\displaystyle n(n+1)}{\displaystyle 2}}\bra{\Phi_0}\ket{0}\bra{0}\ket{\Phi_1}  \qquad \quad \nonumber \\ \times \text{Pf}\left( M_\tau^{(1)} \right)\text{Pf}\left( -M^{(0)*} + M_\tau^{(1)-1} \right).
\end{eqnarray}
We naturally set:
\begin{eqnarray}\label{ctro_88}\begin{cases}
M^{(0)} = \begin{pmatrix}
           0 & - \frac{\displaystyle v^{(0)}}{\displaystyle u^{(0)}} \\
          \frac{\displaystyle v^{(0)}}{\displaystyle u^{(0)}} & 0
          \end{pmatrix}
          \\
           \\
           M_\tau^{(1)} = \begin{pmatrix}
           0 & - \tau^{01}\frac{\displaystyle v^{(0)}}{\displaystyle u^{(0)}}\tau^{01T} \\
          \tau^{01}\frac{\displaystyle v^{(0)}}{\displaystyle u^{(0)}}\tau^{01T} & 0
          \end{pmatrix}\end{cases}.
\end{eqnarray}
Because of the special structure of Eq.(\ref{ctro_88}), similar to the one of the axial and time-reversal invariant case, we write by analogy:
\begin{eqnarray}
\bra{\Phi_0}\ket{\Phi_1} = \bra{\Phi_0}\ket{0}\bra{0}\ket{\Phi_1} \qquad \qquad \qquad \notag \\  \times \text{Det} \left( I + \frac{\displaystyle v^{(0)}}{\displaystyle u^{(0)}}\tau^{(01)}\frac{\displaystyle v^{(1)}}{\displaystyle u^{(1)}} \left(\tau^{(01)}\right)^T \right).
\end{eqnarray}
Replacing the overlaps $\bra{\Phi_0}\ket{0}$ and $\bra{0}\ket{\Phi_1}$ by their expressions leads to:
\begin{eqnarray}
\bra{\Phi_0}\ket{\Phi_1} = \prod_{k'} u_{k'}^{(1)} \prod_k u_k^{(0)} \qquad \qquad \quad \notag \\ \times \text{Det}\left(I + \frac{v^{(0)}}{u^{(0)}}\tau^{(01)}\frac{v^{(1)}}{u^{(1)}}(\tau^{(01)})^T\right).
\end{eqnarray}
Finally, we recover the \enquote{Haider-Gogny} formula:
\begin{eqnarray}
\bra{\Phi_0}\ket{\Phi_1} = \text{Det}\left(\tau^{01}\right) \times \qquad \qquad \qquad \qquad \qquad \\ \nonumber \text{Det}\left( u^{(1)}\left(\tau^{(01)}\right)^{-1}u^{(0)} + v^{(1)}\left(\tau^{(01)}\right)^T v^{(0)} \right).
\end{eqnarray}
This derivation demonstrates that the \enquote{Haider–Gogny} formula already accounts for the phase ambiguity discussed above.

\subsection{The \enquote{V phasis}}

\begin{figure}
\centering
\includegraphics[width=1.0\linewidth]{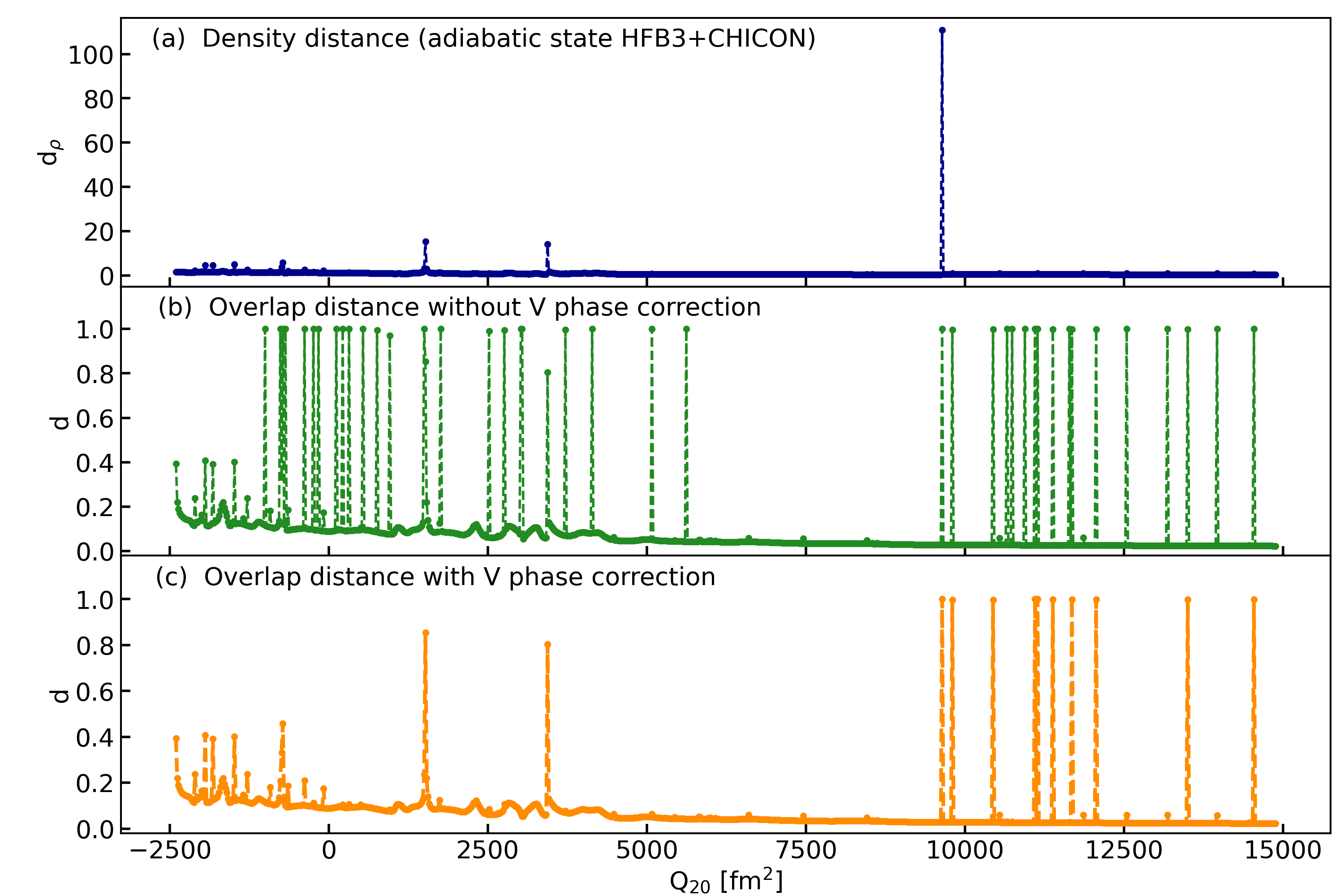}
\caption{Illustration of the unexpected differences between $d$ and d$_\rho$.
Panel (a): the\enquote{density distance} $d_\rho$ between each state and its neighbor on the right.
Panel (b): the overlap distance $d$ between each state and its neighbor on the right without the \enquote{V phase} correction. Panel (c): the overlap distance $d$ between each state and its neighbor on the right with the \enquote{V phase} correction.}
\label{ctro_90}
\end{figure}

In our PES calculations using the $\mathcal{P}_{20}$ procedure, we have observed unusual behavior in the overlap distance. Specifically, some neighboring states with nearly identical multipole moments exhibited a significant overlap distance (close to one). Furthermore, the discontinuities detected by the overlap distance were not reflected in the density distance d$_\rho$ \cite{DRdisco}.
In panel (a) of FIG. \ref{ctro_90}, we display the density distance between each HFB state and its right neighbor for the adiabatic set obtained with the $\mathcal{P}_{20}$ procedure in $^{240}$Pu, plotted as a function of the quadrupole deformation. In panel (b), we show the overlap distance between each HFB state and its right neighbor for the same set, also as a function of the quadrupole deformation.
We clearly observe that the overlap distance detects far more discontinuities than the density distance. Given that these overlap distance discontinuities were excessively numerous and furthermore, unrelated to any discontinuity in the multipole moments, we hypothesized that they must arise from phase issues.
Consequently, we examined in detail the $U$ and $V$ Bogoliubov matrices associated with the problematic states. We ultimately found that the $U$ and $V$ matrices for two neighboring states exhibiting abnormal overlap distance discontinuities were nearly identical, except for sign differences in certain ($\Omega$,$\tau$) sub-blocks.
If $U^{(0)}$, $V^{(0)}$ and $U^{(1)}$, $V^{(1)}$ denote the Bogoliubov matrices associated with two HFB states $\ket{\Phi_0}$ and $\ket{\Phi_1}$ characterized by an unexpected overlap distance discontinuity, we have evaluated the overlap between $\ket{\Phi_0}$ and $\ket{\Phi_1}$ using the formula given in Eq. (\ref{ctro_92}):
\begin{eqnarray}\label{ctro_93}
\bra{\Phi_0}\ket{\Phi_1} = \bra{0}\ket{\Phi_1} \bra{\Phi_0}\ket{0} \times  \qquad \qquad \\ \prod_{(\Omega,\tau)} \text{Det}\left(I + V^{(0)\tau \Omega}\left(U^{(0)\tau \Omega}\right)^{-1}V^{(1)\tau \Omega}\left(U^{(1)\tau \Omega}\right)^{-1}\right). \nonumber 
\end{eqnarray}
We intentionally expanded the ($\Omega$,$\tau$) sub-structure of Eq. (\ref{ctro_93}). Clearly, if we change the sign of a $V^{(1) \tau \Omega}$ matrix while leaving everything else unchanged, the resulting overlap is modified. However, considering only the state $\ket{\Phi_1}$, this sign difference does not affect any one-body observables. This is not surprising, since $\rho = V^{(1)}V^{(1)T}$. When it comes to the total binding energy, the sign difference induces a sign change in the corresponding sub-block of the pairing tensor $\kappa = V^{(1)}U^{(1)T}$. Since the pairing tensor elements are squared in the energy evaluation, this change does not result in any energy variation.

These observations have led us to replace the overlap formula given in Eq. (\ref{ctro_93}) with the following:
\begin{widetext}
\begin{eqnarray}\label{ctro_94}
\bra{\Phi_0}\ket{\Phi_1} = \bra{0}\ket{\Phi_1} \bra{\Phi_0}\ket{0} \prod_{(\Omega,\tau)}\left| \text{max}_{\pm} \right| \left[ \text{Det} \left(I \pm V^{(0)\tau \Omega}(U^{(0)\tau \Omega})^{-1}V^{(1)\tau \Omega}(U^{(1)\tau \Omega})^{-1} \right) \right].
\end{eqnarray}
\end{widetext}
The notation $\left| \text{max}_{\pm} \right|$ signifies that we choose the sign $+$ or $-$ that maximizes the absolute value of the related determinant.

In panel (c) of FIG. \ref{ctro_90}, we have plotted the overlap distance including the \enquote{V phase} correction introduced in Eq.(\ref{ctro_94}).
We observe that most of the discontinuities detected using Eq. (\ref{ctro_93}) disappear when Eq. (\ref{ctro_94}) is applied. Furthermore, the remaining discontinuities on the left-hand side correspond to those already identified by the density distance (see panel (a)). The differences between the corrected overlap distance and the density distance that persist on the right-hand side of FIG. \ref{ctro_90} are not related to the \enquote{V phase} issue. These differences may arise from various causes. We did not investigate them further, as we were not interested in the physics of the states belonging to the $\mathcal{P}_{20}$ fusion valley.

In addition to its role in evaluation, Eq. (\ref{ctro_94}) can be viewed as a prescription for defining the \enquote{V phase} of states within the same set. Furthermore, we recognize that the observed sign changes do affect the pairing gaps (they reverse the signs of the corresponding ($\Omega, \tau$) sub-blocks). Although pairing gaps are not observables, they could provide an interesting prescription for fixing the \enquote{V phase} of a given state without considering its neighbors.
We were unable to identify the numerical origin of the observed sign changes. They most frequently occurred during basis parameter optimization, but we also observed spontaneous changes (e.g., when starting calculations from an adjacent state without basis parameter optimization).
Nevertheless, we have identified the nature of the \enquote{V phase} appearance which is related to the phases of the different particle number subspaces. Thus, the \enquote{V phase} accounts for particle number symmetry breaking. To understand this, one considers a time-even HFB state state $\ket{\Phi}$ constructed with only four QPs:
\begin{eqnarray}\label{ctro_95}
\ket{\Phi} = \mathcal{N}\xi_2 \bar \xi_2 \xi_1 \bar \xi_1 \ket{0}.
\end{eqnarray}
We have analyzed the coefficients associated with the various particle state components of 
$\ket{\Phi}$. From the commutation relations between the particle annihilation and creation 
operators, it is clear that the coefficient of the state $\ket{0}$ corresponds to products 
of exactly two $V$ elements. Meanwhile, the coefficients of the states 
$\ket{1 \bar 1}$, $\ket{2 \bar 1}$, $\ket{\bar 2 1}$ and $\ket{2 \bar 2}$ are associated 
with products of exactly three $V$ elements. Finally, the coefficient of the state 
$\ket{2 \bar 2 1 \bar 1}$ involves products of exactly four $V$ elements.
These observations imply that changing the sign of $V$ results in a sign change for the 
coefficients of the states $\ket{1 \bar 1}$, $\ket{2 \bar 1}$, $\ket{\bar 2 1}$ and 
$\ket{2 \bar 2}$, whereas the coefficients of the states $\ket{0}$ and 
$\ket{2 \bar 2 1 \bar 1}$ remain unaffected.
This example can be generalized to any arbitrary time-even HFB state. If we denote by $2n$  
the total number of QPs in the time-even HFB state and by $p$ the particle 
number associated with an arbitrary particle-number subspace of the HFB state, the rules to follow for the sign changing in the $V$ matrix are:
\begin{widetext}
 \begin{eqnarray}\label{ctro_96}
{n \; \text{even}} \; \Rightarrow \begin{cases}
                                \text{If} \; p \equiv 0 \; (4), \; \text{the sign of the associated subspace does not change.} \\
                                \text{If} \; p \equiv 2 \; (4), \; \text{the sign of the associated subspace changes.}
                                \end{cases}\nonumber
\end{eqnarray} 
\begin{eqnarray}
{n \; \text{odd}} \; \Rightarrow \begin{cases}
                                \text{If} \; p \equiv 0 \; (4), \; \text{the sign of the associated subspace changes.} \\
                                \text{If} \; p \equiv 2 \; (4), \; \text{the sign of the associated subspace does not change.}
                                \end{cases}\nonumber
\end{eqnarray} 
\end{widetext}
We conclude that the \enquote{V phasis} appearance is a quite subtle detail which has to 
be taken into account when using the overlap constraint methods in the HFB context. Indeed, it is essential to ensure that the evaluated overlaps reflect a physical distance between the states under consideration rather than artifacts arising from phase changes. This is particularly relevant for the deflation-based methods, where we must verify if orthogonality conditions are not satisfied merely due to spurious sign flips.

\section{Evaluation of overlap kernels between an HFB vacuum and a 2QP state}\label{overlap2QP}

In this Appendix, we discuss the formula of the overlap kernels involving 
2QP states, which have been used in the overlap constraints.

\subsection{Relevant contractions}

This subsection is devoted to the evaluation of the contractions relevant to the calculation of Hamiltonian kernels. We will consider the general case directly, which includes two different two-center harmonic-oscillator representations. Indeed, the simpler case involving only a single harmonic-oscillator representation can be trivially obtained from this general formulation. 
To the best of our knowledge, our approach to handling contractions with two different harmonic-oscillator representations is entirely novel. Our method allows for an arbitrary overlap matrix $R$ between the bases $\{ 0 \}$ and  $\{ 1 \}$, whereas the approach presented in \cite{2BrBz} requires an invertible overlap matrix. Although it is always possible to extend the sets $\{ 0 \}$ and $\{ 1 \}$ to satisfy this requirement, our new method does not enable the evaluation of quantities that were previously inaccessible. Nevertheless, we believe that the derivations presented in this section are of significant interest, and that our new formulas are particularly convenient.

Throughout this section, we have consistently taken axial and time-reversal symmetries into account, except in the derivations of the matrices defined below $\rho^{01}$,$\kappa^{01}$ and $\bar \kappa^{01}$, which have been kept as general as possible. Indeed, these derivations form the core of our new evaluation method, and we aimed to present them in full detail.

\subsubsection{Expressions of $\rho^{01}$,$\kappa^{01}$ and $\bar \kappa^{01}$}\label{ctro_128}

The expression of three matrices related to the transition density $\rho^{01}$, and pairing tensor $\kappa^{01}$ and $\bar \kappa^{01}$ are investigated:
\begin{eqnarray}
\displaystyle \rho_{\delta \beta}^{01} = \frac{\bra{\Phi_0}c^+_{0,\beta}c_{1,\delta}\ket{\Phi_1}}{\bra{\Phi_0}\ket{\Phi_1}},
\end{eqnarray}
\begin{eqnarray}
\displaystyle \kappa_{\alpha \beta}^{01} = \frac{\displaystyle \bra{\Phi_0}c_{1,\alpha}c_{1,\beta}\ket{\Phi_1}}{\displaystyle \bra{\Phi_0}\ket{\Phi_1}},
\end{eqnarray}
and
\begin{eqnarray}
\bar \kappa_{\alpha \beta}^{01} = \frac{\bra{\Phi_0}c^+_{0,\alpha}c^+_{0,\beta}\ket{\Phi_1}}{\bra{\Phi_0}\ket{\Phi_1}}.
\end{eqnarray}
We start by first defining the function $f: \mathbb{R} \rightarrow \mathbb{C}$ such that:
\begin{widetext}
\begin{eqnarray}\label{ctro_116}
\displaystyle f(x) = \bra{\Phi_0}\ket{\Phi_1(x)}  =
 \frac{\bra{0}e^{\displaystyle \frac{1}{2}\sum_{kk'}\left(V^{(0)}U^{(0)-1}\right)_{kk'}c_{0,k'} c_{0,k}} e^{\displaystyle \frac{1}{2}\sum_{kk'}\left(RV^{(1)}U^{(1)-1}R^T\right)^*_{kk'}c_{0,k}^+c_{0,k'}^+x_{kk'}} \ket{0}}{\displaystyle \bra{0}\ket{\Phi_1} \bra{\Phi_0}\ket{0}},
\end{eqnarray}
\end{widetext}
where the quantity $R$ denotes the reduced overlap matrix between the harmonic-oscillator representations $\{ 0 \}$ and $\{ 1 \}$. 
Besides, the notation $x_{kk'}$ stands for:
\begin{eqnarray}
\qquad \begin{cases}
       (k,k') = (\alpha,\beta) \; \text{or} \; (\beta,\alpha) \Rightarrow x_{kk'} = x. \\
       (k,k') \ne (\alpha,\beta) \; \text{or} \; (\beta,\alpha) \Rightarrow x_{kk'} = 1.
       \end{cases}
\end{eqnarray}
Differentiating the function $f$ provides:
\begin{eqnarray}\label{ctro_115}
f'(x)= \left( RV^{(1)}U^{(1)-1}R^T \right)^*_{\alpha \beta}
\bra{\Phi_0} c^+_{0,\alpha}c^+_{0,\beta}\ket{\Phi_1(x)}.
\end{eqnarray}
Evaluating the function $f'$ at $x=1$ leads to:
\begin{eqnarray}
f'(1) = \bra{\Phi_0}\ket{\Phi_1}\left( M^{(1)}_R(1) \right)_{\alpha \beta}
\bar \kappa^{01}_{\alpha \beta}.
\end{eqnarray}
with
\begin{eqnarray}
(M^{(1)}_R(x))_{kk'} = \left( RV^{(1)}U^{(1)-1}R^T \right)^*_{kk'}x_{kk'}.
\end{eqnarray}
As $M^{(1)}_R(x)$ is a skew-symmetric matrix for all $x$, we directly obtain a new expression of $f$ using previous results:
\begin{eqnarray}
\displaystyle f(x) =(-1)^{\displaystyle \frac{n(n+1)}{2}}\bra{0}\ket{\Phi_1} \bra{\Phi_0}\ket{0} \text{Pf}\left(-M^{(0)*}\right) \nonumber \\ \times \text{Pf}\left(M_R^{(1)}(x)-(M^{(0)*})^{-1} \right). \; \;
\end{eqnarray}
Then, we use the Pfaffian differentiation formula for a given function $A(x)$:
\begin{eqnarray}
\frac{\text{d}~ \text{Pf}(A(x))}{\text{d}x} = \frac{\text{Pf}(A(x))}{2}\text{Tr}\left[ \left( A(x) \right)^{-1}\frac{\text{d}~A(x)}{\text{d}x} \right].
\end{eqnarray}
Replacing the function $A(x)$ by $M_R^{(1)}(x)-(M^{(0)*})^{-1}$ gives:
\begin{eqnarray}\label{ctro_117}
\left(\frac{\text{d}A}{\text{d}x}(x) \right)_{kk'} = \left[ \delta_{(k,k') = (\alpha \beta)} - \delta_{(k,k') = (\beta \alpha)} \right] \nonumber \\ \times  \left(RV^{(1)}U^{(1)-1}R^T \right)^*_{\alpha \beta}.
\end{eqnarray}
Then, Eq.(\ref{ctro_117}) directly implies:
\begin{eqnarray}\label{ctro_118}
\text{Tr}\left[ A(x)^{-1}\frac{\text{d}A}{\text{d}x}(x) \right] = \notag \qquad \qquad \qquad \\ -2 A^{-1}_{\alpha \beta}(x)\left( RV^{(1)}U^{(1)-1}R^T \right)^*_{\alpha \beta}.
\end{eqnarray}
Eq.(\ref{ctro_118}) is useful to obtain an alternative expression of $f'(1)$:
\begin{eqnarray}
f'(1) = \bra{\Phi_0}\ket{\Phi_1}\left( RV^{(1)}U^{(1)-1}R^T \right)^*_{\alpha \beta} \qquad \qquad \nonumber \\ \times \left[ \left(RV^{(1)}U^{(1)-1}R^T \right)^* - \left(V^{(0)}U^{(0)-1}\right)^{-1} \right]_{\alpha \beta}^{-1}. \;
\end{eqnarray}
Finally, we deduce the expression of $\bar \kappa^{01}$:
\begin{eqnarray}
\bar \kappa^{01} = \left[ \left( RV^{(1)}U^{(1)-1}R^T \right)^* - \left( V^{(0)}U^{(0)-1} \right)^{-1} \right]^{-1}.
\end{eqnarray}
As $\kappa^{01} = - \bar \kappa^{10*}$, we directly have:
\begin{eqnarray}
 \kappa^{01} = \left[ -R^TV^{(0)}U^{(0)-1}R + \left( V^{(1)*}U^{(1)*-1} \right)^{-1} \right]^{-1}.
\end{eqnarray}

For the transition density matrix $\rho^{01}$, we begin by expressing it using the Thouless theorem:
\begin{eqnarray}\label{ctro_120}
\displaystyle \rho^{01}_{\delta \beta} = \displaystyle \frac{\bra{0}\ket{\Phi_1}}{\bra{\Phi_0}\ket{\Phi_1}}
\times \qquad \qquad \qquad \qquad \\ \bra{\Phi_0}c_{0,\beta}^+ c_{1,\delta} e^{\displaystyle \frac{1}{2}\sum_{kk'}\left(V^{(1)}U^{(1)-1} \right)^*_{kk'}c_{1,k}^+c_{1,k'}^+} \ket{0}.\nonumber
\end{eqnarray}
Then, remarking the following property:
\begin{widetext}
\begin{eqnarray}\label{ctro_119}
e^{\displaystyle -\frac{1}{2}\sum_{kk'}\left( V^{(1)}U^{(1)-1} \right)^*_{kk'}c_{1,k}^+c_{1,k'}^+} c_{1,\delta} e^{\displaystyle \frac{1}{2}\sum_{kk'}\left( V^{(1)}U^{(1)-1} \right)^*_{kk'}c_{1,k}^+c_{1,k'}^+}
 = c_{1,\delta} - \sum_{k}\left( V^{(1)}U^{(1)-1} \right)^*_{k\delta} c_{1,k}^+ ,
\end{eqnarray}
\end{widetext}
and inserting Eq.(\ref{ctro_119}) into Eq.(\ref{ctro_120}), leads to:
\begin{eqnarray}
\displaystyle \rho^{01}_{\delta \beta} =  - \sum_{k}\left( V^{(1)}U^{(1)-1} \right)^*_{k\delta} \frac{\bra{\Phi_0}c_{0,\beta}^+ c_{1,k}^+ \ket{\Phi_1}}{\bra{\Phi_0}\ket{\Phi_1}}.
\end{eqnarray}
Then, we use the full transformation matrix $\bar R$ between the two harmonic-oscillator representations:
\begin{eqnarray}
\displaystyle \rho^{01}_{\delta \beta} =
-  \sum_{k} \left( V^{(1)}U^{(1)-1} \right)^*_{k\delta} \notag \qquad \qquad \\ \times \frac{\displaystyle \bra{\Phi_0}c_{0,\beta}^+ \sum_l^{+\infty} \bar R_{lk} c_{0,l}^+ \ket{\Phi_1}}{\bra{\Phi_0}\ket{\Phi_1}}.
\end{eqnarray}
All the contributions of the sum labeled by $l \notin \{0\}$ vanish. Therefore:
\begin{eqnarray}
\rho^{01}_{\delta \beta} =
- \sum_l   \sum_{k}R_{lk}\left( V^{(1)}U^{(1)-1} \right)^*_{k\delta} \notag \qquad  \\ \times \frac{\bra{\Phi_0}c_{0,\beta}^+  c_{0,l}^+ \ket{\Phi_1}}{\bra{\Phi_0}\ket{\Phi_1}}.
\end{eqnarray}
We identify the matrix $\bar \kappa^{01}$:
\begin{eqnarray}
\rho^{01}_{\delta \beta} = \sum_l   \sum_{k}R_{lk}\left( V^{(1)}U^{(1)-1} \right)^*_{k\delta} \bar \kappa^{01}_{\beta l}.
\end{eqnarray}
Thus, the transition matrix $\rho^{01}$ has for final expression:
\begin{eqnarray}\label{ctro_121}
\rho^{01}= \left( V^{(1)}U^{(1)-1} \right)^* R^T \bar \kappa^{01},
\end{eqnarray}
or more explicitly:
\begin{eqnarray}
\rho^{01}= \left( V^{(1)}U^{(1)-1} \right)^* R^T \times \qquad \qquad \nonumber \\ \left[ \left( RV^{(1)}U^{(1)-1}R^T \right)^* - \left( V^{(0)}U^{(0)-1} \right)^{-1} \right]^{-1}.
\end{eqnarray}

\subsubsection{Connection with the formulas of L.M. Robledo}

In the following, we show the link between our new formulas and the ones given in Ref.\cite{2BrBz}. We start with $\bar \kappa^{01}$ with its expression given in \cite{2BrBz}:

\begin{eqnarray}\label{ctro_122}
\bar \kappa^{01} =  \left( R^T \right)^{-1} U^{(1)^*} \qquad \qquad \qquad\\ \times \left[ V^{(0)T} R V^{(1)*} + U^{(0)T} \left( R^T \right)^{-1} U^{(1)*} \right]^{-1} V^{(0)T}. \nonumber
\end{eqnarray}
Eq.(\ref{ctro_122}) includes the correction to the sign mistake found in Eq.(19) of \cite{2BrBz}. As $R$ is a square invertible matrix, we write:
\begin{eqnarray}
\bar \kappa^{01} =  \left[ R V^{(1)*}\left( U^{(1)*} \right)^{-1}R^T + \left( V^{(0)T} \right)^{-1}U^{(0)T} \right]^{-1}. \nonumber
\end{eqnarray}
Finally, the use of the skew-symmetry property of the matrix $V^{(0)}U^{(0)-1}$ leads to:
\begin{eqnarray}
\bar \kappa^{01} =  \left[ R V^{(1)*}\left( U^{(1)*} \right)^{-1}R^T - \left( V^{(0)}U^{(0)-1} \right)^{-1} \right]^{-1}, \nonumber
\end{eqnarray}
which recover our formula. \\

Now, we consider the definition of the matrix $\kappa^{01}$ given in Ref. \cite{2BrBz}:
\begin{eqnarray}
\kappa^{01} =   V^{(1)^*} \left[ V^{(0)T} R V^{(1)*} + U^{(0)T} \left( R^T \right)^{-1} U^{(1)*} \right]^{-1} \nonumber \\ \times U^{(0)T} \left( R^{T} \right)^{-1}. \qquad \qquad
\end{eqnarray}
As $R$ is a square invertible matrix, we can write:
\begin{eqnarray}
\kappa^{01} =   \left[ R^T \left( U^{(0)T} \right)^{-1}V^{(0)T} R  +U^{(1)*} \left( V^{(1)*} \right)^{-1} \right]^{-1}. \nonumber 
\end{eqnarray}
Finally, using the skew-symmetry of the matrix $V^{(0)}U^{(0)-1}$ leads to our formula for $\kappa^{01}$:
\begin{eqnarray}
\kappa^{01} =   \left[ - R^T V^{(0)}U^{(0)-1} R  + \left( V^{(1)*}U^{(1)*-1} \right)^{-1} \right]^{-1}. \nonumber
\end{eqnarray}

To conclude, we consider the definition of the transition density matrix $\rho^{01}$ given in Ref. \cite{2BrBz}:
\begin{eqnarray}
\rho^{01} =   V^{(1)^*} \left[V^{(0)T} R V^{(1)*} + \right. \qquad \qquad \qquad \qquad \nonumber \\ \left. U^{(0)T} \left( R^T \right)^{-1} U^{(1)*} \right]^{-1} V^{(0)T}. \;
\end{eqnarray}
In that case, the transformation is very straightforward:
\begin{eqnarray}
\rho^{01} =   \left( V^{(1)}U^{(1)-1} \right)^{*} R^{T} \bar \kappa^{01},
\end{eqnarray}
and leads to our new formula.

\subsubsection{Axial and time-reversal case}

In the following, the explicit expressions of the matrices $\rho^{01}$, $\bar \kappa^{01}$, and $\kappa^{01}$ are provided in the axial and time-reversal case. In this particular case, the Bogoliubov matrices $\tilde U^{(0)}$, $\tilde V^{(0)}$ and $\tilde U^{(1)}$, $\tilde V^{(1)}$ associated with the axial and time-reversal HFB states $\ket{\Phi_0}$ and $\ket{\Phi_1}$ have a particular form:
\begin{eqnarray}\label{ctro_123}
\tilde U^{(i)} =
\begin{pmatrix}
       U^{(i)}    & 0 \\  0 &  U^{(i)}
\end{pmatrix},
\end{eqnarray}
and
\begin{eqnarray}\label{ctro_123_bis}
\tilde V^{(i)} =
\begin{pmatrix}
       0 & - V^{(i)} \\  V^{(i)} & 0
\end{pmatrix}.
\end{eqnarray}
Moreover, the overlap matrix $R$ displays also a special form:
\begin{eqnarray}\label{ctro_124}
\tilde R=
\begin{pmatrix}
       R    & 0 \\  0 &  R
\end{pmatrix}.
\end{eqnarray}
Thanks to Eqs.(\ref{ctro_123}), (\ref{ctro_123_bis}), Eq.(\ref{ctro_124}), and to the results found for the general case, we directly obtain:
\begin{eqnarray}
\bar \kappa^{01} = \left[- RV^{(1)}U^{(1)-1}R^T - \left( V^{(0)}U^{(0)-1} \right)^{-1} \right]^{-1}, \; \;
\end{eqnarray}
\begin{eqnarray}
 \kappa^{01} = \left[ R^TV^{(0)}U^{(0)-1}R + \left( V^{(1)}U^{(1)-1} \right)^{-1} \right]^{-1}, \; \;
\end{eqnarray}
and
\begin{eqnarray}
\rho^{01}= V^{(1)}U^{(1)-1} R^T \bar \kappa^{01}.
\end{eqnarray}
It is important to keep in mind that the matrices $\bar \kappa^{01}$ and $\kappa^{01}$ are the reduced matrices whose elements are the $\bar \kappa_{\alpha \bar \beta}^{01}$ and $ \kappa_{\alpha \bar \beta}^{01}$ with $\alpha, \beta >0$. In addition, the elements of the reduced transition matrix $\rho^{01}$ are the $\rho^{01}_{\alpha \beta}$ with $\alpha, \beta >0$.

\subsection{Expressions of $W$,$\bar W$, $Z$ and $\bar Z$}

\noindent In this part, we give the explicit expressions of the following quantities:
\begin{eqnarray}
 W_{\alpha i} = \frac{\bra{\Phi_0}c_{1,\alpha}\xi_{1,i}^+\ket{\Phi_1}}{\bra{\Phi_0}\ket{\Phi_1}},
\end{eqnarray}
\begin{eqnarray}
 \bar W_{j \beta} = \frac{\bra{\Phi_0}\xi_{0,j}c^+_{0,\beta}\ket{\Phi_1}}{\bra{\Phi_0}\ket{\Phi_1}},
\end{eqnarray}
and
\begin{eqnarray}
 Z_{\alpha \bar i} = \frac{\bra{\Phi_0}c^+_{0,\alpha}\bar \xi_{1,i}^+\ket{\Phi_1}}{\bra{\Phi_0}\ket{\Phi_1}},
\end{eqnarray}
\begin{eqnarray}
 \bar Z_{j \bar \beta} = \frac{\bra{\Phi_0}\xi_{0,j} \bar c_{1,\beta} \ket{\Phi_1}}{\bra{\Phi_0}\ket{\Phi_1}},
\end{eqnarray}
where the sets $\{c_{n, \alpha}, ~c_{n, \alpha}^+  \}$ and $\{\xi_{n, i}, ~\xi_{n, i}^+  \}$ represent particle and QP annihilation and creation operators, respectively. They are labeled by the representation $n$ and the particle state number $\alpha$ or QP state number $i$.

All these quantities are easily deduced from the matrices $\bar \kappa^{01}$, $\kappa^{01}$, and $\rho^{01}$. Therefore, we only present the derivations concerning $W$ explicitly and we simply give the formulas associated with the other quantities. 
To evaluate $W$, we start by expanding the QP operator $\xi^+$ in terms of the particle operators $\{c_{n, \alpha}, ~c_{n, \alpha}^+  \}$:
\begin{eqnarray}\label{ctro_126}
 W_{\alpha i} = -\sum_l U^{(1)}_{li}\frac{\bra{\Phi_0}c_{1,l}^+c_{1,\alpha}\ket{\Phi_1}}{\bra{\Phi_0}\ket{\Phi_1}}  + \left[ \kappa^{01}V^{(1)} \right]_{\alpha i}. \; \; \; 
\end{eqnarray}
Then, we remark that:
\begin{equation}\label{ctro_125}
\begin{array}{lcl}
\displaystyle \frac{\bra{\Phi_0}c_{1,l}^+c_{1,\alpha}\ket{\Phi_1}}{\bra{\Phi_0}\ket{\Phi_1}} &=& \displaystyle \sum_m^{+\infty}\bar R_{ml} \frac{\bra{\Phi_0}c_{0,m}^+ c_{1,\alpha}\ket{\Phi_1}}{\bra{\Phi_0}\ket{\Phi_1}}, \\ &=& \displaystyle \sum_m R_{ml} \frac{\bra{\Phi_0}c_{0,m}^+ c_{1,\alpha}\ket{\Phi_1}}{\bra{\Phi_0}\ket{\Phi_1}}, \\ &=& \displaystyle \left[ \rho^{01}R \right]_{\alpha l}.
\end{array}
\end{equation}
Inserting Eq.(\ref{ctro_125}) into Eq.(\ref{ctro_126}) finally leads to:
\begin{eqnarray}
 W = - \rho^{01}RU^{(1)} + \kappa^{01}V^{(1)}.
\end{eqnarray}

With the same method, we deduce the expression of $\bar W$, $Z$ and $\bar Z$:
\begin{eqnarray}
 \bar W = - U^{(0)T}R\rho^{01} - V^{(0)T}\bar \kappa^{01},
\end{eqnarray}
\begin{eqnarray}
 Z = \bar \kappa^{01}RU^{(1)}- \rho^{01T}V^{(1)},
\end{eqnarray}
and
\begin{eqnarray}
 \bar Z = U^{(0)T} R \kappa^{01} + V^{(0)T}\rho^{01T}.
\end{eqnarray}

\subsection{Expressions of $Y$,$T$ and $S$}

\noindent The use of 2QP excited states introduces the matrices $Y$, $T$, and $S$ defined by:
\begin{eqnarray}
 Y_{j \bar j'} = \frac{\bra{\Phi_0}\xi_{0,j}\bar \xi_{0,j'}\ket{\Phi_1}}{\bra{\Phi_0}\ket{\Phi_1}},
 \end{eqnarray}
\begin{eqnarray}
 T_{i \bar i'} = \frac{\bra{\Phi_0}\xi^+_{1,i}\bar \xi^+_{1,i'}\ket{\Phi_1}}{\bra{\Phi_0}\ket{\Phi_1}},
\end{eqnarray}
and
\begin{eqnarray}
 S_{j i} = \frac{\bra{\Phi_0}\xi_{0,j} \xi^{+}_{1,i}\ket{\Phi_1}}{\bra{\Phi_0}\ket{\Phi_1}}.
\end{eqnarray}

\noindent All these quantities are linked to the $W$, $\bar W$, $Z$, and $\bar Z$ matrices. Indeed, it is easy to show that:
\begin{eqnarray}
 Y = \bar Z R^T U^{(0)} - \bar W V^{(0)},
\end{eqnarray}
\begin{eqnarray}
 T = U^{(1)T}R^T Z + V^{(1)T}W,
\end{eqnarray}
and
\begin{eqnarray}
 S = \bar W U^{(1)} + \bar Z V^{(1)}.
\end{eqnarray}

\section{Hamiltonian kernels between two different HFB states}\label{hamiltonkernel}

This Appendix is dedicated to the evaluation of Hamiltonian kernels between two different HFB states $\ket{\Phi_0}$ and $\ket{\Phi_1}$. These expressions are valid for both the adiabatic states built with the \enquote{Link} and \enquote{Drop} procedures and the variational excited states created with the \enquote{Continuous Deflation} (see second article of the trilogy \cite{trilogy2}) method as they are all HFB vacua. 

We start by defining the Hamiltonian $\hat H^{01}$:
\begin{eqnarray}\label{ctro_1}
\hat H^{01} = \sum_{\alpha \beta} t_{\alpha \beta}c_{0,\alpha}^+ c_{1,\beta} + \frac{1}{4}\sum_{\alpha \beta \gamma \delta} v_{\alpha \beta \gamma \delta}^{(a)} c_{0,\alpha}^+ c_{0,\beta}^+ c_{1,\delta} c_{1,\gamma}, \nonumber 
\end{eqnarray}
where the indices 0 and 1 stand for the two possibly different harmonic-oscillator bases related to the HFB states $\ket{\Phi_0}$ and $\ket{\Phi_1}$. We note $E^{01}_{00}=\bra{\Phi_0}\hat H^{01} \ket{\Phi_1}$ the Hamiltonian kernel associated with the Hamiltonian $\hat H^{01}$, $\ket{\Phi_0}$ and $\ket{\Phi_1}$. Its expression is given by:
\begin{equation}\label{ctro_2}
\begin{array}{lcl}
\displaystyle E^{01}_{00} &=& \displaystyle \bra{\Phi_0}\sum_{\alpha \beta} t_{\alpha \beta}c_{0,\alpha}^+ c_{1,\beta}\ket{\Phi_1} \\ \displaystyle &+& \displaystyle \bra{\Phi_0}\frac{1}{4}\sum_{\alpha \beta \gamma \delta} v_{\alpha \beta \gamma \delta}^{(a)} c_{0,\alpha}^+ c_{0,\beta}^+ c_{1,\delta} c_{1,\gamma}\ket{\Phi_1},
 \end{array}
\end{equation}
where the kinetic and the two-body interaction parts will be noted $E^{01}_{00}(t)$ 
and $E^{01}_{00}(v)$ in the following, respectively. 

\noindent The kinetic part of the Hamiltonian kernel is equal to:
\begin{eqnarray}\label{ctro_3}
E^{01}_{00}(t) = \bra{\Phi_0}\ket{\Phi_1} \sum_{\alpha \beta} t_{\alpha \beta} \rho^{01}_{\beta \alpha}.
\end{eqnarray}
Using the time-reversal properties of the transition density matrix $\rho^{01}_{\beta \alpha}$ in Eq.(\ref{ctro_3}), we obtain:
\begin{eqnarray}
E^{01}_{00}(t) = 2 \bra{\Phi_0}\ket{\Phi_1} \sum_{\alpha \beta>} t_{\alpha \beta} (-1)^{s_\alpha - s_\beta} \rho^{01}_{\beta \alpha},
\end{eqnarray}
where the numbers $s_\alpha$ and $s_\beta$ characterize the phase associated with the intrinsic spin value of the HO representation. As the matrix element $t_{\alpha \beta}$ imposes $s_\alpha = s_\beta$, this part of the Hamiltonian kernel reduces to:
\begin{eqnarray}\label{ctro_4}
E^{01}_{00}(t) = 2 \bra{\Phi_0}\ket{\Phi_1} \sum_{\alpha \beta>} t_{\alpha \beta} \rho^{01}_{\beta \alpha}.
\end{eqnarray}
The expression given in Eq.(\ref{ctro_4}) stands for the kinetic contribution to the collective mean-field. Its detailed expression is given in Appendix K of Ref. \cite{TPaul}. To conclude, it is easy to show that the following relation holds:
\begin{eqnarray}
E^{01}_{00}(t) = E^{10}_{00}(t).
\end{eqnarray}

The two-body part of the Hamiltonian kernel has for formal expression:
\begin{eqnarray}
E^{01}_{00}(v) = \frac{1}{4}\sum_{\alpha \beta \gamma \delta} v_{\alpha \beta \gamma \delta}^{(a)} \bra{\Phi_0} c_{0,\alpha}^+ c_{0,\beta}^+ c_{1,\delta} c_{1,\gamma} \ket{\Phi_1}.
\end{eqnarray}
Then, we use the generalized Wick theorem in such a way that:
\begin{eqnarray}\label{ctro_5}
E^{01}_{00}(v) = \frac{1}{4}\bra{\Phi_0}\ket{\Phi_1}\sum_{\alpha \beta \gamma \delta} v_{\alpha \beta \gamma \delta}^{(a)} \qquad \qquad \qquad \\ \times \left[ \frac{\bra{\Phi_0} c_{0,\alpha}^+ c_{0,\beta}^+ \ket{\Phi_1}}{\bra{\Phi_0}\ket{\Phi_1}}
\frac{\bra{\Phi_0} c_{1,\delta} c_{1,\gamma} \ket{\Phi_1}}{\bra{\Phi_0}\ket{\Phi_1}} \right. \\ \nonumber \left. + \frac{\bra{\Phi_0} c_{0,\alpha}^+ c_{1,\gamma} \ket{\Phi_1}}{\bra{\Phi_0}\ket{\Phi_1}} \frac{\bra{\Phi_0} c_{0,\beta}^+ c_{1,\delta} \ket{\Phi_1}}{\bra{\Phi_0}\ket{\Phi_1}} \right. \\ \left.
- \frac{\bra{\Phi_0} c_{0,\alpha}^+ c_{1,\delta} \ket{\Phi_1}}{\bra{\Phi_0}\ket{\Phi_1}}\frac{\bra{\Phi_0}c_{0,\beta}^+c_{1,\gamma}\ket{\Phi_1}}{\bra{\Phi_0}\ket{\Phi_1}} \right],
\end{eqnarray}
where $v_{\alpha \beta \gamma \delta}^{(a)}$ represents the anti-symmetrized matrix elements of the interaction. Using this exchange property of this anti-symmetrization in Eq.(\ref{ctro_5}) leads to:
\begin{widetext}
\begin{eqnarray}\label{ctro_6}
E^{01}_{00}(v) = \frac{\bra{\Phi_0}\ket{\Phi_1}}{4}\sum_{\alpha \beta \gamma \delta} v_{\alpha \beta \gamma \delta}^{(a)}\frac{\bra{\Phi_0} c_{0,\alpha}^+ c_{0,\beta}^+ \ket{\Phi_1}}{\bra{\Phi_0}\ket{\Phi_1}}
\frac{\bra{\Phi_0} c_{1,\delta} c_{1,\gamma} \ket{\Phi_1}}{\bra{\Phi_0}\ket{\Phi_1}} \qquad \qquad \qquad \qquad \qquad \qquad \\ +
\frac{\bra{\Phi_0}\ket{\Phi_1}}{2}\sum_{\alpha \beta \gamma \delta} v_{\alpha \beta \gamma \delta}^{(a)}\frac{\bra{\Phi_0} c_{0,\alpha}^+ c_{1,\gamma} \ket{\Phi_1}}{\bra{\Phi_0}\ket{\Phi_1}} \frac{\bra{\Phi_0} c_{0,\beta}^+ c_{1,\delta} \ket{\Phi_1}}{\bra{\Phi_0}\ket{\Phi_1}}. \nonumber
\end{eqnarray}
\end{widetext}
Thus:
\begin{eqnarray}
E^{01}_{00}(v) = \frac{\bra{\Phi_0}\ket{\Phi_1}}{2} \left[ \sum_{\alpha \gamma} \bar \Gamma_{\alpha \gamma}\rho^{01}_{\gamma \alpha} -
\sum_{\alpha \beta} \bar \Delta_{\alpha \beta} \bar \kappa_{\alpha \bar \beta}^{01} \right], \; \; 
\end{eqnarray}
where the collective mean- $\bar \Gamma$ and pairing $\bar \Delta$ field have for expression:
\begin{eqnarray}
\bar \Gamma_{\alpha \gamma} = \sum_{\beta \delta} v^{(a)}_{\alpha \beta \gamma \delta} \rho^{01}_{\delta \beta}, 
\end{eqnarray}
\begin{eqnarray}
\bar \Delta_{\alpha \beta} = \frac{1}{2}\sum_{\gamma \delta} \left( -1 \right)^{s_\beta - s_\delta} v^{(a)}_{\alpha \beta \gamma \delta}  \kappa^{01}_{\gamma \bar \delta}.
\end{eqnarray}
Using the time-reversal properties of the collective fields along with those of the matrices $\rho^{01}$ and $\bar \kappa^{01}$, one deduces:
\begin{eqnarray}
E^{01}_{00}(v) = \bra{\Phi_0}\ket{\Phi_1} \left[ \sum_{\alpha \gamma>}\bar \Gamma_{\alpha \gamma}\rho^{01}_{\gamma \alpha} -
\sum_{\alpha \beta>} \bar \Delta_{\alpha \bar \beta} \bar \kappa_{\alpha \bar \beta}^{01} \right]. \; \;
\end{eqnarray}
The detailed expressions of the collective mean- and pairing fields are given in the Appendices E to J of Ref.\cite{TPaul} in the case of the D1S Gogny interaction. To conclude, it is easy to show that the following relation holds:
\begin{eqnarray}
E^{01}_{00}(v) = E^{10}_{00}(v).
\end{eqnarray}

\section{Separation method for the evaluation of fragment proton and neutron distributions}\label{z-separation}

To perform this analysis, we have chosen to use the method introduced in Refs. \cite{zSep,zSepVer}, which relies on separating in space the orthonormal particle basis wave functions $\{ \varphi_k \}$. Specific modifications to the original method have been performed in order to adapt it to the two-center representation. More details can be found in Ref. \cite{TPaul}.

To operate this separation, we first define a $z_{neck}$ abscissa that divides the space into two parts. In practice, there exist various methods for defining $z_{neck}$. For instance, the abscissa $z$ can be chosen as the one that minimizes the local density $\rho(\vec{r})$ along the $z$-axis between both fragments. In this work, the definition already implemented in the HFB3 code has been selected, which consists in choosing the $z$ abscissa that minimizes $Q_{neck}$.
Once $z_{neck}$ is defined, the $\varphi_k$ wave functions can be rewritten as:
\begin{eqnarray}\label{ctwo_163}
\displaystyle \varphi_k (\vec{r}) = \varphi_k(\vec{r}) \delta_{(z < z_{neck})} + \varphi_k(\vec{r}) \delta_{(z \geq z_{neck})}.
\end{eqnarray}
Then, the squared norms of the functions appearing on the right hand side of Eq.(\ref{ctwo_163}) are calculated:
\begin{eqnarray}\label{ctwo_164}\begin{cases}
\displaystyle (c_k^{(l)})^2 = || \varphi_k \delta_{(z < z_{neck})} ||^2
 = \int d \vec{r}_\perp \int_{-\infty}^{z_{neck}}  \varphi^*_k (\vec{r}) \varphi_k (\vec{r}) \\
\displaystyle  (c_k^{(r)})^2 = || \varphi_k \delta_{(z \geq z_{neck})} ||^2
 = \int d \vec{r}_\perp \int^{+\infty}_{z_{neck}}  \varphi^*_k (\vec{r}) \varphi_k (\vec{r})
 \end{cases}
\end{eqnarray}
In practice, we evaluate both $(c_k^{(l)})^2$ and $(c_k^{(r)})^2$ numerically. Thanks to Eq.(\ref{ctwo_164}), the left and right normalized wave functions $\varphi^{(l)}_k$ and $\varphi^{(r)}_k$ associated with $\varphi_k$ are defined:
\begin{eqnarray}
 \label{ctwo_165}\begin{cases}
\displaystyle \varphi^{(l)}_k(\vec{r}) = \frac{1}{c^{(l)}_k} \varphi_k(\vec{r}) \delta_{(z < z_{neck})} \\
\displaystyle \varphi^{(r)}_k(\vec{r}) = \frac{1}{c^{(r)}_k} \varphi_k(\vec{r}) \delta_{(z \geq z_{neck})}
 \end{cases}.
\end{eqnarray}
Using Eq.(\ref{ctwo_165}) in Eq.(\ref{ctwo_163}) leads to:
\begin{eqnarray}
\displaystyle \varphi_k (\vec{r}) = c^{(l)}_k \varphi^{(l)}_k(\vec{r})  + c^{(r)}_k \varphi^{(r)}_k(\vec{r}).
\end{eqnarray}
The left and right sets $\{ \varphi^{(l)}_k \}$ and $\{ \varphi^{(r)}_k \}$ associated with the left and right normalized wave functions are not orthonormal. These sets are ortho-normalized using the same method as for the 2-center representations presented in section \ref{twocenter}.
After this ortho-normalization process, we obtain two orthonormal bases $\{ \tilde \varphi^{(l)}_i \}$ and $\{ \tilde \varphi^{(r)}_j \}$. 
The total orthonormal basis $\{ \tilde \varphi^{(lr)}_\alpha \}$, which is the direct sum of $\{ \tilde \varphi^{(l)}_i \}$ and $\{ \tilde \varphi^{(r)}_j \}$, allows to represent all the wave functions of the initial set $\{ \varphi_k \}$. The matrices $\Theta^{(l)}$, $\Theta^{(r)}$ and $\Theta^{(lr)}$ are called the transformation matrices between these latter bases in such a way that:
\begin{eqnarray}\label{ctwo_171} \begin{cases}
\displaystyle \Theta^{(l)}_{ki} = \int d\vec{r} \varphi^*_k(\vec{r}) \tilde \varphi_i^{(l)}(\vec{r}) \\
\displaystyle \Theta^{(r)}_{kj} = \int d\vec{r} \varphi^*_k(\vec{r}) \tilde \varphi_j^{(r)}(\vec{r})
\end{cases},
\end{eqnarray}
and:
\begin{eqnarray}\label{ctwo_172}
\Theta^{(lr)} = \begin{pmatrix}\Theta^{(l)} & \Theta^{(r)}\end{pmatrix}.
\end{eqnarray}
As the bases under consideration are orthonormal, they can be associated to particle creation and annihilation operators $\{ a^{(l)+}_i \}$,$\{ a^{(l)}_i \}$, $\{ a^{(r)+}_j \}$, $\{ a^{(r)}_j \}$, $\{ a^{(lr)+}_\alpha \}$ and $\{ a^{(lr)}_\alpha \}$. 
The following step is to rewrite the HFB wave functions using these new creation and annihilation operators.

In the code, the canonical particle basis has been chosen as the starting point of the method. This choice has been done not only because it is more convenient to work with, but also because certain QP states associated with small $v_k$ can be eliminated. In practice, only the QP annihilation operators $\eta_k$ associated with $|v_k|> 10^{-4}$ have been kept. These operators read:
\begin{eqnarray}\label{ctwo_170}
 \eta_k = u_k a_k + v_k \bar a_k^+.
\end{eqnarray}
Eq.(\ref{ctwo_170}) can be transformed using the matrices defined in Eq.(\ref{ctwo_171}):
\begin{eqnarray}\label{ctwo_167}
\displaystyle \eta_k = u_k (\sum_i \Theta^{(l)}_{ki} a_i^{(l)} + \sum_j \Theta^{(r)}_{kj} a_i^{(r)}) \quad \quad \quad \quad \nonumber \\  \displaystyle + v_k (\sum_i \Theta^{(l)}_{ki} \bar a_i^{(l)+} + \sum_j \Theta^{(r)}_{kj} \bar a_i^{(r)+}).
\end{eqnarray}
Using Eq.(\ref{ctwo_172}), Eq.(\ref{ctwo_167}) can be written in a more compact form:
\begin{eqnarray}\label{ctwo_166}
 \begin{pmatrix}
  \eta \\ \bar \eta^+
 \end{pmatrix}
 = \begin{pmatrix}
    u & v \\
    - v & u
   \end{pmatrix}
   \begin{pmatrix}
   \Theta^{(lr)} &   0 \\
   0 &   \Theta^{(lr)}
   \end{pmatrix}
   \begin{pmatrix}
  a^{(lr)} \\  \bar a^{(lr)+}
 \end{pmatrix}.
\end{eqnarray}
We would like the transformation displayed in Eq.(\ref{ctwo_166}) to be an HFB transformation. However, in general, it is not the case as the matrix $\Theta^{(lr)}$ is rectangular. Therefore, we complete $\Theta^{(lr)}$ into a square matrix adding a basis of its null space to it. Calling $k'$ the index spanning the additional space, we extend the diagonal matrices $u$ and $v$ setting $u_{k'}=1$ and $v_{k'}=0$. 
Doing so, a new set of \enquote{ghost} QP annihilation operators $\{ \eta_{k'} \}$
has been created that do not change the content of the HFB wave function but enable to consider Eq.(\ref{ctwo_166}) as an HFB transformation. In the following, this operation is assumed to have been made, and $\Theta^{(lr)}$, $u$, and $v$ designed the completed matrices.

Now, the projection of a given HFB wave function $\ket{\Phi}$ onto a specific fragmentation can be achieved. As both isospins are treated independently, they are suppressed in the notations in the following to simplify the derivations. One starts 
by defining the left and right particle number operators $\hat N^{(l)}$ et $\hat N^{(r)}$:
\begin{eqnarray}\begin{cases}\label{ctwo_200}
\displaystyle \hat N^{(l)} = \sum_i a^{(l)+}_i a^{(l)}_i \\
\displaystyle  \hat N^{(r)} = \sum_j a^{(r)+}_j a^{(r)}_j
\end{cases}.
\end{eqnarray}

\noindent Thanks to the operators $\hat N^{(l)}$ and $\hat N^{(r)}$, we can define the left and right particle number projectors $\hat P^{(l)}$ and $\hat P^{(r)}$:

\begin{eqnarray}\begin{cases}
\displaystyle \hat P^{(l)}_{N^{(l)}_0} = \frac{1}{2\pi}\int_0^{2\pi} e^{-i \varphi(\hat N^{(l)} - N^{(l)}_0)}d\varphi  \\
\displaystyle \hat P^{(r)}_{N^{(r)}_0} = \frac{1}{2\pi}\int_0^{2\pi} e^{-i \varphi(\hat N^{(r)} - N^{(r)}_0)}d\varphi
\end{cases}.
\end{eqnarray}
The projector $\hat P^{(l)}_{N^{(l)}_0}$ projects the left part of the HFB wave function onto its subspace associated with the particle number $N^{(l)}_0$, and $\hat P^{(r)}_{N^{(r)}_0}$ projects the right part of the HFB wave function onto its subspace associated with the particle number $N^{(r)}_0$.

\noindent In the following, one notes the true total particle number of the nucleus by $N_0$. The objective is to find the probabilities $Y(N^{(l)}_0,N^{(r)}_0)$ associated with all fragmentations such as $N_0^{(l)} + N^{(r)}_0 = N_0$. Calling $\hat P_{N_0}$ the customary particle number projector:
\begin{eqnarray}\label{ctwo_173}
Y(N^{(l)}_0,N^{(r)}_0) = \left(\frac{\bra{\Phi} \hat P^{(l)}_{N^{(l)}_0} \hat P^{(r)}_{N^{(r)}_0} \ket{\Phi}}{\bra{\Phi} \hat P_{N_0} \ket{\Phi}}\right)^{2}.
\end{eqnarray}
After discretization, the numerator in Eq.(\ref{ctwo_173}) reads as:
\begin{widetext}
\begin{eqnarray}\label{ctwo_174}
\displaystyle \bra{\Phi} \hat P^{(l)}_{N^{(l)}_0} \hat P^{(r)}_{N^{(r)}_0} \ket{\Phi}
= \frac{1}{n_{\varphi_l}}\frac{1}{n_{\varphi_r}}\sum_{\varphi_l=1}^{n_{\varphi_l}}\sum_{\varphi_l=1}^{n_{\varphi_r}}e^{\displaystyle 2 i \frac{(n_{\varphi_l}-\varphi_l)}{n_{\varphi_l}} \pi N^{(l)}_0}  \times\displaystyle e^{\displaystyle 2 i \frac{(n_{\varphi_r}-\varphi_r)}{n_{\varphi_r}} \pi N^{(r)}_0}   \quad \quad \quad \quad \\ \times
\displaystyle \bra{\Phi}e^{\displaystyle - 2 i \frac{(n_{\varphi_l}-\varphi_l)}{n_{\varphi_l}} \pi \hat N^{(l)}} e^{\displaystyle - 2 i \frac{(n_{\varphi_r}-\varphi_r)}{n_{\varphi_r}} \pi \hat N^{(r)}} \ket{\Phi}. \notag
\end{eqnarray}
\end{widetext}

\noindent The difficult part in evaluating Eq.(\ref{ctwo_174}) is to treat the following quantities (factors in the exponentials have been intentionally omitted for simplification purposes):
\begin{eqnarray}
\displaystyle \bra{\Phi} \ket{\Phi(\varphi_l,\varphi_r)} = \bra{\Phi}e^{\displaystyle -i\varphi_l \hat N^{(l)}}e^{\displaystyle -i\varphi_r \hat N^{(r)}} \ket{\Phi}.
\end{eqnarray}
Using Eq.(\ref{ctwo_167}), one can describe how the exponentials operate on the QP annihilation operators $\eta_k$:
\begin{eqnarray}
 \eta_k(\varphi_l,\varphi_r) = e^{\displaystyle -i\varphi_l \hat N^{(l)}}e^{\displaystyle -i\varphi_r \hat N^{(r)}} \eta_k e^{i\varphi_l \hat N^{(l)}}e^{\displaystyle i\varphi_r \hat N^{(r)}}
\end{eqnarray}
that is to say:
\begin{eqnarray}\label{ctwo_168}
\eta_k(\varphi_l,\varphi_r) = u_k (\sum_\alpha \Theta^{(l)}_{ki} e^{i \varphi_l}a_i^{(l)}
+ \sum_\beta \Theta^{(r)}_{kj} e^{i \varphi_r} a_i^{(r)}) \nonumber
\\ + v_k (\sum_\alpha \Theta^{(l)}_{ki} e^{-i \varphi_l}\bar a_i^{(l)+}
+ \sum_\beta \Theta^{(r)}_{kj} e^{i \varphi_r}\bar a_i^{(r)+}). \; \quad
\end{eqnarray}
Thus, the matrix $\Theta^{(lr)}$ can be defined by:
\begin{eqnarray}\label{ctwo_175}
\Theta^{(lr)}(\varphi_l,\varphi_r) = \begin{pmatrix}\Theta^{(l)}e^{i\varphi_l}  & \Theta^{(r)}e^{i\varphi_r} \\
\Theta^{(cl)} e^{i\varphi_l} & \Theta^{(cr)} e^{i\varphi_r}
\end{pmatrix}.
\end{eqnarray}
Here, $\Theta^{(cl)}$ and $\Theta^{(cr)}$ stand for the vectors introduced to complete $\Theta^{(lr)}$ into a square matrix. Thanks to Eq.(\ref{ctwo_175}), the HFB transformation between the QP operators $\{ \eta^+_k(\varphi_l,\varphi_r) \}$ and $\{ \eta_k(\varphi_l,\varphi_r) \}$ and the particle operators $\{ a^{+(lr)}_k \}$ and $\{ a^{(lr)}_k \}$ can be defined as:
\begin{eqnarray}
 \begin{pmatrix}
  \eta(\varphi_l,\varphi_r) \\ \bar \eta^+(\varphi_l,\varphi_r)
 \end{pmatrix}
 = \begin{pmatrix}
    u & v \\
    -v & u
   \end{pmatrix}
   \begin{pmatrix}
   \Theta^{(lr)}(\varphi_l,\varphi_r) &   0 \\
   0 &   \Theta^{(lr)}(\varphi_l,\varphi_r)
   \end{pmatrix} \nonumber \\ \times 
   \begin{pmatrix}
  a^{(lr)} \\  \bar a^{(lr)+}
 \end{pmatrix}. \quad \quad \quad \quad \quad \quad 
\end{eqnarray}
Finally, we write:
\begin{eqnarray}\label{ctwo_176}
\bra{\Phi} \ket{\Phi(\varphi_l,\varphi_r)} = \quad \quad \quad \quad \quad \quad \quad \quad \quad \quad \quad \quad \quad \quad \nonumber \\ 
\text{det}(u \Theta^{(lr)}(\Theta^{(lr)}(\varphi_l,\varphi_r))^T u (\Theta^{(lr)}(\varphi_l,\varphi_r))^* \nonumber \\ + v \Theta^{(lr)} (\Theta^{(lr)}(\varphi_l,\varphi_r))^+ v (\Theta^{(lr)}(\varphi_l,\varphi_r))^*). \quad
\end{eqnarray}
With Eq.(\ref{ctwo_176}), it is straightforward to calculate the fragmentation probabilities $Y(N_0^{(l)},N_0^{(r)})$ defined in Eq.(\ref{ctwo_173}).

\section{QP rotation for disentangling fragments in the adiabatic path}\label{rotation}

In the version of the separation method we propose in this work, we assumes that the $C$ matrix of the Bloch-Messiah theorem (see Eq.\eqref{blochmessiah}) is a relevant rotation for separating the QP operators. In other words, the method consists in separating the QP operators $\{ \eta_k \}$ associated with the so-called \enquote{canonical representation}. The original intuition comes from the fact that it simplifies naturally the QP operator structures. Therefore, we expect this representation to provide us with rather small values of the separation index $s$. The choice of the \enquote{canonical representation} is really efficient ansatz in practice. Moreover, in the \enquote{canonical representation}, the QP annihilation operators and the particle creation operators are labeled by the same index. Thanks to that, it is possible to give an interpretation of the separation results in terms of single particle orbitals, which we find very interesting. Besides, it is this special feature that allows to couple this separation method version with the particle number projection formalism.

In the canonical representation, the sub-matrices $\rho^k$ and $\kappa^k$ labeled by the QP indices read as follows:
\begin{eqnarray}\begin{cases}
 \rho^{k}_{\alpha \beta} = D_{\alpha k} D_{\beta k} v_k^{2}   \\
 \kappa^{k}_{\alpha \beta} = D_{\alpha k} D_{\beta k} v_k u_k
 \end{cases}.
\end{eqnarray}
To decide if a QP operator $\eta_k$ belongs to the left subset $\{ \eta_{k_l} \}$ or to the right one $\{ \eta_{k_l} \}$, we still use their contribution  $\rho^{k}(r)$ to the local density. The latter reads:
\begin{equation}\label{ctwo_186}
\begin{array}{lcl}
\rho^{k}(\vec{r}) &=& \displaystyle v_k^{2} \sum_{\alpha \beta} \delta_{s_\alpha s_\beta} D_{\beta k}D_{\alpha k}  \psi^*_\alpha(\vec{r}) \psi_\beta(\vec{r}) \\ &=&
\displaystyle v_k^{2} |\varphi_k (\vec{r})|^2.
\end{array}
\end{equation}
The wave functions $\varphi_k$ are the wave functions \eqref{ctwo_163} associated with the canonical particle orthonormal basis.
To characterize the squared norm of the left and right parts of the wave functions $\varphi_k$, we use Eq. \eqref{ctwo_164}.
Thus,
\begin{eqnarray}\label{ctwo_187}\begin{cases}
(v^{(l)}_{k})^2= v_k^{2} (c^{(l)}_k)^2\\
(v^{(r)}_{k})^2 = v_k^{2} (c^{(r)}_k)^2
\end{cases}.
\end{eqnarray}
Thanks to Eq.(\ref{ctwo_187}), we can define the $s$ index standing for 
the separation method:
\begin{eqnarray}
s = 2 \sum_k v_k^{2} \text{min}\big[(c^{(l)}_{k})^2,(c^{(r)}_{k})^2\big]  = \sum_k s_k.
\end{eqnarray}
The results shown in FIG. \ref{ctwo_188} are rather frustrating. While the overall decreasing trends align with our expectations, specific anomalies disrupt these otherwise positive results.
This observation led us to investigate the cause of the anomalies. The Bloch-Messiah matrix $D$ (see Eq. \eqref{blochmessiah}) is defined up to rotations within the degenerate subspaces of both $v_k$ and $u_k$, provided that only $\rho$ and $\kappa$ are considered. These rotations occur among the particle states. To preserve the structure of the$U$ and $V$ matrices, the same rotation must also be applied to the $C$ matrix, which defines a rotation among the QP states. Clearly, the entire process preserves the properties of the canonical representation.\\

If we denote the two-particle state wave functions associated with the anomaly as $\varphi_1$ and $\varphi_2$, the previous discussion implies that these wave functions can be replaced by their rotated counterparts, $\tilde \varphi_1(\theta)$ and $\varphi_2(\theta)$, while preserving the properties of the canonical representation. These rotated wave functions can be expressed as:
\begin{eqnarray}
\begin{cases}
                \tilde \varphi_1(\theta) = \text{cos}(\theta)\varphi_1 - \text{sin}(\theta)\varphi_2 \\
                \tilde \varphi_2(\theta) = \text{sin}(\theta)\varphi_1 + \text{cos}(\theta)\varphi_2
                \end{cases}.
\end{eqnarray}
In light of these observations, we have enhanced the separation method by adding a rotation step, the procedure for which is described above:
\begin{itemize}
 \item We search for paired states $\varphi_{1}$ and $\varphi_2$ belonging to the same ($\Omega,\tau$) subspace, such that $s_1 + s_2 > 0.005$  and $|v_1^2 - v^2_2|<10^{-4}$.
 \item We find the rotation angle $\theta \in$[$0,\pi$] such that the value $d_{12} = \tilde s_1 + \tilde s_2$ is minimized.
 \item We add the resulting rotation characterized by the angle $\theta$ to the matrix $D$.
\end{itemize}
As $v_{1}^2$ and $v_{2}^2$ are not exactly equal, the additional rotations may induce a small change in the $\rho$ and $\kappa$ matrices. In practice, we have never found a difference between the initial and the rotated $\rho$ and $\kappa$ matrices whose order of magnitude was greater that $10^{-4}$ (with respect to the Frobenius norm).

\bibstyle{apsrev4-2}

\bibliography{static1_SCIM_final}

\end{document}